\documentclass[manuscript]{aastex62}

\usepackage{txfonts}
\usepackage{latexsym,bm}
\usepackage{lineno}
\usepackage{url}
\usepackage{color}
\usepackage{colortbl}
\usepackage{multirow}

\definecolor{mygray}{gray}{.9}

\usepackage[titletoc,title]{appendix}

\shorttitle{INVARIANCE}
\shortauthors{WANG AND QIN}

\begin{document}
\arraycolsep 0pt

\title{THE INVARIANCE OF THE DIFFUSION COEFFICIENT WITH 
THE ITERATIVE OPERATIONS OF CHARGED PARTICLES' TRANSPORT 
EQUATION}

\correspondingauthor{G. Qin}
\email{qingang@hit.edu.cn}

\author[0000-0002-9586-093X]{J. F. Wang}
\affiliation{School of Science, Harbin Institute of Technology, Shenzhen,
518055, China; qingang@hit.edu.cn}

\author[0000-0002-3437-3716]{G. Qin}
\affiliation{School of Science, Harbin Institute of Technology, Shenzhen,
518055, China; qingang@hit.edu.cn}

\begin{abstract}
The Spatial Parallel Diffusion Coefficient (SPDC)
is one of the important quantities describing energetic charged
particle transport.
There are three different definitions for the SPDC, i.e.,
the Displacement Variance definition
$\kappa_{zz}^{DV}=\lim_{t\rightarrow t_{\infty}}d\sigma^2/(2dt)$,
the Fick's Law definition
$\kappa_{zz}^{FL}=J/X$ with $X=\partial{F}/\partial{z}$,
and the TGK formula definition $\kappa_{zz}^{TGK}=\int_0^{\infty}dt
\langle v_z(t)v_z(0) \rangle$.
For constant mean magnetic field, the three different
definitions of the 
SPDC give the same result.
However,  for focusing field
it is demonstrated that the results of 
the different definitions are not the same.
In this paper, from the Fokker-Planck equation
we find that different methods, e.g.,
the general Fourier expansion and perturbation theory,
can give the different Equations of the
Isotropic Distribution Function (EIDFs).
But it is shown that one EIDF can be 
transformed into another by some
Derivative Iterative Operations (DIOs).
If one definition of the SPDC is invariant for the DIOs,
it is clear that the definition
is also an invariance for different EIDFs, therewith
it is an invariant quantity for the different
Derivation Methods of EIDF (DMEs).
For the focusing field we suggest that
the TGK definition $\kappa_{zz}^{TGK}$
is only the approximate formula,
and the Fick's Law definition $\kappa_{zz}^{FL}$
is not invariant to some DIOs.
However, at least for the special condition, 
in this paper we show that the definition $\kappa_{zz}^{DV}$
is the invariant quantity to the kinds of the DIOs.
Therefore, for spatially varying field the
displacement variance definition
$\kappa_{zz}^{DV}$, rather than 
the Fick's law definition $\kappa_{zz}^{FL}$
and TGK formula definition $\kappa_{zz}^{TGK}$, 
is the most appropriate definition
of the SPDCs.

\end{abstract}


\section{INTRODUCTION}

The behavior of charged energetic particles
in the turbulent magnetic field superposed
on the mean magnetic field
is a long-standing important problem
in astrophysics, e.g.,
cosmic ray physics, astrophysical
plasmas,
space weather research and fusion plasma physics
\citep{Jokipii1966,
Schlickeiser2002, MatthaeusEA2003,
ShalchiEA2005, ShalchiEA2006, Qin2007,
HauffEA2008, Shalchi2009,
Shalchi2010, QinEA2014, WangEA14, QinAWang15}.
For the collisionless limit, 
the complicated interaction between
energetic charged particles 
and turbulent magnetic fields, which
takes the place of two-body Coulomb collisions
as the principal scattering agent,
leads to the stochastic particle motion, so
we have to use the statistical method to
describe the complicated transport
of charged particles
\citep{Earl1974a, Earl1976,
BeeckEA1986, SchlickeiserEA2007, Shalchi2011,
Litvinenko2012a,
Litvinenko2012b, ShalchiEA2013, HeEA2014,
WangEA2017a, WangEA2017b, wq2018}.

The complete models describing the evolution
of the phase-space density are based on
the Fokker-Planck type transport equation
which is deduced from the Master
equation in phase space.
For the analytical treatment one need to simplify
the Fokker-Planck equation and obtain
the spatial diffusion type equations.
Accordingly, the specific forms of the
spatial diffusion coefficient
have to be defined in terms
of different physical processes.
So far, there are three different definitions
for the Spatial Parallel
Diffusion Coefficient (SPDC, the acronyms used
in this paper is listed in Table
\ref{The table of the acronyms}), i.e.,
the Fick's law definition $\kappa_{zz}^{FL}$,
the displacement variance definition
$\kappa_{zz}^{DV}$,
and the TGK formula definition
$\kappa_{zz}^{TGK}$.

In response to the particle concentration
gradient expressed as the change
in concentration due to a change
in position along the $z$ direction,
i.e., $\partial F/\partial z$
with particle concentration $F(z,t)$,
the local rule for movement
of flux $J(z,t)$ is given by
the Fick's first law
\begin{equation}
J(z,t)=-\kappa_{zz}
\frac{\partial F}{\partial z}
\end{equation}
with diffusion coefficient
$\kappa_{zz}$. Therefore, the spatial diffusion
coefficient can be defined as
the flux per unit area per unit time
per concentration gradient
\begin{equation}
\kappa_{zz}^{FL}=-\frac{J(z,t)}
{\partial F/\partial z}.
\label{Fick law definition}
\end{equation}
This is called the Fick's law definition
of the spatial parallel diffusion coefficient
(SPDC) in this paper.
Accordingly, all the terms that can be written as
$\kappa\partial^2{F}/\partial{z^2}$ 
in the equations of the
isotropic distribution function (EIDFs)
are called the diffusion ones, and the
corresponding parameters $\kappa$ 
are the SPDCs with the Fick's
law definition.

For a non-vanishing mean motion of
the charged energetic particles,
the variance $\sigma^2$ of
the particle random displacement
$\Delta z=z-z_0$ is written as
\begin{equation}
\sigma^2=\langle (\Delta z)^2
\rangle-\langle (\Delta z) \rangle^2.
\end{equation}
The parallel diffusion coefficient
defined by the displacement variance
is then given by
\begin{equation}
\kappa_{zz}^{DV}=\frac{1}{2}
\lim_{t\rightarrow t_{\infty}}
\frac{d\sigma^2}{dt}.
\label{displacement variance definition}
\end{equation}
Here, $t_{\infty}\gg t_d$ in which
the particles
approach diffusive behavior
with the characteristic
time-scale $t_d$.
It should be emphasized that the quantity
$t_{\infty}$ is not
infinite because any physical process
lasts for a finite time interval.
In this paper, Equation 
(\ref{displacement variance definition})
is called
the displacement variance
definition of the SPDC.

A useful tool to compute the diffusion coefficient
is the so-called Taylor-Green-Kubo formula
\citep{Taylor1922,Green1951, Kubo1962},
which is the time integral over the velocity
correlation function,
\begin{equation}
\kappa_{zz}^{TGK}=\int_0^{\infty}dt
\langle v_z(t)v_z(0) \rangle,
\end{equation}
where  $v_z$ is the z-component of
particle speed $v$.
This is called the TGK definition
of the SPDC.

When particles transport in the
constant mean magnetic field,
the displacement variance definition
$\kappa_{zz}^{DV}$
is equivalent to the TGK definition
$\kappa_{zz}^{TGK}$
 \citep{Shalchi2009}.
\citet{wq2019} also analytically 
demonstrated that
the variance definition $\kappa_{zz}^{DV}$
is equal to the Fick's law definition
$\kappa_{zz}^{FL}$
for this scenario.
Therefore, for constant field
the three different definitions
of the SPDC are equivalent.

For some scenarios, e.g.,
the position is close to 
the Sun or in the solar corona,
we expect that the non-uniformity of
the mean field has to be considered.
The spatially varying background field
has an impact on the transport of
particles, i.e., the so-called adiabatic focusing
of energetic charged particles appears.
Many researchers investigated the influence
of this effect on the parallel
and perpendicular diffusion
of the charged particles \citep{Roelof1969,
Earl1976,Kunstmann1979,BeeckEA1986,BieberEA1990,
Kota2000, SchlickeiserEA2008, Shalchi2011,
ShalchiEA2013,Litvinenko2012a,Litvinenko2012b,
ShalchiEA2013,DanosEA2013,HeEA2014,WangEA2016,
WangEA2017b, wq2018, wq2019}.
For focusing field, \citet{wq2019}
found the relationship
between the Fick's law definition
$\kappa_{zz}^{FL}$
and the displacement variance definition
$\kappa_{zz}^{DV}$ as
\begin{equation}
\kappa_{zz}^{DV}=\frac{1}{2}
\lim_{t\rightarrow t_{\infty}}
\frac{d\sigma^2}{dt}=\kappa_{zz}^{FL}
-\kappa_z\kappa_{tz},
\label{Wang and Qin formula 2019}
\end{equation}
where $\kappa_z$ and
$\kappa_{tz}$ are the
coefficients of
convection term
and the first-order spatial 
and temporal derivative term
in the EIDF, respectively.
Because $\kappa_z$ and
$\kappa_{tz}$ are not equal to zero
for the spatially varying mean magnetic field,
Equation (\ref{Wang and Qin formula 2019})
demonstrates that
the displacement
variance definition $\kappa_{zz}^{DV}$
and the Fick's law definition $\kappa_{zz}^{FL}$
give different results. In addition, it is suggested
that the TGK definition
$\kappa_{zz}^{TGK}$
is not equivalent to the variance definition
$\kappa_{zz}^{DV}$
for focusing field 
\citep{DanosEA2013,LitvinenkoEA2013,LasuikEA2017}.
Thus, the three definitions of the spatial
parallel diffusion coefficient (SPDCs),
i.e., the Fick's law definition $\kappa_{zz}^{FL}$,
the displacement
variance definition $\kappa_{zz}^{DV}$, 
and the TGK formula definition
$\kappa_{zz}^{TGK}$,
are not equivalent for focusing field.

Starting from the Fokker-Planck equation 
with the simple BGK collision operator 
or relaxation-time
approximation \citep{GombosiEA1993,ZankEA2000}, 
using the Legendre polynomial
expansion method which belongs 
to the general Fourier expansion,
\citet{GombosiEA1993} provided 
a derivation method of EIDF by
obtaining the Equation of
the Isotropic Distribution Function (EIDF) which
contains infinite-order temporal derivative terms.
According to the scaling analysis theory,
the first- and second-order EIDFs
were obtained. It was shown that the first-order
EIDF is the well-known diffusion equation, and
the second-order EIDF belongs to
the telegraph equation.

In this paper, we
investigate whether the SPDC, i.e.,
the Fick's law definition $\kappa_{zz}^{FL}$,
the displacement variance definition 
$\kappa_{zz}^{DV}$, and
the TGK formulation definition $\kappa_{zz}^{TGK}$,
are changed by the iterative operations of
derivation.
The remainder of this
paper is organized as follows:
In Section \ref{Relationship of the general Fourier
expansion method and the perturbation one},
we demonstrate that the EIDFs with different
forms derived by different methods
can be transformed by the derivative iterative operations.
In Section \ref{The differential equation
of the isotropic distribution function},
starting from the modified Fokker-Planck equation
with adiabatic focusing effect,
we derive the differential EIDF.
In Section \ref{The Displacement Variance
definition and the Fick's law one for PzI operation},
we demonstrate the displacement variance
definition $\kappa_{zz}^{DV}$
is an invariant for the
partial derivative over z iterative
operation but
the Fick's law definition $\kappa_{zz}^{FL}$
is not.
In Section 
\ref{The Variance definition and 
the Fick's law one for PtzI operation},
we explore the change of the displacement
variance definition $\kappa_{zz}^{DV}$
for the
partial derivative over t and z iterative
operation.
In Section \ref{The Variance definition and
the Fick's law one for PtI operation},
we investigate the change of the
displacement variance definition
$\kappa_{zz}^{DV}$
for the
partial derivative over t iterative
operation.
In Section \ref{TGK definition FOR FOCUSING FIELD},
we show the TGK formulation is an approximate formula
for the focusing field.
In Section \ref{EVALUATING THE displacement
VARIANCE DEFINITION}, we evaluate
the parallel diffusion coefficients
$\kappa_{zz}^{DV}$.
We conclude and summarize
our results in Section \ref{SUMMARY AND CONCLUSION}.

\section{The Derivative Iterative Operations}
\label{Relationship of the general Fourier
expansion method and the perturbation one}

\subsection{The derivation methods of DIEF}

There are different methods to derive 
EIDFs indicated as the Derivation 
Methods of EIDF (DMEs).

\subsubsection{The first- and second-order 
EIDFs obtained by Gombosi et al (1993)}
\label{The first and second order isotropic
distribution function equations obtained by
Gombosi et al (1993)}
In this subsection we briefly introduce the
DME provided by \citet{GombosiEA1993} because
it is the basis of this section.
The propagation of energetic charged particles
in magnetic turbulence superposed on the
large scale magnetic field
is described by the Fokker-Planck equation 
with the simple BGK collision operator 
or relaxation-time
approximation \citep{GombosiEA1993,ZankEA2000}
\begin{eqnarray}
\frac{\partial{f(z,\mu,t)}}{\partial{t}}+v\mu
\frac{\partial{f(z,\mu,t)}}{\partial{z}}=
-\frac{f(z,\mu,t)-F(z,t)}{\tau},
\label{Boltzmann equation}
\end{eqnarray}
which describes the particle transport
in the phase space.
Here, $f(z,\mu,t)$ is the distribution
function of energetic charged particles,
$t$ is time, $z$
is the distance
along the background
magnetic field,
$F(z,t)$ is the isotropic part of the
distribution function $f(z,\mu,t)$,
$\tau$ is the pitch-angle scattering
 mean free time,
$v$ is particle velocity which is conserved,
$\mu$ is the pitch-angle cosine.

According to scale analysis theory,
by setting the following relation of the terms
\begin{eqnarray}
\frac{\partial{F}}{\partial{t}}\sim
\frac{\partial^2{F}}{\partial{z^2}}\sim O(1),
\label{order of the terms}
\end{eqnarray}
\citet{GombosiEA1993} obtained the first-
and second-order EIDFs
\begin{eqnarray}
&&\frac{\partial{F}}{\partial{t}}
-\frac{v\lambda}{3}\frac{\partial^2{F}}{\partial{z^2}}=0,
\label{1st eq in Gombaosi}\\
&&\frac{\partial{F}}{\partial{t}}+\frac{\tau}{5}
\frac{\partial^2{F}}{\partial{t^2}}
-\frac{v\lambda}{3}\frac{\partial^2{F}}{\partial{z^2}}=0,
\label{2nd eq in Gombaosi}
\end{eqnarray}
which are Equation (28) and (29) in
\citet{GombosiEA1993}, respectively.
Obviously, the first-order equation is
the well-known diffusion equation, and
the second-order equation belongs to
the telegraph equation since it
contains the second-order temporal
derivative term $\partial^2{F}/\partial{t^2}$.

\subsubsection{The first- and second-order 
	EIDFs obtained by the method of Wang and Qin (2018)}
\label{The governing equation obtained by
using the method in Wang and Qin (2018)}

Here, we also start from the Fokker-Planck equation
(see Equation (\ref{Boltzmann equation}))
to provide a new DME
by employing the method in \citet{wq2018} 
which belongs to the perturbation theory.

If the gyrotropic
cosmic-ray phase space
density $f(z, \mu, t)$ due to dominating
pitch-angle diffusion adjusts very quickly to
a quasi-equilibrium through pitch-angle diffusion,
the gyrotropic cosmic ray phase space density
can be divided into
its average $F(z,t)$ and the
anisotropic part $g(z,\mu,t)$
\citep[see, e.g.,][]
{SchlickeiserEA2007,
SchlickeiserEA2008, HeEA2014,
WangEA2017b, wq2018, wq2019}
\begin{equation}
f(z, \mu, t)=F(z, t)+g(z, \mu, t)
\label{f=F+g}
\end{equation}
with
\begin{equation}
F(z, t)=\frac{1}{2}\int_{-1}^1 d\mu f(z, \mu, t)
\label{F and f}
\end{equation}
and
\begin{equation}
\int_{-1}^1 d\mu g(z, \mu, t)=0.
\label{integrate g over mu=0}
\end{equation}

Inserting Equation (\ref{f=F+g}) into
Equation (\ref{Boltzmann equation}) gives
\begin{eqnarray}
\frac{\partial{F}}{\partial{t}}+
\frac{\partial g}{\partial{t}}
+v\mu \frac{\partial{F}}{\partial{z}}
+v\mu \frac{\partial{g}}{\partial{z}}=
-\frac{g}{\tau}.
\label{Boltzmann equation with F and g}
\end{eqnarray}
From Equation (\ref{Boltzmann equation with F and g}) 
the formula of the anisotropic
distribution function can be found
\begin{eqnarray}
g=-\tau\left(\frac{\partial{F}}{\partial{t}}
+\frac{\partial g}{\partial{t}}
+v\mu \frac{\partial{F}}{\partial{z}}+v\mu
\frac{\partial{g}}{\partial{z}}\right).
\label{g of Boltzmann equation}
\end{eqnarray}
Because there exists the anisotropic distribution
function $g(z,\mu,t)$ on the right-hand side of
the latter equation,
the formula of $g(z,\mu,t)$ is an iterative
function, actually.

Integrating Equation (\ref{Boltzmann equation})
over $\mu$ from $-1$ to $1$ gives
\begin{eqnarray}
\frac{\partial{F}}{\partial{t}}+\frac{v}{2}
\frac{\partial }{\partial{z}}\int_{-1}^1 d\mu\mu g=0.
\label{F equation with g}
\end{eqnarray}
To proceed, the
integral $v\partial/(2\partial{z})\int_{-1}^1 d\mu\mu g$
has to be obtained.
Using Equation (\ref{g of Boltzmann equation}) 
we can find
\begin{eqnarray}
\frac{v}{2}\frac{\partial }{\partial{z}}\int_{-1}^1 d\mu\mu g
=-\frac{v}{2}\tau\left(\int_{-1}^1d\mu\mu
\frac{\partial^2{g}}{\partial{t}\partial{z}}
+\frac{2v}{3}\frac{\partial^2{F}}{\partial{z^2}}
+v\int_{-1}^1d\mu\mu^2\frac{\partial^2{g}}{\partial{z^2}} \right).
\label{relation-1}
\end{eqnarray}
Similarily, by defining the relation $\lambda=v\tau$
we can obtain the following integral formulas 
from Equation (\ref{g of Boltzmann equation})
\begin{eqnarray}
&&\frac{\lambda}{2}\int_{-1}^1 d\mu\mu
\frac{\partial^2{g}}{\partial{t}\partial{z}}
=-\frac{\lambda^2}{3}\frac{\partial^3{F}}
{\partial{t}\partial{z^2}}
-\frac{\tau\lambda}{2}\int_{-1}^1 d\mu
\mu \frac{\partial^3{g}}{\partial{t^2}\partial{z}}
-\frac{\lambda^2}{2}\int_{-1}^1 d\mu\mu^2
\frac{\partial^3{g}}{\partial{t}\partial{z^2}},\\
&&\frac{v\lambda}{2}\int_{-1}^1 d\mu\mu^2
\frac{\partial^2{g}}{\partial{z^2}}
=-\frac{\lambda^2}{3}\frac{\partial^3{F}}
{\partial{t}\partial{z^2}}
-\frac{\lambda^2}{2}\int_{-1}^1 d\mu\mu^2
\frac{\partial^3{g}}{\partial{t}\partial{z^2}}
-\frac{v\lambda^2}{2}\int_{-1}^1 d\mu\mu^3
\frac{\partial^3{g}}{\partial{z^3}},\\
&&-\frac{\tau\lambda}{2}\int_{-1}^1 d\mu\mu
\frac{\partial^3{g}}{\partial{t^2}\partial{z}}
=\frac{\tau\lambda^2}{3}\frac{\partial^4{F}}
{\partial{t^2}\partial{z^2}}
+\frac{\tau^2\lambda}{2}\int_{-1}^1 d\mu\mu
\frac{\partial^4{g}}{\partial{t^3}\partial{z}}
+\frac{\tau\lambda^2}{2}\int_{-1}^1 d\mu\mu^2
\frac{\partial^4{g}}{\partial{t^2}\partial{z^2}},\\
&&-\lambda^2\int_{-1}^1 d\mu\mu^2\frac{\partial^3{g}}
{\partial{t}\partial{z^2}}
=\frac{2\tau\lambda^2}{3}\frac{\partial^4{F}}
{\partial{t^2}\partial{z^2}}
+\tau\lambda^2\int_{-1}^1 d\mu\mu^2 \frac{\partial^4{g}}
{\partial{t^2}\partial{z^2}}
+\lambda^3\int_{-1}^1 d\mu\mu^3 \frac{\partial^4{g}}
{\partial{t}\partial{z^3}},\\
&&-\frac{v\lambda^2}{2}\int_{-1}^1 d\mu\mu^3
\frac{\partial^3{g}}{\partial{z^3}}
=\frac{v\lambda^3}{5}\frac{\partial^4{F}}{\partial{z^4}}
+\frac{\lambda^3}{2}\int_{-1}^1 d\mu\mu^3
\frac{\partial^4{g}}{\partial{t}\partial{z^3}}
+\frac{v\lambda^3}{2}\int_{-1}^1 d\mu\mu^4
\frac{\partial^4{g}}{\partial{z^4}}.
\label{relation-2}
\end{eqnarray}
Combining Equations (\ref{F equation with g})
and (\ref{relation-1})-(\ref{relation-2})
yields
\begin{eqnarray}
\frac{\partial{F}}{\partial{t}}
=\frac{v\lambda}{3}
\frac{\partial^2{F}}{\partial{z^2}}
-\frac{2\lambda^2}{3}\frac{\partial^3{F}}
{\partial{t}\partial{z^2}}
+\tau\lambda^2\frac{\partial^4{F}}
{\partial{t^2}\partial{z^2}}
+\frac{v\lambda^3}{5}\frac{\partial^4{F}}
{\partial{z^4}}+H(g)
\label{F equation with g-2}
\end{eqnarray}
with
\begin{eqnarray}
H(g)\equiv&&\frac{\lambda^3}{2}\int_{-1}^1 d\mu\mu^3
\frac{\partial^4{g}}{\partial{t}\partial{z^3}}
+\frac{v\lambda^3}{2}\int_{-1}^1 d\mu\mu^4
\frac{\partial^4{g}}{\partial{z^4}}\nonumber
+\tau\lambda^2\int_{-1}^1 d\mu\mu^2
\frac{\partial^4{g}}{\partial{t^2}\partial{z^2}}
+\lambda^3\int_{-1}^1 d\mu\mu^3
\frac{\partial^4{g}}{\partial{t}\partial{z^3}}\\
&&+\frac{\tau^2\lambda}{2}\int_{-1}^1 d\mu\mu
\frac{\partial^4{g}}{\partial{t^3}\partial{z}}
+\frac{\tau\lambda^2}{2}\int_{-1}^1 d\mu\mu^2
\frac{\partial^4{g}}{\partial{t^2}\partial{z^2}}
\end{eqnarray}
Analogous to Equations
(\ref{relation-1})-(\ref{relation-2}),
the integrals in the latter equation
can also
be derived by employing
the anisotropic distribution function $g(z,\mu,t)$
(see Equation (\ref{g of Boltzmann equation})).
By inserting Equations (\ref{g of Boltzmann equation})
into the right-hand side of Equation
(\ref{g of Boltzmann equation})
again and again,
we find that the anisotropic distribution function
$g(\mu,z,t)$ can be written as the
series of the derivative terms of the
isotropic distribution function $F(z,t)$
\begin{equation}
g(\mu,t)=\sum_{m, n}
\Theta_{m,n}(\mu)
\frac{\partial^{m+n}{}}
{\partial {t^m}
\partial{z}^n}  F.
\label{expanded g of Boltzmann equation}
\end{equation}
Here, $\Theta_{m,n}(\mu)$ is the corresponding
coefficient for the derivative term
$\partial^{m+n}{F}/(\partial {t^m}\partial{z}^n)$
with $m,n=0,1,2,3,\cdots$ except for $m=n=0$. Therefore,
combining Equations (\ref{F equation with g-2})
and (\ref{expanded g of Boltzmann equation}) can give
the equation of the isotropic distribution
function (EIDF)
\begin{equation}
\frac{\partial{F}}{\partial{t}}=\sum_{m, n}\Xi_{m,n}
\frac{\partial^{m+n}{}}
{\partial {t^m}\partial{z}^n}  F,
\label{the isotropic distribution equation
from Boltzmann equation}
\end{equation}
where $\Xi_{m,n}$ is the coefficient of
the derivative term
$\partial^{m+n}{F}/(\partial {t^m}\partial{z}^n)$ with
$m,n=0,1,2,3,\cdots$ but except for $m=n=0$.

In order to obtain the first- and second-order
EIDFs according to relation (\ref{order of the terms}),
the form of Equation (\ref{F equation with g-2})
is enough and the terms in $H(g)$
do not need to be expanded anymore as done in Equations
(\ref{relation-1})-(\ref{relation-2}).
From Equation (\ref{F equation with g-2}),
considering the relation (\ref{order of the terms})
we can easily find that
the first-order equation is identical with
Equation (\ref{1st eq in Gombaosi}).
The second-order equation can be obtained as
\begin{eqnarray}
\frac{\partial{F}}{\partial{t}}=&&
\frac{v\lambda}{3}\frac{\partial^2{F}}{\partial{z^2}}
-\frac{2\lambda^2}{3}\frac{\partial^3{F}}
{\partial{t}\partial{z^2}}
+\frac{v\lambda^3}{5}\frac{\partial^4{F}}{\partial{z^4}},
\label{2nd eq}
\end{eqnarray}
which is different from the second-order equation in
\citet{GombosiEA1993} (see Equation
(\ref{2nd eq in Gombaosi})
in subsection
\ref{The first and second order isotropic distribution
function equations obtained by Gombosi et al (1993)}).
It is obvious that the higher-order governing
equations of $F(z,t)$
from the DME of \citet{wq2018}
also have the different forms
from that obtained in
\citet{GombosiEA1993}.

From the above investigation we find that
the different DMEs give the different EIDFs,
between which there may exist some relationship.

\subsection{The transformation between the
EIDFs}

\subsubsection{The Derivative Iterative Operations}

Multiplying Equation (\ref{2nd eq})
by the differential operator
$\partial^{m+n}{}/(\partial{t^n}\partial{z^m})$
with $n=0,1,2,3,\cdots,
m=0,1,2,3,\cdots$, but $(0,0)\notin (n,m)$, we obtain the
following equation
\begin{eqnarray}
\frac{\partial^{n+m+1}{F}}{\partial{t^{n+1}}
	\partial{z^m}}=
&&\frac{v\lambda}{3}\frac{\partial^{n+m+2}{F}}
{\partial t^{n}\partial{z^{m+2}}}
-\frac{2\lambda^2}{3}\frac{\partial^{n+m+3}{F}}
{\partial{t^{n+1}}\partial{z^{m+2}}}
+\frac{v\lambda^3}{5}\frac{\partial^{n+m+4}{F}}
{\partial t^{n}\partial{z^{m+4}}}.
\label{eq n+m EDIOs}
\end{eqnarray}
By pulling out one term, e.g., 
$\partial^{n+m+2}F/(\partial t^n\partial{z^{m+2}})$,
from Equation (\ref{eq n+m EDIOs}) 
and putting it on the left-hand side of the equal sign,
and then leaving all the other terms on 
the right-hand side, 
\begin{eqnarray}
\frac{\partial^{n+m+2}{F}}{\partial t^{n}
	\partial{z^{m+2}}}
=\frac{3}{v\lambda}\frac{\partial^{n+m+1}{F}}
{\partial{t^{n+1}}\partial{z^m}}
+\frac{2\lambda}{v}\frac{\partial^{n+m+3}{F}}
{\partial{t^{n+1}}\partial{z^{m+2}}}
-\frac{3\lambda^2}{5}\frac{\partial^{n+m+4}{F}}
{\partial t^{n}\partial{z^{m+4}}}.
\label{eq C-2}
\end{eqnarray}
Of course, similarily, we can also obtain
the equations in which
the term
$\partial^{n+m+3}F/(\partial{t^{n+1}}\partial{z^{m+2}})$ or
$\partial^{n+m+4}F/(\partial{t^{n}}\partial{z^{m+4}})$ 
is on the left-hand
side and the rest ones on the right-hand side.
Inserting Equation (\ref{eq C-2}) into Equation
(\ref{the isotropic distribution equation from
Boltzmann equation})
we find another new EIDF.
The above method shows the derivative operation 
with iteration indicated as
the Derivative Iterative Operation (DIO)
which is equivalent substitution without 
any approximation or undetermined constants
introduced, and is
 called the
$(n+m)$th-order Partial derivative over t and z 
Iterative (PtzI)
operation in this paper.
In Equation (\ref{eq n+m EDIOs}),
if $n=0$ and $m=1,2,3,\cdots$, i.e.,
\begin{eqnarray}
\frac{\partial^{m+1}{F}}{\partial{t}\partial{z^m}}=
&&\frac{v\lambda}{3}\frac{\partial^{m+2}{F}}
{\partial{z^{m+2}}}
-\frac{2\lambda^2}{3}\frac{\partial^{m+3}{F}}
{\partial{t}\partial{z^{m+2}}}
+\frac{v\lambda^3}{5}\frac{\partial^{m+4}{F}}
{\partial{z^{m+4}}},
\label{eq B}
\end{eqnarray}
and operating the similar iterative method
as done in the previous paragraphs,
we can also obtain the EIDFs
with different forms.
These manipulations are called
the $m$th-order Partial derivative over z
Iterative (PzI) operation.
Similarily, for $n=1,2,3,\cdots$ and $m=0$,
we obtain the $n$th-order Partial derivative 
over time t
Iterative (PtI) operation.

In the following subsection, we demonstrate
that the DIOs
can convert Equation (\ref{2nd eq}) derived
by the perturbation theory
into Equation (\ref{2nd eq in Gombaosi})
obtained by the general Fourier expansion method.

\subsubsection{The transformation 
between the second-order
EIDFs with different forms}

Firstly, by employing the second-order
PzI operation, i.e.,
taking the two-order derivative of
Equation (\ref{2nd eq}) over z,
we can obtain
\begin{eqnarray}
\frac{\partial^3{F}}{\partial{t}
	\partial{z^2}}
=&&\frac{v\lambda}{3}
\frac{\partial^4{F}}{\partial{z^4}}
-\frac{2\lambda^2}{3}
\frac{\partial^5{F}}{\partial{t}\partial{z^4}}
+\frac{v\lambda^3}{5}
\frac{\partial^6{F}}{\partial{z^6}}.
\end{eqnarray}
The latter equation can be rewritten as
\begin{eqnarray}
\frac{\partial^4{F}}{\partial{z^4}}=&&\frac{3}
{v\lambda}\frac{\partial^3{F}}{\partial{t}\partial{z^2}}
+\frac{2\lambda}{v}\frac{\partial^5{F}}
{\partial{t}\partial{z^4}}
+\frac{3\lambda^2}{5}\frac{\partial^6{F}}{\partial{z^6}}.
\end{eqnarray}
Replacing the term $\partial^4{F}/\partial{z^4}$
in Equation (\ref{2nd eq}) by the latter equation yields
\begin{eqnarray}
\frac{\partial{F}}{\partial{t}}=&&\frac{v\lambda}{3}
\frac{\partial^2{F}}{\partial{z^2}}
-\frac{\lambda^2}{15}\frac{\partial^3{F}}
{\partial{t}\partial{z^2}}.
\label{2nd eq with PzI}
\end{eqnarray}
Here, we only retain the first- and second-order
terms, and neglect the higher-order ones in
the latter equation.

To proceed, we have to employ the first-order 
PtI operation.
Multiplying Equation (\ref{2nd eq}) by the
first-order differential operator 
$\partial{}/\partial{t}$ gives
\begin{eqnarray}
\frac{\partial^2{F}}{\partial{t^2}}
=&&\frac{v\lambda}{3}
\frac{\partial^3{F}}{\partial{t}\partial{z^2}}
-\frac{2\lambda^2}{3}\frac{\partial^4{F}}
{\partial{t^2}\partial{z^2}}
+\frac{v\lambda^3}{5}\frac{\partial^5{F}}
{\partial{t}\partial{z^4}}.
\end{eqnarray}
The latter equation can be rewritten as
\begin{eqnarray}
\frac{\partial^3{F}}{\partial{t}\partial{z^2}}
=\frac{3}{v\lambda}\frac{\partial^2{F}}
{\partial{t^2}}
+\frac{2\lambda}{v}\frac{\partial^4{F}}
{\partial{t^2}\partial{z^2}}
-\frac{3\lambda^2}{5}\frac{\partial^5{F}}
{\partial{t}\partial{z^4}}
\end{eqnarray}
By inserting the latter equation into Equation
(\ref{2nd eq with PzI}) and neglecting the third-
and higher-order terms
we can obtain
\begin{eqnarray}
\frac{\partial{F}}{\partial{t}}
+\frac{\tau}{5}\frac{\partial^2{F}}
{\partial{t^2}}-\frac{v\lambda}{3}
\frac{\partial^2{F}}{\partial{z^2}}=0,
\end{eqnarray}
which is identical with the second-order equation
obtained in \citet{GombosiEA1993}.
It is obvious that the transformation from
Equation (\ref{2nd eq in Gombaosi}) into
Equation (\ref{2nd eq}) can also be found,
but for saving of space we neglect this content.

The above investigation demonstrates that the EIDFs
derived from the DME with 
the perturbation theory method and 
from the DME with
the general Fourier expansion method 
can be converted into each other by the DIOs.
In fact, the EIDFs can give rise to all kinds of 
transformations by the DIOs.
The purpose of this paper is to explore 
whether the SPDC
has the same form for different DMEs 
which give many EIDFs, for the cases with
adiabatic focusing effect and 
the small angle scattering.
If the SPDC is not changed by the DIOs, 
it is also invariant quantity for
different EIDFs since which can be transformed  
by DIOs, consequently, the SPDC is
invariant for different DMEs since 
which give many kinds of EIDFs.

\section{The differential equation
of the isotropic distribution
function}
\label{The differential equation
of the isotropic distribution
function}

\subsection{The differential equation
with isotropic distribution
function $F(z,t)$ and anisotropic distribution
function $g(z,\mu,t)$}
\label{The differential equation
with isotropic distribution
function and anisotropic distribution function}

Because the background plasmas in the solar
system and the interstellar space are highly
conducting,
the large-scale electric fields can be ignored.
Throughout this paper, we only consider
magnetic fluctuations which is superposed
on the background magnetic field.
When charged particles propagate close to the
sun or in the solar corona,
we expect that the non-uniformity of
the mean field is important.
The spatially varying mean magnetic field
gives rise to the so-called particle adiabatic focusing.
For mathematical tractability, we neglect
the perpendicular diffusion and only consider
the parallel one in this paper.
In these circumstances,
the evolution of the two-dimensional
distribution function $f(z,\mu,t)$ of
the energetic charged particles is described by
the following
two-dimensional modified Fokker-Planck equation with
the effects of the along-field
adiabatic focusing effect and the small pitch-angle
scattering
\citep{Kunstmann1979,Litvinenko2012a,Litvinenko2012b,
ShalchiEA2013,HeEA2014,WangEA2016,
WangEA2017b, wq2018, wq2019}
\begin{equation}
\frac{\partial{f}}{\partial{t}}
+ v\mu
\frac{\partial{f}}
{\partial{z}}=
\frac{\partial{}}{\partial{\mu}}
\left[D_{\mu \mu}(\mu)
\frac{\partial{f}}
{\partial{\mu}}-\frac{v}{2L}
(1-\mu^2)f \right],
\label{modified Fokker-Planck equation}
\end{equation}
which satisfies the conservation of
the particle number.
Here, $f=f_0(z,\mu,t)/B_0(z)$ is 
the modified distribution function
of charged energetic particles 
with $f_0(z,\mu,t)$ being the distribution
function of charged energetic particles
and $B_0(z)$ being the background 
magnetic field,
$t$ is time, $z$
is the distance
along the background
magnetic field,
$\mu=v_z /v$ is the pitch-angle
cosine with particle
speed $v$ and
its z-component $v_z$,
$D_{\mu \mu}(\mu)$
is the
pitch-angle diffusion
coefficient which is only
the function of the pitch-angle cosine 
$\mu$ in this paper,
$L(z)=-B_0 (z)/ (dB_0 (z) / dz)$
is the focusing length of the
large-scale magnetic field $B_0(z)$.
For simplification, in this paper
the focusing length is assumed as a constant.
The terms related to source and 
momentum are ignored
in Equation (\ref{modified Fokker-Planck equation}).
In the following, we refer $f$ as 
distribution function for simplicity
purpose. The more complete form of 
the Fokker-Planck equation
can be found in \citet{Schlickeiser2002}.

The distribution function $f(z,\mu,t)$
can be divided into
the isotropic part and the anisotropic
one as Equation (\ref{f=F+g}).
The formulas (\ref{F and f}) and
(\ref{integrate g over mu=0}) are also
satisfied.
Integrating Equation
(\ref{modified Fokker-Planck equation})
over $\mu$ yields
\begin{equation}
\frac{\partial{F}}{\partial{t}}
+ \frac{v}{2}
\frac{\partial{}}{\partial{z}}\int_{-1}^{1}
\mu g d\mu=0.
\label{Equation of F with g}
\end{equation}
In the following, by integrating Equation
(\ref{modified Fokker-Planck equation})
over $\mu$ from $-1$ to $\mu$,
we can find
\begin{eqnarray}
\frac{\partial{F}}
{\partial{t}}(\mu+1)
&&+ \frac{\partial{}}
{\partial{t}}
\int_{-1}^{\mu}d\nu g
+\frac{v(\mu^2-1)}{2}
\frac{\partial{F}}
{\partial{z}}
+v\frac{\partial{}}
{\partial{z}}
\int_{-1}^{\mu}
d\nu \nu g 
=D_{\mu\mu}(\mu)
\frac{\partial{g}}
{\partial{\mu}}
-\frac{v(1-\mu^2)}{2L}F
-\frac{v(1-\mu^2)}{2L}g.
\label{integrate
	from -1 to mu}
\end{eqnarray}
Here, the regularity
$D_{\mu\mu}(\mu=\pm 1)=0$ is used.

Subtracting Equation
(\ref{Equation of F with g})
from
(\ref{integrate from -1 to mu})
gives
\begin{eqnarray}
\frac{\partial{F}}
{\partial{t}}\mu
+ \frac{\partial{}}
{\partial{t}}
\int_{-1}^{\mu}d\nu g
&&+\frac{v(\mu^2-1)}{2}
\frac{\partial{F}}
{\partial{z}}
+v\frac{\partial{}}
{\partial{z}}
\int_{-1}^{\mu}
d\nu \nu g 
-\frac{v}{2}
\frac{\partial{}}
{\partial{z}}\int_{-1}^{1}
\mu g d\mu \nonumber\\
&&=D_{\mu\mu}(\mu)
\frac{\partial{g}}
{\partial{\mu}}
-\frac{v(1-\mu^2)}{2L}F
-\frac{v(1-\mu^2)}{2L}g.
\label{integrate from
	-1 to mu -2}
\end{eqnarray}
After a straightforward
algebra, Equation
(\ref{integrate from -1 to mu -2})
reduces to the following form
as presented in \citet{wq2018}
\begin{equation}
\frac{\partial{}}{\partial{\mu}}
\Bigg\{\Bigg[g(\mu,t)
-L\left(\frac{\partial{F}}
{\partial{z}}
-\frac{F}{L} \right)
\Bigg]e^{-M(\mu,t)}\Bigg\}
=e^{-M(\mu,t)}\Phi(\mu,t)
\label{HS-like}
\end{equation}
with
\begin{equation}
M(\mu)=\frac{v}{2L}
\int_{-1}^{\mu} d\nu
\frac{1-\nu^2}{D_{\nu \nu}(\nu)}
\label{M(mu)}
\end{equation}
and
\begin{eqnarray}
\Phi(\mu,t)=
\frac{1}{D_{\mu\mu}(\mu)}
\Bigg[\left(\frac{\partial{F}}
{\partial{t}}\mu
+\frac{\partial{}}{\partial{t}}
\int_{-1}^{\mu}gd\nu\right)\\
+\frac{v}{2}\frac{\partial{}}
{\partial{z}}
\left(2\int_{-1}^{\mu}d\nu \nu g-
\int_{-1}^{1}d\mu \mu g\right)\Bigg].
\label{Phi}
\end{eqnarray}

As shown in \citet{wq2018},
the anisotropic
distribution function $g(z,\mu,t)$
can be obtained
from Equation (\ref{HS-like})
\begin{equation}
g(z,\mu,t)=L\left(\frac{\partial{F}}
{\partial{z}}
-\frac{F}{L}\right)\left[1-
\frac{2e^{M(\mu)}}{\int_{-1}^{1}
	d\mu e^{M(\mu) }}\right]
+e^{M(\mu)}\left[R(\mu,t)
-\frac{\int_{-1}^{1}d\mu
	e^{M(\mu)}R(\mu,t)}
{\int_{-1}^{1}d\mu
	e^{M(\mu) }}\right]
\label{g}
\end{equation}
with
\begin{eqnarray}
R(\mu,t)&=&\int_{-1}^{\mu} d\nu
e^{-M(\nu)}\Phi(\nu,t).
\label{R(mu)}
\end{eqnarray}
By combining Equations (\ref{Phi}),
(\ref{g}) and  (\ref{R(mu)}),
the iterated function of 
$g(z,\mu,t)$ can be obtained
as following
\begin{eqnarray}
g(z,\mu,t)&=&L\left(\frac{\partial{F}}
{\partial{z}}
-\frac{F}{L}\right)\left[1-
\frac{2e^{M(\mu)}}{\int_{-1}^{1}
d\mu e^{M(\mu) }}\right]
+e^{M(\mu)}\Bigg\{\int_{-1}^{\mu}
d\nu
e^{-M(\nu)}
\frac{1}{D_{\nu\nu}(\nu)}
\Bigg[\left(\frac{\partial{F}}
{\partial{t}}\nu
+\frac{\partial{}}{\partial{t}}
\int_{-1}^{\nu}
g(z,\rho,t)d\rho\right)
\nonumber \\
&&+\frac{v}{2}\frac{\partial{}}
{\partial{z}}
\left(2\int_{-1}^{\nu}d\rho \rho
g(z,\rho,t)-
\int_{-1}^{1}d\mu \mu
g(z,\mu,t)\right)
\Bigg]\nonumber\\
&&-\frac{1}
{\int_{-1}^{1}d\mu
	e^{M(\mu) }}\int_{-1}^{1}d\mu
e^{M(\mu)}\int_{-1}^{\mu} d\nu
e^{-M(\nu)}
\frac{1}{D_{\nu\nu}(\nu)}
\Bigg[\left(\frac{\partial{F}}
{\partial{t}}\nu
+\frac{\partial{}}{\partial{t}}
\int_{-1}^{\nu}g(z,\rho,t)
d\rho\right)\nonumber\\
&&+\frac{v}{2}
\frac{\partial{}}
{\partial{z}}
\left(2\int_{-1}^{\mu}
d\nu \nu g(z,\nu,t)-
\int_{-1}^{1}d\mu \mu
g(z,\mu,t)\right)\Bigg]\Bigg\}.
\label{g2}
\end{eqnarray}

Using Equation (\ref{g2})  
we can obtain the following formula
\begin{eqnarray}
\frac{\partial{}}{\partial{z}}
\int_{-1}^{1}&&d\mu \mu g(z,\mu,t)
=2\left(\frac{\partial{F}}{\partial{z}}
-L\frac{\partial^2{F}}{\partial{z^2}}\right)
\frac{\int_{-1}^{1}d\mu \mu e^{M(\mu)}}
{\int_{-1}^{1}
	d\mu e^{M(\mu) }}\nonumber\\
&&+\int_{-1}^{1}d\mu \mu e^{M(\mu)}
\int_{-1}^{\mu}d\nu
e^{-M(\nu)}
\frac{1}{D_{\nu\nu}(\nu)}
\Bigg[\left(\frac{\partial^2{F}}
{\partial{t}\partial{z}}\nu
+\frac{\partial^2{}}{\partial{t}\partial{z}}
\int_{-1}^{\nu}
g(z,\rho,t)d\rho\right)
\nonumber \\
&&+\frac{v}{2}\frac{\partial^2{}}
{\partial{z^2}}
\left(2\int_{-1}^{\nu}d\rho \rho
g(z,\rho,t)-
\int_{-1}^{1}d\mu \mu
g(z,\mu,t)\right)
\Bigg]\nonumber\\
&&-\frac{\int_{-1}^{1}d\mu \mu e^{M(\mu)}}
{\int_{-1}^{1}d\mu
	e^{M(\mu) }}\int_{-1}^{1}d\mu
e^{M(\mu)}\int_{-1}^{\mu} d\nu
e^{-M(\nu)}
\frac{1}{D_{\nu\nu}(\nu)}
\Bigg[\left(\frac{\partial^2{F}}
{\partial{t}\partial{z}}\nu
+\frac{\partial^2{}}{\partial{t}\partial{z}}
\int_{-1}^{\nu}g(z,\rho,t)
d\rho\right)\nonumber\\
&&+\frac{v}{2}
\frac{\partial^2{}}
{\partial{z^2}}
\left(2\int_{-1}^{\mu}
d\nu \nu g(z,\nu,t)-
\int_{-1}^{1}d\mu \mu
g(z,\mu,t)\right)\Bigg].
\label{int mu g2}
\end{eqnarray}
In order to find the differential equation
with isotropic  
and anisotropic distribution functions,
one inserts Equation (\ref{int mu g2}) 
into Equation
(\ref{Equation of F with g}) to obtain
\begin{eqnarray}
\frac{\partial{F}}{\partial{t}}
&+& \frac{\partial{F}}{\partial{z}}
v\frac{\int_{-1}^{1}d\mu \mu e^{M(\mu)}}
{\int_{-1}^{1}
	d\mu e^{M(\mu) }}=\frac{\partial^2{F}}
{\partial{z^2}}vL\frac{\int_{-1}^{1}d\mu \mu
e^{M(\mu)}}{\int_{-1}^{1}
d\mu e^{M(\mu) }}\nonumber\\
&&-\frac{v}{2}\int_{-1}^{1}d\mu \mu
e^{M(\mu)}\int_{-1}^{\mu}d\nu
e^{-M(\nu)}
\frac{1}{D_{\nu\nu}(\nu)}
\Bigg[\left(\frac{\partial^2{F}}
{\partial{t}\partial{z}}\nu
+\frac{\partial^2{}}{\partial{t}\partial{z}}
\int_{-1}^{\nu}
g(z,\rho,t)d\rho\right)
\nonumber \\
&&+\frac{v}{2}\frac{\partial^2{}}
{\partial{z^2}}
\left(2\int_{-1}^{\nu}d\rho \rho
g(z,\rho,t)-
\int_{-1}^{1}d\mu \mu
g(z,\mu,t)\right)
\Bigg]\nonumber\\
&&+\frac{v}{2}\frac{\int_{-1}^{1}d\mu
\mu e^{M(\mu)}}
{\int_{-1}^{1}d\mu
	e^{M(\mu) }}\int_{-1}^{1}d\mu
e^{M(\mu)}\int_{-1}^{\mu} d\nu
e^{-M(\nu)}
\frac{1}{D_{\nu\nu}(\nu)}
\Bigg[\left(\frac{\partial^2{F}}
{\partial{t}\partial{z}}\nu
+\frac{\partial^2{}}{\partial{t}\partial{z}}
\int_{-1}^{\nu}g(z,\rho,t)
d\rho\right)\nonumber\\
&&+\frac{v}{2}
\frac{\partial^2{}}
{\partial{z^2}}
\left(2\int_{-1}^{\mu}
d\nu \nu g(z,\nu,t)-
\int_{-1}^{1}d\mu \mu
g(z,\mu,t)\right)\Bigg].
\label{Equation of F with g2}
\end{eqnarray}
The latter equation gives the
relationship between
the isotropic distribution
function $F(z,t)$
and the anisotropic distribution
function $g(z,\mu,t)$.

\subsection{The differential equation of
the isotropic distribution
function $F(z,t)$}
\label{The differential equation of
the isotropic distribution
function F}

From Equation (\ref{g2}) we can find
that the anisotropic distribution function
$g(\mu,t)$ is an iteration function.
Thus, by iterating operation
the anisotropic distribution function
can be written as the function of the
temporal and spatial derivative of
the isotropic distribution function
$F(z,t)$ \citep{wq2018}
\begin{equation}
g(\mu,t)=\sum_{m, n}
\epsilon_{m,n}(\mu)
\frac{\partial^{m+n}{}}
{\partial {t^m}
\partial{z}^n}  F,
\label{expanded g}
\end{equation}
Here, $\epsilon_{m,n}(\mu)$ is the
coefficients with $m,n=0,1,2,3,\cdots$.
By inserting Equation (\ref{expanded g})
into Equation 
(\ref{Equation of F with g2})
we can obtain the EIDF
as shown in Appendix in
\citet{wq2019}
\begin{eqnarray}
\frac{\partial{F}}{\partial{t}}=
&&\left(-\kappa_z\frac{\partial{F}}
{\partial{z}}+\kappa_{zz}
\frac{\partial^2{F}}{\partial{z^2}}
+\kappa_{zzz}
\frac{\partial^3{F}}{\partial{z^3}}
+\kappa_{zzzz}
\frac{\partial^4{F}}{\partial{z^4}}
+\cdots\right)
+\left( \kappa_{tz}
\frac{\partial^2{F}}{\partial{t}
	\partial{z}}
+ \kappa_{ttz}\frac{\partial^3{F}}
{\partial{t^2}\partial{z}}
+ \kappa_{tttz}\frac{\partial^4{F}}
{\partial{t^3}\partial{z}}
+\cdots\right)\nonumber\\
&&+\left( \kappa_{tzz}
\frac{\partial^3{F}}{\partial{t}
	\partial{z^2}}
+ \kappa_{ttzz}\frac{\partial^4{F}}
{\partial{t^2}\partial{z^2}}
+ \kappa_{tttzz}\frac{\partial^5{F}}
{\partial{t^3}\partial{z^2}}
+\cdots\right)
+\cdots.
\label{equation of F with constant coefficient}
\end{eqnarray}
This is a constant coefficient
linear differential EIDF
with infinite number of derivative terms.
From the latter equation, 
we can find the SPDC with
the Fick's law definition
\begin{eqnarray}
\kappa_{zz}^{FL}&=&\kappa_{zz},
\label{kzzfl0}
\end{eqnarray}
and with the displacement variance 
definition \citep{wq2019}
\begin{eqnarray}
\kappa_{zz}^{DV}&=&\frac{1}{2}
\lim_{t\rightarrow t_{\infty}}
\frac{d\sigma^2}{dt}=\kappa_{zz}
-\kappa_z\kappa_{tz}.
\label{kzzvd0}
\end{eqnarray}

The coefficients of the terms in Equation
(\ref{equation of F with
constant coefficient}) can be obtained
by using the method in the papers of
\citet{wq2018, wq2019}.
For example, with the method 
we can obtain the formula of the cross-term
coefficient $\kappa_{tz}$ as
\begin{eqnarray}
\kappa_{tz}=&&\frac{v}{2}\frac{\int_{-1}^{1}
	d\mu \mu e^{M(\mu)}}
{\int_{-1}^{1}d\mu
	e^{M(\mu)}}
\int_{-1}^{1}d\mu e^{M(\mu)}\int_{-1}^{\mu}
d\nu \frac{e^{-M(\nu)}}{D_{\nu\nu}(\nu)}
\left(2\frac{\int_{-1}^{\nu}d\rho
	e^{M(\rho)}}{\int_{-1}^{1}d\mu e^{M(\mu)}}
-1\right)\nonumber\\
&&-\frac{v}{2}\int_{-1}^{1}d\mu \mu e^{M(\mu)}
\int_{-1}^{\mu}d\nu
\frac{e^{-M(\nu)}}{D_{\nu\nu}(\nu)}
\left(2\frac{\int_{-1}^{\nu}d\rho
	e^{M(\rho)}}{\int_{-1}^{1}d\mu e^{M(\mu)}}
-1\right)
\label{ktz in the main text}
\end{eqnarray}
with the details of the derivation shown 
in Appendix \ref{The accurate formula of ktz}.
The coefficient $\kappa_{tzz}$
is also shown in Appendix 
\ref{The accurate formula of ktz} 
(see Equation (\ref{ktzz})).

\section{Two definitions for PzI operation}
\label{The Displacement Variance definition
and the Fick's law one for PzI operation}

On the right-hand side of
Equation (\ref{equation of F with constant coefficient}),
there are a lot of spatial derivative terms,
e.g., $\kappa_z\partial {F}/\partial {z}$,
$\kappa_{zz}\partial^2 {F}/\partial {z^2}$,
$\kappa_{zzz}\partial^3 {F}/\partial {z^3}$,
$\cdots$, and spatial and 
temporal cross derivative terms
(hereafter abbreviated
as cross terms), e.g., 
$\kappa_{zt}\partial^2 {F}/(\partial {z}\partial {t})$,
$\kappa_{ttz}\partial^3 {F}/(\partial {t^2}\partial {z})$, 
$\cdots$,
$\kappa_{tzz}\partial^3 {F}/(\partial {t}\partial {z^2})$,
$\kappa_{ttzz}\partial^4 {F}/(\partial {t^2}\partial {z^2})$,
$\cdots$.
As shown in Section \ref{Relationship of the
general Fourier expansion method 
and the perturbation one},
the DMEs with different methods, i.e., 
the general Fourier
expansion and the perturbation theory,
give different EIDFs, which, however,
can be interconverted by the DIOs.
Because EIDFs, no matter how different they are, 
actually describe the same
transport process of particles, 
their corresponding physical
quantities should be invariant. 
The SPDC
has three different definitions, i.e.,
the Fick's law definition $\kappa_{zz}^{FL}$,
the displacement variance definition 
$\kappa_{zz}^{DV}$,
and the TGK formula definition $\kappa_{zz}^{TGK}$.
If one definition of the SPDC is invariant for the
DIOs,
it is more reasonable and should be used 
in the computer simulations, data analysis,
and theoretical research than that 
which are changed.

\subsection{The PzI operation}
\label{The PzI operation}

We firstly multiply Equation
(\ref{equation of F with constant coefficient})
by the differential operator
$\partial^m/\partial{z^m}$ with
$m=1,2,3,\cdots$, and find
\begin{eqnarray}
\frac{\partial^{m+1 }{F}}
{\partial{t}\partial{z^m}}=
&&\left(-\kappa_z\frac{\partial^{m+1}{F}}
{\partial{z^{m+1}}}+\kappa_{zz}
\frac{\partial^{m+2}{F}}{\partial{z^{m+2}}}
+\kappa_{zzz}
\frac{\partial^{m+3}{F}}{\partial{z^{m+3}}}
+\kappa_{zzzz}
\frac{\partial^{m+4}{F}}{\partial{z^{m+4}}}
+\cdots\right)\nonumber\\
&&+\left(\kappa_{tz}
\frac{\partial^{m+2}{F}}{\partial{t}
	\partial{z^{m+1}}}
+\kappa_{ttz}\frac{\partial^{m+3}{F}}
{\partial{t^2}\partial{z^{m+1}}}
+\kappa_{tttz}\frac{\partial^{m+4}{F}}
{\partial{t^3}\partial{z^{m+1}}}
+\cdots\right)\nonumber\\
&&+\left(\kappa_{tzz}
\frac{\partial^{m+3}{F}}{\partial{t}
	\partial{z^{m+2}}}
+\kappa_{ttzz}\frac{\partial^{m+4}{F}}
{\partial{t^2}\partial{z^{m+2}}}
+\kappa_{tttzz}\frac{\partial^{m+5}{F}}
{\partial{t^3}\partial{z^{m+2}}}
+\cdots\right)
+\cdots.
\label{mth PzE}
\end{eqnarray}
After drawing any term 
on the right-hand side of Equation (\ref{mth PzE}),
putting it on the left-hand side and
the other terms on the right-hand side,
we obtain a new equation. 
Thereafter, inserting
this new equation into
Equation (\ref{equation of F with constant coefficient})
we obtain a new EIDF and 
call as Equation {$\mathcal A$}.
Mathematically speaking, 
Equation {$\mathcal A$} is equivalent
to Equation 
(\ref{equation of F with constant coefficient}).

The above manipulation is the
$m$th-order Partial derivative over z 
Iterative ($m$th PzI)
operation.
For $m=1$, Equation
(\ref{mth PzE}) becomes
\begin{eqnarray}
\frac{\partial^2{F}}{\partial{t}\partial{z}}=
&&\left(-\kappa_z\frac{\partial^2{F}}{\partial{z^2}}
+\kappa_{zz}
\frac{\partial^3{F}}{\partial{z^3}}
+\kappa_{zzz}
\frac{\partial^4{F}}{\partial{z^4}}
+\kappa_{zzzz}
\frac{\partial^5{F}}{\partial{z^5}}
+\cdots\right)
+\left(\kappa_{tz}
\frac{\partial^3{F}}{\partial{t}
	\partial{z^2}}
+\kappa_{ttz}\frac{\partial^4{F}}
{\partial{t^2}\partial{z^2}}
+\kappa_{tttz}\frac{\partial^5{F}}
{\partial{t^3}\partial{z^2}}
+\cdots\right)\nonumber\\
&&+\left(\kappa_{tzz}
\frac{\partial^4{F}}{\partial{t}
	\partial{z^3}}
+\kappa_{ttzz}\frac{\partial^5{F}}
{\partial{t^2}\partial{z^3}}
+\kappa_{tttzz}\frac{\partial^6{F}}
{\partial{t^3}\partial{z^3}}
+\cdots\right)
+\cdots,
\label{equation of F with constant coefficient
by d/dz-0}
\end{eqnarray}
which is the equation of the first-order PzI operation.
In this subsection, we explore the Fick's law definition
$\kappa_{zz}^{FL}$
and the displacement variance definitions $\kappa_{zz}^{DV}$
for the first-order PzI operation.

\subsubsection{$R_{tz}$ of the first-order PzI operation}
\label{Fick's law definition kzzFL and Displacement Variance
definition kzzVD for Rtz of the first PzI operation}

Firstly, multiplying Equation 
(\ref{equation of F with
constant coefficient by d/dz-0})
by the parameter $\kappa_{tz}$,
we can find
\begin{eqnarray}
\kappa_{tz}\frac{\partial^2{F}}
{\partial{t}\partial{z}}=
&&\left(-\kappa_{tz}\kappa_z\frac{\partial^2{F}}
{\partial{z^2}}+\kappa_{tz}\kappa_{zz}
\frac{\partial^3{F}}{\partial{z^3}}
+\kappa_{tz}\kappa_{zzz}
\frac{\partial^4{F}}{\partial{z^4}}
+\kappa_{tz}\kappa_{zzzz}
\frac{\partial^5{F}}{\partial{z^5}}
+\cdots\right)
+\left(\kappa_{tz}^2
\frac{\partial^3{F}}{\partial{t}
	\partial{z^2}}
+ \kappa_{tz}\kappa_{ttz}\frac{\partial^4{F}}
{\partial{t^2}\partial{z^2}}
+ \kappa_{tz}\kappa_{tttz}\frac{\partial^5{F}}
{\partial{t^3}\partial{z^2}}
+\cdots\right)\nonumber\\
&&+\left( \kappa_{tz}\kappa_{tzz}
\frac{\partial^4{F}}{\partial{t}
	\partial{z^3}}
+ \kappa_{tz}\kappa_{ttzz}\frac{\partial^5{F}}
{\partial{t^2}\partial{z^3}}
+ \kappa_{tz}\kappa_{tttzz}\frac{\partial^6{F}}
{\partial{t^3}\partial{z^3}}
+\cdots\right)
+\cdots.
\label{equation of F with constant coefficient by d/dz}
\end{eqnarray}

Replacing $\kappa_{tz}\partial^2 F/(\partial t\partial z)$
in Equation (\ref{equation of F with
constant coefficient}) by the latter equation
gives
\begin{eqnarray}
\frac{\partial{F}}{\partial{t}}=
&&-\kappa_z\frac{\partial{F}}
{\partial{z}}+\left(\kappa_{zz}
-\kappa_{tz}\kappa_z\right)
\frac{\partial^2{F}}{\partial{z^2}}
+\left(\kappa_{zzz}+\kappa_{tz}\kappa_{zz}\right)
\frac{\partial^3{F}}{\partial{z^3}}
+\left(\kappa_{zzzz}+\kappa_{tz}\kappa_{zzz}\right)
\frac{\partial^4{F}}{\partial{z^4}}
+\cdots
+\kappa_{ttz}\frac{\partial^3{F}}
{\partial{t^2}\partial{z}}
+ \kappa_{tttz}\frac{\partial^4{F}}
{\partial{t^3}\partial{z}}
+\cdots\nonumber\\
&&+\left(\kappa_{tzz}+\kappa_{tz}^2\right)
\frac{\partial^3{F}}{\partial{t}
	\partial{z^2}}
+ \left(\kappa_{ttzz}+\kappa_{tz}
\kappa_{ttz}\right)\frac{\partial^4{F}}
{\partial{t^2}\partial{z^2}}
+ \left(\kappa_{tttzz}+\kappa_{tz}\kappa_{tttz}\right)
\frac{\partial^5{F}}
{\partial{t^3}\partial{z^2}}
+\cdots\nonumber\\
&&+\left(\kappa_{t3z}+\kappa_{tz}\kappa_{tzz}\right)
\frac{\partial^4{F}}{\partial{t}\partial{z^3}}
+\left(\kappa_{2t3z}+\kappa_{tz}\kappa_{ttzz}\right)
\frac{\partial^5{F}}{\partial{t^2}\partial{z^3}}
+\left(\kappa_{3t3z}+\kappa_{tz}\kappa_{tttzz}\right)
\frac{\partial^6{F}}{\partial{t^3}\partial{z^3}}
+\cdots.
\label{equation of F for the first interation}
\end{eqnarray}
Here, the subscript $2t3z$ of $\kappa_{2t3z}$ in
the latter equation denotes that
there are two letters $t$ and three letters $z$, i.e.,
$\kappa_{ttzzz}$.
In the same way, the subscript $ntmz$ of
$\kappa_{ntmz}$ in the latter equation
presents $n$ letters $t$ and $m$ letters $z$.
The other cases denote the same meaning.
This notation is used throughout this paper.
The above manipulation is called as the $R_{tz}$
of the 1st PzI operation.

From Equation 
(\ref{equation of F for the first interation})
we can easily find that
the SPDC of the Fick's law definition
is $\kappa_{zz}^{FL}=\kappa_{zz}-\kappa_{tz}\kappa_z$,
which is different from Equation (\ref{kzzfl0}).
Therefore,
the Fick's law definition $\kappa_{zz}^{FL}$
is changed, or in other words, is not invariant, 
by the $R_{tz}$ of the 1st PzI operation.

By multiplying Equation
(\ref{equation of F for the first interation})
with
$ \Delta z$ and
integrating the result over z, one can  find
\begin{eqnarray}
\frac{d}{dt}\langle (\Delta z)\rangle
&=&\int_{-\infty}^\infty dz (\Delta z)\frac{\partial{F}}
{\partial{t}}\nonumber\\
&=&\int_{-\infty}^\infty dz
(\Delta z)\Bigg[-\kappa_z\frac{\partial{F}}
{\partial{z}}+\left(\kappa_{zz}-\kappa_{tz}\kappa_z\right)
\frac{\partial^2{F}}{\partial{z^2}}
+\left(\kappa_{zzz}+\kappa_{tz}\kappa_{zz}\right)
\frac{\partial^3{F}}{\partial{z^3}}
+\left(\kappa_{zzzz}+\kappa_{tz}\kappa_{zzz}\right)
\frac{\partial^4{F}}{\partial{z^4}}
+\cdots\nonumber\\
&&+\kappa_{ttz}\frac{\partial^3{F}}
{\partial{t^2}\partial{z}}
+ \kappa_{tttz}\frac{\partial^4{F}}
{\partial{t^3}\partial{z}}
+\cdots
+\left(\kappa_{tzz}+\kappa_{tz}^2\right)
\frac{\partial^3{F}}{\partial{t}\partial{z^2}}
+ \left(\kappa_{ttzz}+\kappa_{tz}
\kappa_{ttz}\right)\frac{\partial^4{F}}
{\partial{t^2}\partial{z^2}}
+ \left(\kappa_{tttzz}+\kappa_{tz}
\kappa_{tttz}\right)\frac{\partial^5{F}}
{\partial{t^3}\partial{z^2}}
+\cdots\nonumber\\
&&+\left(\kappa_{t3z}+\kappa_{tz}
\kappa_{tzz}\right)
\frac{\partial^4{F}}{\partial{t}
	\partial{z^3}}
+ \left(\kappa_{2t3z}+\kappa_{tz}
\kappa_{ttzz}\right)\frac{\partial^5{F}}
{\partial{t^2}\partial{z^3}}
+\left(\kappa_{3t3z}+\kappa_{tz}
\kappa_{tttzz}\right)\frac{\partial^6{F}}
{\partial{t^3}\partial{z^3}}
+\cdots\Bigg].
\label{ddt Delta z for Rtz of 1st PzI operation}
\end{eqnarray}
The latter Equation can be formally rewritten as
\begin{eqnarray}
\frac{d}{dt}\langle (\Delta z)\rangle=\sum_{i}\int_{-\infty}^\infty
dz (\Delta z)E_i
\end{eqnarray}
where $E_i$ denotes the terms 
on the right-hand side of Equation
(\ref{equation of F for the first interation}).

By using the following regularities
\begin{eqnarray}
&&F(z=\pm\infty)=0,
\label{regularity-1}\\
&&\frac{\partial^n{F}}
{\partial{z^n}}(z=\pm\infty)=0
\hspace{0.5cm} n=1,2,3,\cdots,
\end{eqnarray}
and employing the following formulas
\begin{eqnarray}
&&\int_{-\infty}^\infty
dz (\Delta z)\left(-\kappa_z
\frac{\partial{F}}{\partial{z}}\right)
=-\kappa_z\int_{-\infty}^\infty
dz (\Delta z)\frac{\partial{F}}
{\partial{z}}=\kappa_z=constant,
\label{1-integration by parts}\\
&&\int_{-\infty}^\infty dz (\Delta z)
\left[\left(\kappa_{n z}
+\kappa_{tz}\kappa_{(n-1)z}\right)
\frac{\partial^n{F}}{\partial{z^n}}\right]
=\left(\kappa_{n z}
+\kappa_{tz}\kappa_{(n-1)z}\right)
\int_{-\infty}^\infty
dz (\Delta z)
\frac{\partial^n{F}}{\partial{z^n}}
=0 \hspace{0.5cm}n=2,3,\cdots,
\label{1-zn}\\
&&\int_{-\infty}^\infty dz
(\Delta z)\left[\left(\kappa_{n z}+\kappa_{tz}
\kappa_{nt(m-1)z}\right)
\frac{\partial^{n+m}{F}}{\partial{t^m}
	\partial{z^n}}\right]
=\left(\kappa_{n z}
+\kappa_{tz}\kappa_{nt(m-1)z}\right)
\frac{d^m}{dt^m}\int_{-\infty}^\infty
dz (\Delta z)
\frac{\partial^n{F}}{\partial{z^n}}
=0 \nonumber\\
&&\hspace{5cm}
n=1,2,3,\cdots,
\hspace{0.2cm}m=1,2,3,\cdots,
\label{1-tmzn}\\
&&\int_{-\infty}^\infty dz
(\Delta z)^2\left(-\kappa_z
\frac{\partial{F}}{\partial{z}}
\right)
=-\kappa_z\int_{-\infty}^\infty
dz (\Delta z)^2\frac{\partial{F}}{\partial{z}}
=2\kappa_z \langle (\Delta z)
\rangle,
\label{2-z}\\
&&\int_{-\infty}^\infty dz(\Delta z)^2
\left[\left(\kappa_{zz}-\kappa_{tz}\kappa_z\right)
\frac{\partial^2{F}}{\partial{z^2}}\right]
=\left(\kappa_{zz}-\kappa_{tz}\kappa_z\right)
\int_{-\infty}^\infty
dz (\Delta z)^2\frac{\partial^2{F}}{\partial{z^2}}
=2\left(\kappa_{zz}-\kappa_{tz}\kappa_z\right),
\label{2-zz}\\
&&\int_{-\infty}^\infty dz (\Delta z)^2
\left[\left(\kappa_{n z}
+\kappa_{tz}\kappa_{(n-1)z}\right)
\frac{\partial^n{F}}{\partial{z^n}}\right]
=\left(\kappa_{n z}
+\kappa_{tz}\kappa_{(n-1)z}\right)
\int_{-\infty}^\infty dz
(\Delta z)^2\frac{\partial^n{F}}
{\partial{z^n}}
=0 \hspace{0.5cm}n=3, 4, 5,
\cdots,\\
&&\int_{-\infty}^\infty dz(\Delta z)^2
\left(\kappa_{n tz}
\frac{\partial^{n+1}{F}}
{\partial{t^n}\partial{z}}\right)
=\kappa_{n tz}\frac{d^n}{dt^n}
\int_{-\infty}^\infty dz(\Delta z)^2
\frac{\partial{F}}{\partial{z}}
=-2\kappa_{n tz}
\frac{d^n}{dt^n}\langle
(\Delta z) \rangle=0\nonumber\\
&&\hspace{5cm}n=1,2,3,\cdots,
\label{1-tzn}\\
&&\int_{-\infty}^\infty dz(\Delta z)^2
\left[\left(\kappa_{ntmz}+\kappa_{tz}
\kappa_{nt(m-1)z}\right)
\frac{\partial^{n+m}{F}}
{\partial{t^n}\partial{z^m}}\right]
=\left(\kappa_{ntmz}+\kappa_{tz}
\kappa_{nt(m-1)z}\right)
\frac{d^n}{dt^n}
\int_{-\infty}^\infty
dz (\Delta z)^2
\frac{\partial{F^m}}
{\partial{z^m}}=0\nonumber\\
&&\hspace{5cm}n=1,2,3,
\cdots, \hspace{0.2cm}m=2,3,\cdots,
\label{last-integration by parts}
\end{eqnarray}
we can find that only the term 
$\partial{F}/\partial{z}$
on the right-hand side of Equation 
(\ref{ddt Delta z for Rtz of 1st PzI operation}) 
is left.
That is,
\begin{eqnarray}
\frac{d}{dt}\langle (\Delta z)\rangle=\sum_i
\int_{-\infty}^\infty dz 
(\Delta z)E_i=\int_{-\infty}^\infty dz
(\Delta z)\left(-\kappa_z\frac{\partial{F}}
{\partial{z}}\right).
\label{symbol-1}
\end{eqnarray}
Using integration by parts, 
we can rewrite the latter
equation as
\begin{eqnarray}
\frac{d}{dt}\langle (\Delta z)
\rangle=\kappa_z.
\label{delta z-1}
\end{eqnarray}
Similarily, we can find the following formula
\begin{eqnarray}
\frac{d}{dt}\langle (\Delta z)^2
\rangle =\sum_i\int_{-\infty}^\infty dz (\Delta z)^2E_i
=2\kappa_z \langle (\Delta z)
\rangle+2\kappa_{zz}
-2\kappa_{tz}\kappa_z.
\label{delta z2-1}
\end{eqnarray}
Replacing $\kappa_z$ 
on the right-hand side of the latter
equation by Equation (\ref{delta z-1}) 
and considering the relation
\begin{eqnarray}
\sigma^2=\langle
(\Delta z)^2 \rangle
-\langle (\Delta z)
\rangle^2,
\end{eqnarray}
we can easily find that 
the displacement variance definition
of the SPDCs
as following
\begin{eqnarray}
\kappa_{zz}^{DV}=\frac{1}{2}\lim_{t\rightarrow
t_{\infty}}\frac{d\sigma^2}{dt}
=\kappa_{zz}-\kappa_{tz}\kappa_z,
\end{eqnarray}
which is identical with Equation 
(\ref{kzzvd0}).
Therefore, the displacement 
variance definition
$\kappa_{zz}^{DV}$ is not 
changed by the $R_{tz}$ of the 1st PzI operation.

From the above investigation 
we can find the following conclusion:
The Fick's law definition $\kappa_{zz}^{FL}$ 
is changed by
$R_{tz}$ of the first-order PzI 
operation which is one of DIOs.
However, the formula 
$\lim _{t\rightarrow t_{\infty}}d\sigma^2/(2dt)
=\kappa_{zz}-\kappa_{tz}\kappa_z$
is invariant for this manipulation.
That is, the displacement variance definition
$\kappa_{zz}^{DV}
=\lim _{t\rightarrow t_{\infty}}d\sigma^2/(2dt)$
is an invariance for the $R_{tz}$ 
of the 1st PzI operation.

\subsubsection{$R_{zzz}$ of the first-order PzI operation}

By drawing $\partial^3F /\partial z^3$ from
Equation (\ref{equation of F 
with constant coefficient
by d/dz-0}),
we can rewrite the equation as
\begin{eqnarray}
\frac{\partial^3{F}}{\partial{z^3}}
&=&\frac{1}{\kappa_{zz}}\frac{\partial^2{F}}
{\partial{t}\partial{z}}
+\left(\frac{\kappa_z}{\kappa_{zz}}\frac{\partial^2{F}}
{\partial{z^2}}
-\frac{\kappa_{zzz}}{\kappa_{zz}}\frac{\partial^4{F}}
{\partial{z^4}}
-\frac{\kappa_{zzzz}}{\kappa_{zz}}
\frac{\partial^5{F}}{\partial{z^5}}
-\cdots\right)
-\left(\frac{\kappa_{tz}}{\kappa_{zz}}
\frac{\partial^3{F}}{\partial{t}\partial{z^2}}
+\frac{\kappa_{ttz}}{\kappa_{zz}}\frac{\partial^4{F}}
{\partial{t^2}\partial{z^2}}
+\frac{\kappa_{tttz}}{\kappa_{zz}}\frac{\partial^5{F}}
{\partial{t^3}\partial{z^2}}
+\cdots\right)\nonumber\\
&&-\left(\frac{\kappa_{tzz}}{\kappa_{zz}}
\frac{\partial^4{F}}{\partial{t}
	\partial{z^3}}
+\frac{\kappa_{ttzz}}{\kappa_{zz}}\frac{\partial^5{F}}
{\partial{t^2}\partial{z^3}}
+\frac{\kappa_{tttzz}}{\kappa_{zz}}\frac{\partial^6{F}}
{\partial{t^3}\partial{z^3}}
+\cdots\right)+\cdots.
\label{d3F/dz3}
\end{eqnarray}
Multiplying Equation (\ref{d3F/dz3}) by $\kappa_{zzz}$
and inserting the result into Equation
(\ref{equation of F with constant coefficient})
gives
\begin{eqnarray}
\frac{\partial{F}}{\partial{t}}=
&&\left[-\kappa_z\frac{\partial{F}}
{\partial{z}}+\left(\kappa_{zz}
+\frac{\kappa_{zzz}\kappa_z}
{\kappa_{zz}}\right)
\frac{\partial^2{F}}{\partial{z^2}}
+\left(
\kappa_{zzzz}-\frac{\kappa_{zzz}^2}{\kappa_{zz}}\right)
\frac{\partial^4{F}}{\partial{z^4}}+
\left(\kappa_{5z}-\frac{\kappa_{zzz}\kappa_{4z}}
{\kappa_{zz}}\right)
\frac{\partial^5{F}}{\partial{z^5}}
-\cdots\right]\nonumber\\
&&+\left[\left(\frac{\kappa_{zzz}}{\kappa_{zz}}
+\kappa_{tz}\right)
\frac{\partial^2{F}}{\partial{t}
	\partial{z}}
+ \kappa_{ttz}\frac{\partial^3{F}}
{\partial{t^2}\partial{z}}
+ \kappa_{tttz}\frac{\partial^4{F}}
{\partial{t^3}\partial{z}}
+\cdots\right]\nonumber\\
&&+\left[\left(\kappa_{tzz}
-\frac{\kappa_{zzz}\kappa_{tz}}{\kappa_{zz}}\right)
\frac{\partial^3{F}}{\partial{t}\partial{z^2}}
+ \left(\kappa_{ttzz}
-\frac{\kappa_{zzz}\kappa_{ttz}}
{\kappa_{zz}}\right)\frac{\partial^4{F}}
{\partial{t^2}\partial{z^2}}
+ \left(\kappa_{tttzz}
-\frac{\kappa_{zzz}\kappa_{tttz}}
{\kappa_{zz}}\right)\frac{\partial^5{F}}
{\partial{t^3}\partial{z^2}}
+\cdots\right]\nonumber\\
&&+\left[\left(\kappa_{t3z}
-\kappa_{zzz}\frac{\kappa_{tzz}}
{\kappa_{zz}}\right)\frac{\partial^4{F}}
{\partial{t}\partial{z^3}}
+\left(\kappa_{2t3z}-\frac{\kappa_{zzz}
\kappa_{ttzz}}{\kappa_{zz}}\right)
\frac{\partial^5{F}}{\partial{t^2}
	\partial{z^3}}
+\left(\kappa_{3t3z}-\frac{\kappa_{zzz}
	\kappa_{tttzz}}{\kappa_{zz}}\right)
\frac{\partial^6{F}}{\partial{t^3}\partial{z^3}}
+\cdots\right]+\cdots.
\label{equation of F for the first interation-2}
\end{eqnarray}
The above manipulation, which is one of the DIOs, is
called $R_{zzz}$ of the 1st PzI operation.

In Equation (\ref{equation of F for the first interation-2})
the parallel diffusion coefficient,
which is in front of $\partial^2{F}/\partial{z^2}$,
is $\kappa_{zz}^{FL}=\kappa_{zz}
+\kappa_{zzz}\kappa_z/\kappa_{zz}$.
However, it is $\kappa_{zz}^{FL}=\kappa_{zz}$
in Equations (\ref{equation of F with
constant coefficient}).
Therefore, the Fick's law definition 
$\kappa_{zz}^{FL}$
is variant for $R_{zzz}$ of the first PzI operation.

As done in \ref{Fick's law definition 
kzzFL and Displacement
Variance definition kzzVD for 
Rtz of the first PzI operation},
from Equation 
(\ref{equation of F for the first interation-2})
the derivative of the first- 
and second-order moments of parallel displacement
over time can be obtained as
\begin{eqnarray}
&&\frac{d}{dt}\langle (\Delta z)
\rangle=\kappa_z,
\label{delta z1-3}\\
&&\frac{d}{dt}\langle (\Delta z)^2
\rangle =2\kappa_z \langle (\Delta z)
\rangle+2\left(\kappa_{zz}
+\frac{\kappa_{zzz}\kappa_z}{\kappa_{zz}}\right)
-2\left(\frac{\kappa_{zzz}}
{\kappa_{zz}}+\kappa_{tz}\right)
\frac{d}{dt}\langle \Delta z \rangle.
\label{delta z2-3}
\end{eqnarray}

Combining Equations (\ref{delta z1-3}) 
and (\ref{delta z2-3}) gives
\begin{equation}
\kappa_{zz}^{DV}=\frac{1}{2}
\lim_{t\rightarrow t_{\infty}}
\frac{d\sigma^2}{dt}=\frac{1}{2}\frac{d}{dt}\langle
\left(\Delta z\right)^2\rangle-\langle\Delta z
\rangle\frac{d}{dt}\langle\Delta z\rangle
=\kappa_{zz}-\kappa_z\kappa_{tz},
\label{Wang and Qin formula 2019-3}
\end{equation}
which is the same as Equation (\ref{kzzvd0}).
Therefore,
the displacement variance definition $\kappa_{zz}^{DV}$ is
invariant for $R_{zzz}$ of the first PzI operation.

\subsubsection{$R_{it(j+1)z}$ of the first-order PzI operation}

In this part, we explore the influence of other kinds of DIOs, i.e.,
$R_{it(j+1)z}$ ($(i,j)
\in N^2$ but $(0,0)\notin N^2$) of the first-order PzI operation,
on the Fick's law and the displacement variance definitions.
Here, $N^2=N\times N$ denotes 
the ordered pairs of all the natural numbers.
As the manipulation in the previous parts, 
the EIDF corresponding to
$R_{it(j+1)z}$ of the first-order PzI operation
can be obtained
\begin{eqnarray}
\frac{\partial{F}}{\partial{t}}=&&
\left[-\kappa_z\frac{\partial{F}}{\partial{z}}
+\left(\kappa_{zz}+\frac{\kappa_{it(j+1)z}}{\kappa_{itjz}}
\kappa_z\right)\frac{\partial^2{F}}{\partial{z^2}}
+\left(\kappa_{3z}-\frac{\kappa_{it(j+1)z}}{\kappa_{itjz}}
\kappa_{zz}\right)\frac{\partial^3{F}}{\partial{z^3}}
+\left(\kappa_{4z}-\frac{\kappa_{it(j+1)z}}{\kappa_{itjz}}
\kappa_{3z}\right)\frac{\partial^4{F}}{\partial{z^4}}
+\cdots\right]\nonumber\\
&&+\left[\left(\kappa_{tz}+\frac{\kappa_{it(j+1)z}}{\kappa_{itjz}}\right)
\frac{\partial^2{F}}{\partial{t}\partial{z}}
+\kappa_{ttz}\frac{\partial^3{F}}{\partial{t^2}\partial{z}}
+\kappa_{tttz}\frac{\partial^4{F}}
{\partial{t^3}\partial{z}}
+\cdots\right]\nonumber\\
&&+\left[\left(\kappa_{tzz}-\frac{\kappa_{it(j+1)z}}{\kappa_{itjz}}
{\kappa_{tz}}\right)\frac{\partial^3{F}}{\partial{t}\partial{z^2}}
+\left(\kappa_{ttzz}-\frac{\kappa_{it(j+1)z}}{\kappa_{itjz}}
{\kappa_{2tz}}\right)\frac{\partial^4{F}}{\partial{t^2}\partial{z^2}}
+\left(\kappa_{tttzz}-\frac{\kappa_{it(j+1)z}}{\kappa_{itjz}}
{\kappa_{3tz}}\right)\frac{\partial^5{F}}{\partial{t^3}\partial{z^2}}
+\cdots\right]\nonumber\\
&&+\left[\left(\kappa_{t3z}-\frac{\kappa_{it(j+1)z}}{\kappa_{itjz}}
\kappa_{tzz}\right)\frac{\partial^4{F}}{\partial{t}\partial{z^3}}
+\left(\kappa_{2t3z}-\frac{\kappa_{it(j+1)z}}{\kappa_{itjz}}
\kappa_{2tzz}\right)\frac{\partial^5{F}}{\partial{t^2}\partial{z^3}}
+\left(\kappa_{3t3z}-\frac{\kappa_{it(j+1)z}}{\kappa_{itjz}}
\kappa_{3tzz}\right)\frac{\partial^6{F}}{\partial{t^3}\partial{z^3}}
+\cdots\right]\nonumber\\
&&+\cdots.
\label{Rit(j+1)z equation for the first PzI}
\end{eqnarray}

Equation (\ref{Rit(j+1)z equation for the first PzI})
shows the Fick's law definition of the SPDC is 
$\kappa_{zz}^{FL}=\kappa_{zz}
+\kappa_{it(j+1)z}\kappa_z/\kappa_{itjz}$, 
which is dfferent from Equation (\ref{kzzfl0}), 
to demonstrate that
the Fick's law definition 
$\kappa_{zz}^{FL}$ is variant for
$R_{it(j+1)z}$ of the first-order PzI operation.

The formulas of the first- and second-order moments of
the parallel displacement
can be obtained from Equation
(\ref{Rit(j+1)z equation for the first PzI})
\begin{eqnarray}
&&\frac{d}{dt}\langle (\Delta z)
\rangle=\kappa_z=constant,\\
&&\frac{d}{dt}\langle (\Delta z)^2\rangle =2\kappa_z \langle
(\Delta z)\rangle
+2\left(\kappa_{zz}+\frac{\kappa_{it(j+1)z}}{\kappa_{itjz}}
\kappa_z\right)
-2\left(\kappa_{tz}+\frac{\kappa_{it(j+1)z}}{\kappa_{itjz}}\right)
\frac{d}{dt}\langle (\Delta z) \rangle.
\end{eqnarray}
To proceed, we can easily derive the following formula
from the latter two equations
\begin{equation}
\kappa_{zz}^{DV}=\frac{1}{2}\lim_{t\rightarrow t_{\infty}}
\frac{d\sigma^2}{dt}=\frac{1}{2}\frac{d}{dt}\langle
\left(\Delta z\right)^2\rangle-\langle\Delta z
\rangle\frac{d}{dt}\langle\Delta z\rangle
=\kappa_{zz}-\kappa_z\kappa_{tz}.
\end{equation}
The latter formula is identical with Equation (\ref{kzzvd0})
to show that the displacement variance definition
$\kappa_{zz}^{DV}$ is not changed by
$R_{it(j+1)z}$ of the first-order PzI operation.

\subsection{Second-order PzI operation}
For $n=2$, Equation (\ref{mth PzE}) becomes
\begin{eqnarray}
\frac{\partial^{3}{F}}{\partial{t}\partial{z^2}}=
&&\left(-\kappa_z\frac{\partial^{3}{F}}
{\partial{z^{3}}}+\kappa_{zz}
\frac{\partial^{4}{F}}{\partial{z^{4}}}
+\kappa_{zzz}
\frac{\partial^{5}{F}}{\partial{z^{5}}}
+\kappa_{zzzz}
\frac{\partial^{6}{F}}{\partial{z^{6}}}
+\cdots\right)\
+\left(\kappa_{tz}
\frac{\partial^{4}{F}}{\partial{t}
	\partial{z^{3}}}
+\kappa_{ttz}\frac{\partial^{5}{F}}
{\partial{t^2}\partial{z^{3}}}
+\kappa_{tttz}\frac{\partial^{6}{F}}
{\partial{t^3}\partial{z^{3}}}
+\cdots\right)\nonumber\\
&&+\left(\kappa_{tzz}
\frac{\partial^{5}{F}}{\partial{t}
	\partial{z^{4}}}
+\kappa_{ttzz}\frac{\partial^{6}{F}}
{\partial{t^2}\partial{z^{4}}}
+\kappa_{tttzz}\frac{\partial^{7}{F}}
{\partial{t^3}\partial{z^{4}}}
+\cdots\right)
+\cdots,
\label{2nd PzE}
\end{eqnarray}
which is the equation of the second-order PzI operation. 
Because there exist a limitless variety 
of terms in the latter equation,
combining Equations (\ref{equation of F with
constant coefficient})
and (\ref{2nd PzE})
can give countless kinds of new
EIDFs 
by numerous types of the DIOs, which are the
replacement manipulations
for the second-order PzI operation.
In the following part, we explore the influence of
these manipulations on the Fick's law and displacement
variance definitions of the SPDC.

\subsubsection{$R_{tzz}$ of the second-order PzI operation}
\label{Two definitions for Rtzz
of the second-order PzI operation}

Multiplying Equation (\ref{2nd PzE}) by $\kappa_{tzz}$ yields
\begin{eqnarray}
\kappa_{tzz}\frac{\partial^{3}{F}}{\partial{t}\partial{z^2}}=
&&\left(-\kappa_{tzz}\kappa_z\frac{\partial^{3}{F}}
{\partial{z^{3}}}+\kappa_{tzz}\kappa_{zz}
\frac{\partial^{4}{F}}{\partial{z^{4}}}
+\kappa_{tzz}\kappa_{zzz}
\frac{\partial^{5}{F}}{\partial{z^{5}}}
+\kappa_{tzz}\kappa_{zzzz}
\frac{\partial^{6}{F}}{\partial{z^{6}}}
+\cdots\right)\nonumber\\
&&+\left(\kappa_{tzz}\kappa_{tz}
\frac{\partial^{4}{F}}{\partial{t}\partial{z^{3}}}
+\kappa_{tzz}\kappa_{ttz}\frac{\partial^{5}{F}}
{\partial{t^2}\partial{z^{3}}}
+\kappa_{tzz}\kappa_{tttz}\frac{\partial^{6}{F}}
{\partial{t^3}\partial{z^{3}}}
+\cdots\right)\nonumber\\
&&+\left(\kappa_{tzz}\kappa_{tzz}
\frac{\partial^{5}{F}}{\partial{t}\partial{z^{4}}}
+\kappa_{tzz}\kappa_{ttzz}\frac{\partial^{6}{F}}
{\partial{t^2}\partial{z^{4}}}
+\kappa_{tzz}\kappa_{tttzz}\frac{\partial^{7}{F}}
{\partial{t^3}\partial{z^{4}}}
+\cdots\right)
+\cdots.
\end{eqnarray}
Inserting the latter equation into Equation
(\ref{equation of F with constant coefficient})
we can find
\begin{eqnarray}
\frac{\partial{F}}{\partial{t}}=
&&-\kappa_z\frac{\partial{F}}{\partial{z}}+\kappa_{zz}
\frac{\partial^2{F}}{\partial{z^2}}
+\left(\kappa_{3z}-\kappa_{tzz}\kappa_z\right)
\frac{\partial^3{F}}{\partial{z^3}}
+\left(\kappa_{4z}+\kappa_{tzz}\kappa_{zz}\right)
\frac{\partial^{4}{F}}{\partial{z^{4}}}
+\left(\kappa_{5z}+\kappa_{tzz}\kappa_{3z}\right)
\frac{\partial^{5}{F}}{\partial{z^{5}}}
+\cdots\nonumber\\
&&+\left( \kappa_{tz}
\frac{\partial^2{F}}{\partial{t}\partial{z}}
+ \kappa_{ttz}\frac{\partial^3{F}}
{\partial{t^2}\partial{z}}
+ \kappa_{tttz}\frac{\partial^4{F}}{\partial{t^3}\partial{z}}
+ \kappa_{4tz}\frac{\partial^5{F}}{\partial{t^4}\partial{z}}
+ \kappa_{5tz}\frac{\partial^6{F}}{\partial{t^5}\partial{z}}
+\cdots\right)\nonumber\\
&&+\left(\kappa_{ttzz}\frac{\partial^4{F}}{\partial{t^2}\partial{z^2}}
+\kappa_{3tzz}\frac{\partial^5{F}}{\partial{t^3}\partial{z^2}}
+\kappa_{4tzz}\frac{\partial^6{F}}{\partial{t^4}\partial{z^2}}
+\kappa_{5tzz}\frac{\partial^7{F}}{\partial{t^5}\partial{z^2}}
+\cdots\right)\nonumber\\
&&+\left(\left(\kappa_{t3z}+\kappa_{tzz}\kappa_{tz}\right)
\frac{\partial^{4}{F}}{\partial{t}\partial{z^{3}}}
+\left(\kappa_{tt3z}+\kappa_{tzz}\kappa_{ttz}\right)
\frac{\partial^{5}{F}}{\partial{t^2}\partial{z^{3}}}
+\left(\kappa_{3t3z}+\kappa_{tzz}\kappa_{3tz}\right)
\frac{\partial^{6}{F}}{\partial{t^3}\partial{z^{3}}}
+\left(\kappa_{4t3z}+\kappa_{tzz}\kappa_{4tz}\right)
\frac{\partial^{7}{F}}{\partial{t^4}\partial{z^{3}}}
+\cdots\right)\nonumber\\
&&+\left(\left(\kappa_{t4z}+\kappa_{tzz}\kappa_{tzz}\right)
\frac{\partial^{5}{F}}{\partial{t}\partial{z^{4}}}
+\left(\kappa_{tt4z}+\kappa_{tzz}\kappa_{ttzz}\right)
\frac{\partial^{6}{F}}{\partial{t^2}\partial{z^{4}}}
+\left(\kappa_{3t4z}+\kappa_{tzz}\kappa_{3tzz}\right)
\frac{\partial^{7}{F}}{\partial{t^3}\partial{z^{4}}}
+\cdots\right)+\cdots.
\label{Rtzz of second-order PzI operation}
\end{eqnarray}
The manipulation in this subsection is called 
$R_{tzz}$ of the second-order PzI operation
which is one of the DIOs. 

From Equation (\ref{Rtzz of second-order PzI operation})
we can find that the Fick's law definition $\kappa_{zz}^{FL}$
of the SPDC is equal to $\kappa_{zz}$,
which is identical with Equation (\ref{kzzfl0}).
That is, the Fick's law definition is not changed by
the $R_{tzz}$ of the second-order PzI operation. 
In the following part, we explore
the displacement variance definition $\kappa_{zz}^{DV}$
for $R_{tzz}$ of the second PzI operation.

Using the method in subsection
\ref{Fick's law definition kzzFL and Displacement Variance
definition kzzVD for Rtz of the first PzI operation},
we can obtain
\begin{eqnarray}
&&\frac{d}{dt}\langle (\Delta z)
\rangle=\kappa_z,
\label{delta z1-Rtzz of 2nd PzI}\\
&&\frac{d}{dt}\langle (\Delta z)^2
\rangle =2\kappa_z \langle (\Delta z)
\rangle+2\kappa_{zz}-2\kappa_{tz}
\frac{d}{dt}\langle \Delta z \rangle.
\label{delta z2-Rtzz of 2nd PzI}
\end{eqnarray}
Replacing $d\langle \Delta z \rangle/dt$ 
on the right-hand side of
 Equation (\ref{delta z2-Rtzz of 2nd PzI})
by Equation (\ref{delta z1-Rtzz of 2nd PzI}) yields
\begin{eqnarray}
\frac{d}{dt}\langle (\Delta z)^2
\rangle =2\kappa_z \langle (\Delta z)\rangle
+2\kappa_{zz}-2\kappa_{tz}\kappa_{z}.
\end{eqnarray}
Inserting Equation 
(\ref{delta z1-Rtzz of 2nd PzI}) into
the latter equation gives
\begin{equation}
\kappa_{zz}^{DV}=\frac{1}{2}
\lim_{t\rightarrow t_{\infty}}\frac{d\sigma^2}{dt}
=\kappa_{zz}-\kappa_z\kappa_{tz}.
\label{Wang and Qin formula 2019-Rtzz of 2nd PzI}
\end{equation}
Here, the formula $\sigma^2=\langle
(\Delta z)^2 \rangle
-\langle (\Delta z)
\rangle^2$ is used.
Obviously, Equation 
(\ref{Wang and Qin formula 2019-Rtzz of 2nd PzI})
is exactly identical with Equation (\ref{kzzvd0}).
The investigation in this part demonstrates
the displacement variance definition $\kappa_{zz}^{DV}$
is invariant for $R_{tzz}$ of 
the second-order PzI operation.

\subsubsection{$R_{itjz}$ of the second-order PzI operation}
\label{Two definitions for Ritjz of the second-order PzI operation}

Here, we derive the EIDF for the another new DIO, i.e.,
$R_{itjz}$ with $i=0,1,2,\cdots$ and $j=3,4,5,\cdots$ 
of the second-order PzI operation.
Analogous to the previous parts,
one can obtain
\begin{eqnarray}
\frac{\partial{F}}{\partial{t}}
=&&-\kappa_z\frac{\partial{F}}{\partial{z}}
+\kappa_{zz}\frac{\partial^2{F}}{\partial{z^2}}
+\left(\kappa_{zzz}+\kappa_z\frac{\kappa_{itjz}}
{\kappa_{it(j-2)z}}\right)\frac{\partial^3{F}}{\partial{z^3}}
+\left(\kappa_{4z}-\kappa_{zz}\frac{\kappa_{itjz}}
{\kappa_{it(j-2)z}}\right)\frac{\partial^4{F}}{\partial{z^4}}
+\cdots+\kappa_{tz}\frac{\partial^2{F}}
{\partial{t}\partial{z}}\nonumber\\
&&+\kappa_{ttz}\frac{\partial^3{F}}{\partial{t^2}\partial{z}}
+\kappa_{3tz}\frac{\partial^4{F}}{\partial{t^3}\partial{z}}
+\kappa_{4tz}\frac{\partial^5{F}}{\partial{t^4}\partial{z}}
+\cdots \nonumber\\
&&+\left(\kappa_{tzz}+\frac{\kappa_{itjz}}
{\kappa_{it(j-2)z}}\right)\frac{\partial^3{F}}
{\partial{t}\partial{z^2}}
+\kappa_{ttzz}\frac{\partial^4{F}}{\partial{t^2}\partial{z^2}}
+\kappa_{3tzz}\frac{\partial^5{F}}{\partial{t^3}\partial{z^2}}
+\kappa_{4tzz}\frac{\partial^6{F}}{\partial{t^4}\partial{z^2}}
+\cdots \nonumber\\
&&+\left(\kappa_{tzzz}-\kappa_{tz}\frac{\kappa_{itjz}}
{\kappa_{it(j-2)z}}\right)\frac{\partial^4{F}}{\partial{t}\partial{z^3}}
+\left(\kappa_{ttzzz}-\kappa_{ttz}\frac{\kappa_{itjz}}
{\kappa_{it(j-2)z}}\right)\frac{\partial^5{F}}{\partial{t^2}\partial{z^3}}
+\left(\kappa_{3t3z}-\kappa_{3tz}\frac{\kappa_{itjz}}
{\kappa_{it(j-2)z}}\right)\frac{\partial^6{F}}{\partial{t^3}\partial{z^3}}
+\cdots \nonumber\\
&&+\left(\kappa_{t4z}-\kappa_{tzz}\frac{\kappa_{itjz}}
{\kappa_{it(j-2)z}}\right)\frac{\partial^5{F}}{\partial{t}\partial{z^4}}
+\left(\kappa_{tt4z}-\kappa_{ttzz}\frac{\kappa_{itjz}}
{\kappa_{it(j-2)z}}\right)\frac{\partial^6{F}}{\partial{t^2}\partial{z^4}}
+\cdots \nonumber\\
&&+\left(\kappa_{t5z}-\kappa_{t3z}\frac{\kappa_{itjz}}
{\kappa_{it(j-2)z}}\right)\frac{\partial^6{F}}{\partial{t}\partial{z^5}}
+\cdots.
\label{Ritjz equation for the second-order PzI}
\end{eqnarray}

From Equation (\ref{Ritjz equation for the second-order PzI})
we can find that the Fick's law definition $\kappa_{zz}^{FL}$
of the SPDC is equal to $\kappa_{zz}$,
which is identical with Equation (\ref{kzzfl0}).
So the Fick's law definition is invariant for
$R_{itjz}$ of the second-order PzI operation.
In what follows, we investigate the influence of
$R_{itjz}$ of the second-order PzI operation
on the displacement variance definition of the SPDC.

Analogous to the method in subsection
\ref{Fick's law definition kzzFL and Displacement Variance
definition kzzVD for Rtz of the first PzI operation},
we can find
\begin{eqnarray}
&&\frac{d}{dt}\langle (\Delta z)\rangle=\kappa_z,
\label{delta z1-Ritjz of 2nd PzI}\\
&&\frac{d}{dt}\langle (\Delta z)^2
\rangle =2\kappa_z \langle (\Delta z)\rangle
+2\kappa_{zz}-2\kappa_{tz}
\frac{d}{dt}\langle \Delta z \rangle.
\label{delta z2-Ritjz of 2nd PzI}
\end{eqnarray}
Similar to subsection \ref{Two definitions for Rtzz
of the second-order PzI operation}, 
we can obtain obviously
\begin{equation}
\kappa_{zz}^{DV}=\frac{1}{2}
\lim_{t\rightarrow t_{\infty}}\frac{d\sigma^2}{dt}
=\kappa_{zz}-\kappa_z\kappa_{tz},
\label{Wang and Qin formula 2019-Ritjz of 2nd PzI}
\end{equation}
which has the same form with Equation (\ref{kzzvd0}).
Thus, the displacement variance definition
is an invariance quantity for
$R_{itjz}$ of the second-order PzI operation.

In fact, we can easily find that there is no the second-order
spatial derivative term 
in Equation (\ref{2nd PzE}). Thus,
all the DIOs, i.e.,
any iteration operation by inserting the deformations of
Equation (\ref{2nd PzE})
into Equation (\ref{equation of F with constant coefficient}) 
cannot
change the parallel diffusion coefficient in the results.
Therefore, the Fick's law definition 
$\kappa_{zz}^{FL}$ is invariant for
the manipulations of the
second-order PzI operation.

Every term in Equation  (\ref{2nd PzE})
can be  represented by $E$, and
the following formula 
\begin{eqnarray}
\int_{-\infty}^\infty dz (\Delta z)E=0
\end{eqnarray}
holds.
Therefore, the contributions of all of terms 
in Equation (\ref{2nd PzE})
to the first moment is equal to zero.
Thus,
the first-order moment of the EIDFs 
for any EIDF derived by the DIOs of 
the 2nd-order PzI operation
only comes from Equation
(\ref{equation of F with constant coefficient}).
Similarily, there is no contribution 
from any term in Equation (\ref{2nd PzE})
to the second-order moment
\begin{eqnarray}
\int_{-\infty}^\infty dz (\Delta z)^2 E=0.
\end{eqnarray}
Consequently, the displacement 
variance definition, as the function of
$\langle (\Delta z)\rangle$ and 
$\langle (\Delta z)^2\rangle$,
is only determined by Equation 
(\ref{equation of F with constant coefficient}),
from which, actually already  
\citet{wq2019} obtained the following formula
\begin{equation}
\kappa_{zz}^{DV}=\frac{1}{2}
\lim_{t\rightarrow t_{\infty}}\frac{d\sigma^2}{dt}
=\kappa_{zz}-\kappa_z\kappa_{tz}.
\label{kzzvd for Rtzz of 2nd PzI}
\end{equation}
The latter formula
is identical with Equation
(\ref{kzzvd0}). Therefore, 
the displacement variance 
definition is invariant for $R_{itjz}$ 
of the second-order PzI operation. 

\subsection{Two definitions for the third- 
and  higher-order PzI operation}
\label{Displacement Variance definition 
and Fick's law definition
for the higher-order PzI operation}

It is obvious that
there is no the second-order spatial derivative term
in Equation (\ref{mth PzE}) for $m\ge 3$.
Any manipulation of the third- 
and higher-order PzI operation cannot
change the Fick's law definition in the new EIDF
or
influence the first- and second-order moments 
of the parallel displacement.
Therefore, as the discussion in subsection
\ref{Two definitions for Ritjz of 
the second-order PzI operation},
the Fick's law definition $\kappa_{zz}^{FL}$ 
and the displacement variance definition 
$\kappa_{zz}^{DV}$ are
all invariant quantities for the third- and
higher-order PzI operation.

\section{Two definitions
for PtzI operation}
\label{The Variance definition and 
the Fick's law one for PtzI operation}

In the above sections, we find that,
for the SPDC, 
the displacement
variance definition $\kappa_{zz}^{DV}$
is an invariance for the PzI operation, 
while the Fick's law definition 
$\kappa_{zz}^{FL}$ is not. 
 In what follows, we explore whether
 the Fick's law definition $\kappa_{zz}^{FL}$ and
the displacement
variance definition $\kappa_{zz}^{DV}$
are invariant for
the Partial derivative over t and z Iterative (PtzI) operation.

Firstly, we multiply Equation (\ref{equation of F with
constant coefficient}) by the differential operator
$\partial^{n+m}/(\partial{t^n}\partial{z^m})$ with
$n,m=1,2,3,\cdots$, and find
\begin{eqnarray}
\frac{\partial^{n+m+1}{F}}{\partial{t^{n+1}}\partial{z^{m}}}=
&&\left(-\kappa_z\frac{\partial^{n+m+1}{F}}
{\partial{t^n}\partial{z^{m+1}}}+\kappa_{zz}
\frac{\partial^{n+m+2}{F}}{\partial{t^n}\partial{z^{m+2}}}
+\kappa_{zzz}
\frac{\partial^{n+m+3}{F}}{\partial{t^n}\partial{z^{m+3}}}
+\kappa_{zzzz}
\frac{\partial^{n+m+4}{F}}{\partial{t^n}\partial{z^{m+4}}}
+\cdots\right)\nonumber\\
&&+\left( \kappa_{tz}
\frac{\partial^{n+m+2}{F}}{\partial{t^{n+1}}
\partial{z^{m+1}}}
+ \kappa_{ttz}\frac{\partial{t^{n+m+3}}{F}}
{\partial{t^{n+2}}\partial{z^{m+1}}}
+ \kappa_{tttz}\frac{\partial^{n+m+4}{F}}
{\partial{t^{n+3}}\partial{z^{m+1}}}
+\cdots\right)\nonumber\\
&&+\left(\kappa_{tzz}
\frac{\partial^{n+m+3}{F}}{\partial{t^{n+1}}
\partial{z^{m+2}}}
+ \kappa_{ttzz}\frac{\partial^{n+m+4}{F}}
{\partial{t^{n+2}}\partial{z^{m+2}}}
+ \kappa_{tttzz}\frac{\partial^{n+m+5}{F}}
{\partial{t^{n+3}}\partial{z^{m+2}}}
+\cdots\right)
+\cdots,
\label{nmth PtzE}
\end{eqnarray}
which is the equation of the nth-order 
temporal and mth-order spatial PtzI operation. 
For the lowest order case, i.e., $n=1$ and $m=1$,
the latter equation becomes
\begin{eqnarray}
\frac{\partial^{3}{F}}{\partial{t^{2}}\partial{z}}=
&&\left(-\kappa_z\frac{\partial^{3}{F}}
{\partial{t}\partial{z^{2}}}
+\kappa_{zz}
\frac{\partial^{4}{F}}{\partial{t}\partial{z^{3}}}
+\kappa_{zzz}
\frac{\partial^{5}{F}}{\partial{t}\partial{z^{4}}}
+\kappa_{zzzz}
\frac{\partial^{6}{F}}{\partial{t}\partial{z^{5}}}
+\cdots\right)\nonumber\\
&&+\left( \kappa_{tz}
\frac{\partial^{4}{F}}{\partial{t^{2}}
\partial{z^{2}}}
+ \kappa_{ttz}\frac{\partial{t^{5}}{F}}
{\partial{t^{3}}\partial{z^{2}}}
+ \kappa_{tttz}\frac{\partial^{6}{F}}
{\partial{t^{4}}\partial{z^{2}}}
+\cdots\right)\nonumber\\
&&+\left(\kappa_{tzz}
\frac{\partial^{5}{F}}{\partial{t^{2}}
\partial{z^{3}}}
+ \kappa_{ttzz}\frac{\partial^{6}{F}}
{\partial{t^{3}}\partial{z^{3}}}
+ \kappa_{tttzz}\frac{\partial^{7}{F}}
{\partial{t^{4}}\partial{z^{3}}}
+\cdots\right)
+\cdots,
\label{11th PtzE}
\end{eqnarray}
which is the equation of the first-order 
temporal and first-order spatial PtzI operation.  

Combining Equations (\ref{equation of F with
constant coefficient}) and (\ref{11th PtzE}),
employing the DIOs
we can obtain a lot of new EIDFs.
As shown in subsection 
\ref{Two definitions for Ritjz 
of the second-order PzI operation},
the contribution of Equation (\ref{11th PtzE}) 
to the formulas
of the first- and second-order moments
of displacement
are all equal to zero, and
the displacement variance definitions 
$\kappa_{zz}^{DV}$ is invariant
for the lowest order PtzI operation.
At the same time, since there is no 
second-order spatial derivative term in
Equation (\ref{11th PtzE}),
the Fick's law definition $\kappa_{zz}^{Fl}$ 
is not changed by any DIO, i.e.,
any
replacement manipulation of the lowest order PtzI operation.
Similarily, the same results can be obtained 
for the manipulations of the higher-order PtzI operations. 
Therefore, the displacement variance definition 
$\kappa_{zz}^{DV}$ and the Fick's law definition 
$\kappa_{zz}^{Fl}$
are all invariant quantities
for the DIOs of the PtzI operations.

\section{Two definitions for PtI operations}
\label{The Variance definition and 
the Fick's law one for PtI operation}

\subsection{The PtI operations}
\label{The PtI operation}

By multiplying Equation (\ref{equation of F
with constant coefficient})
by the partial differential operator $\partial^n{}/\partial{t^n}$
with $n=1,2,3,\cdots$,
we can obtain the following equation
\begin{eqnarray}
	\frac{\partial^{n+1}{F}}{\partial{t^{n+1}}}=
	&&\left(-\kappa_z\frac{\partial^{n+1}{F}}
	{\partial{t^n}\partial{z}}+\kappa_{zz}
	\frac{\partial^{n+2}{F}}{\partial{t^n}\partial{z^2}}
	+\kappa_{zzz}
	\frac{\partial^{n+3}{F}}{\partial{t^n}\partial{z^3}}
	+\kappa_{zzzz}
	\frac{\partial^{n+4}{F}}{\partial{t^n}\partial{z^4}}
	+\cdots\right)
	+\left( \kappa_{tz}
	\frac{\partial^{n+2}{F}}{\partial{t^{n+1}}
		\partial{z}}
	+ \kappa_{ttz}\frac{\partial^{n+3}{F}}
	{\partial{t^{n+2}}\partial{z}}
	+ \kappa_{tttz}\frac{\partial^{n+4}{F}}
	{\partial{t^{n+3}}\partial{z}}
	+\cdots\right)\nonumber\\
	&&+\left( \kappa_{tzz}
	\frac{\partial^{n+3}{F}}{\partial{t^{n+1}}
		\partial{z^2}}
	+ \kappa_{ttzz}\frac{\partial^{n+4}{F}}
	{\partial{t^{n+2}}\partial{z^2}}
	+ \kappa_{tttzz}\frac{\partial^{n+5}{F}}
	{\partial{t^{n+3}}\partial{z^2}}
	+\cdots\right)
	+\cdots.
	\label{n PtE}
\end{eqnarray}

\subsection{The first-order PtI operation}
\label{The 1st PtI operation}

For $n=1$, Equation (\ref{n PtE}) becomes
\begin{eqnarray}
	\frac{\partial^2{F}}{\partial{t^2}}=
	&&\left(-\kappa_z\frac{\partial^2{F}}
	{\partial{t}\partial{z}}+\kappa_{zz}
	\frac{\partial^3{F}}{\partial{t}\partial{z^2}}
	+\kappa_{zzz}
	\frac{\partial^{4}{F}}{\partial{t}\partial{z^3}}
	+\kappa_{zzzz}
	\frac{\partial^{5}{F}}{\partial{t}\partial{z^4}}
	+\cdots\right)
	+\left( \kappa_{tz}
	\frac{\partial^3{F}}{\partial{t^2}
		\partial{z}}
	+ \kappa_{ttz}\frac{\partial^4{F}}
	{\partial{t^3}\partial{z}}
	+ \kappa_{tttz}\frac{\partial^{5}{F}}
	{\partial{t^{4}}\partial{z}}
	+\cdots\right)\nonumber\\
	&&+\left( \kappa_{tzz}
	\frac{\partial^{4}{F}}{\partial{t^{2}}
		\partial{z^2}}
	+ \kappa_{ttzz}\frac{\partial^{5}{F}}
	{\partial{t^{3}}\partial{z^2}}
	+ \kappa_{tttzz}\frac{\partial^{6}{F}}
	{\partial{t^{4}}\partial{z^2}}
	+\cdots\right)
	+\cdots,
	\label{first PtE}
\end{eqnarray}
which is the governing equation of 
the first-order PtI operation.
After drawing one term from the latter equation,
by putting
it on the left-hand side of the equal sign
and the other ones
on the right-hand side,
we can obtain a new deformation equation.
Obviously, we can obtain a lot of
deformation equations from Equation (\ref{first PtE}),
by one of which,
replacing the corresponding term in Equation 
(\ref{equation of
F with constant coefficient}), 
as done in the previous subsections, 
yields a new EIDF. 
In such way, 
a lot of new EIDFs can be obtained. 
All the latter replacement manipulations, one of which is a DIO,  
are called the
first-order Partial derivative over 
time $t$ Iterative (PtI) operation.

\subsubsection{$R_{tz}$ of the first-order PtI operation}
\label{Displacement Variance definition kzzDV
for  Rtz of the 1st PtI operation}

Equation (\ref{first PtE}) can be rewritten as
\begin{eqnarray}
	\frac{\partial^2{F}}
	{\partial{t}\partial{z}}=\frac{1}{\kappa_z}\Bigg[
	&&-\frac{\partial^2{F}}{\partial{t^2}}+\left(\kappa_{zz}
	\frac{\partial^3{F}}{\partial{t}\partial{z^2}}
	+\kappa_{zzz}
	\frac{\partial^{4}{F}}{\partial{t}\partial{z^3}}
	+\kappa_{zzzz}
	\frac{\partial^{5}{F}}{\partial{t}\partial{z^4}}
	+\cdots\right)
	+\left( \kappa_{tz}
	\frac{\partial^3{F}}{\partial{t^2}
		\partial{z}}
	+ \kappa_{ttz}\frac{\partial^4{F}}
	{\partial{t^3}\partial{z}}
	+ \kappa_{tttz}\frac{\partial^{5}{F}}
	{\partial{t^{4}}\partial{z}}
	+\cdots\right)\nonumber\\
	&&+\left( \kappa_{tzz}
	\frac{\partial^{4}{F}}{\partial{t^{2}}
		\partial{z^2}}
	+ \kappa_{ttzz}\frac{\partial^{5}{F}}
	{\partial{t^{3}}\partial{z^2}}
	+ \kappa_{tttzz}\frac{\partial^{6}{F}}
	{\partial{t^{4}}\partial{z^2}}
	+\cdots\right)
	+\cdots\Bigg].
	\label{first PtE-2}
\end{eqnarray}
Multplying the latter equation by parameter $\kappa_{tz}$
we can easily obtain
\begin{eqnarray}
	\kappa_{tz}\frac{\partial^2{F}}
	{\partial{t}\partial{z}}=
	&&-\frac{\kappa_{tz}}{\kappa_z}\frac{\partial^2{F}}
{\partial{t^2}}+\left(\kappa_{zz}\frac{\kappa_{tz}}{\kappa_z}
	\frac{\partial^3{F}}{\partial{t}\partial{z^2}}
	+\kappa_{zzz}\frac{\kappa_{tz}}{\kappa_z}
	\frac{\partial^{4}{F}}{\partial{t}\partial{z^3}}
	+\kappa_{zzzz}\frac{\kappa_{tz}}{\kappa_z}
	\frac{\partial^{5}{F}}{\partial{t}\partial{z^4}}
	+\cdots\right) \nonumber\\
	&&+\left( \kappa_{tz}\frac{\kappa_{tz}}{\kappa_z}
	\frac{\partial^3{F}}{\partial{t^2}
		\partial{z}}
	+ \kappa_{ttz}\frac{\kappa_{tz}}{\kappa_z}\frac{\partial^4{F}}
	{\partial{t^3}\partial{z}}
	+ \kappa_{tttz}\frac{\kappa_{tz}}{\kappa_z}\frac{\partial^{5}{F}}
	{\partial{t^{4}}\partial{z}}
	+\cdots\right)\nonumber\\
	&&+\left( \kappa_{tzz}\frac{\kappa_{tz}}{\kappa_z}
	\frac{\partial^{4}{F}}{\partial{t^{2}}
		\partial{z^2}}
	+ \kappa_{ttzz}\frac{\kappa_{tz}}{\kappa_z}\frac{\partial^{5}{F}}
	{\partial{t^{3}}\partial{z^2}}
	+ \kappa_{tttzz}\frac{\kappa_{tz}}{\kappa_z}\frac{\partial^{6}{F}}
	{\partial{t^{4}}\partial{z^2}}
	+\cdots\right)
	+\cdots.
	\label{first PtE-3}
\end{eqnarray}
Replacing $\kappa_{tz}\partial^2{F}/(\partial{t}\partial{z})$
in Equation (\ref{equation of F with
constant coefficient}) by Equation (\ref{first PtE-3})
gives
\begin{eqnarray}
\frac{\partial{F}}{\partial{t}}=
&&\left(-\kappa_z\frac{\partial{F}}
	{\partial{z}}+\kappa_{zz}
	\frac{\partial^2{F}}{\partial{z^2}}
	+\kappa_{zzz}
	\frac{\partial^3{F}}{\partial{z^3}}
	+\kappa_{zzzz}
	\frac{\partial^4{F}}{\partial{z^4}}
	+\cdots\right)-\frac{\kappa_{tz}}{\kappa_z}\frac{\partial^2{F}}
{\partial{t^2}}\nonumber\\
	&&+\left(\kappa_{tzz}+\frac{\kappa_{tz}\kappa_{zz}}{\kappa_z}\right)
\frac{\partial^3{F}}{\partial{t}\partial{z^2}}
	+\left(\kappa_{tzzz}+\frac{\kappa_{tz}\kappa_{zzz}}{\kappa_z}\right)
\frac{\partial^4{F}}{\partial{t}\partial{z^3}}
	+\cdots
	\nonumber\\
	&&+\left(\kappa_{ttz}+\frac{\kappa_{tz}^2}{\kappa_z}\right)
\frac{\partial^3{F}}{\partial{t^2}\partial{z}}
	+\left(\kappa_{tttz}+\frac{\kappa_{tz}\kappa_{ttz}}{\kappa_z}\right)
\frac{\partial^4{F}}{\partial{t^3}\partial{z}}
	+\left(\kappa_{ttttz}+\frac{\kappa_{tz}\kappa_{tttz}}{\kappa_z}\right)
\frac{\partial^5{F}}{\partial{t^4}\partial{z}}+\cdots\nonumber\\
	&&+\left(\kappa_{ttzz}+\frac{\kappa_{tz}\kappa_{tzz}}{\kappa_z}\right)
\frac{\partial^4{F}}{\partial{t^2}\partial{z^2}}
	+\left(\kappa_{tttzz}+\frac{\kappa_{tz}\kappa_{ttzz}}{\kappa_z}\right)
\frac{\partial^5{F}}{\partial{t^3}\partial{z^2}}
	+\left(\kappa_{ttttzz}+\frac{\kappa_{tz}\kappa_{tttzz}}{\kappa_z}\right)
\frac{\partial^6{F}}{\partial{t^4}\partial{z^2}}+\cdots\nonumber\\
	&&+\left(\kappa_{ttzzz}+\frac{\kappa_{tz}\kappa_{tzzz}}{\kappa_z}\right)
\frac{\partial^5{F}}{\partial{t^2}\partial{z^3}}
	+\left(\kappa_{tttzzz}+\frac{\kappa_{tz}\kappa_{ttzzz}}{\kappa_z}\right)
\frac{\partial^6{F}}{\partial{t^3}\partial{z^3}}+\cdots\nonumber\\
	&&+\left(\kappa_{ttzzzz}+\frac{\kappa_{tz}\kappa_{tzzzz}}{\kappa_z}\right)
\frac{\partial^6{F}}{\partial{t^2}\partial{z^4}}+\cdots
	+\left(\kappa_{ttzzzzz}+\frac{\kappa_{tz}\kappa_{tzzzzz}}{\kappa_z}\right)
\frac{\partial^7{F}}{\partial{t^2}\partial{z^5}}+\cdots.
	\label{Rst of 1st PtE}
\end{eqnarray}
The above replacement manipulation, which is one of the DIOs,  
is called $R_{tz}$ of the first-order PtI 
operation in this paper.

From Equation (\ref{Rst of 1st PtE}),
we find that the Fick's law definition $\kappa_{zz}^{FL}$ 
is equal to $\kappa_{zz}$, $\kappa_{zz}^{FL}
=\kappa_{zz}$, which is identical with Equation (\ref{kzzfl0}).
Thus, the DIOs, i.e., $R_{tz}$ of the 
first-order PtI operation, cannot change
the Fick's law definition
$\kappa_{zz}^{FL}$.
In the following, we explore the invariance  
of the displacement variance
definition $\kappa_{zz}^{DV}$ for
$R_{tz}$ of the first-order PtI operation.

Using the method in subsection 
\ref{Fick's law definition kzzFL and
Displacement Variance definition kzzVD 
for Rtz of the first PzI operation},
from Equation (\ref{Rst of 1st PtE}) we can find
\begin{eqnarray}
\frac{d}{dt}\langle (\Delta z)
\rangle&=&\kappa_z-\frac{\kappa_{tz}}{\kappa_z}
\frac{d^2}{dt^2}\langle (\Delta z)\rangle,
\label{dzdt Rst of 1st PtE}\\
\frac{d}{dt}\langle (\Delta z)^2\rangle&
=&2\kappa_z\langle (\Delta z)
\rangle+2\kappa_{zz}-\frac{\kappa_{tz}}{\kappa_z}
\frac{d^2}{dt^2}\langle (\Delta z)^2\rangle-2T_{tz}.
\label{dz2dt Rst of 1st PtE}
\end{eqnarray}
Here the term $T_{tz}$ is as follows
\begin{eqnarray}
	T_{tz}=&&\left(\kappa_{ttz}+\frac{\kappa_{tz}^2}{\kappa_z}\right)
\frac{d^2}{dt^2}\langle (\Delta z)\rangle+
	\left(\kappa_{tttz}+\frac{\kappa_{ttz}\kappa_{tz}}{\kappa_z}\right)
\frac{d^3}{dt^3}\langle (\Delta z)\rangle
	+\left(\kappa_{4tz}+\frac{\kappa_{tttz}\kappa_{tz}}{\kappa_z}\right)
\frac{d^4}{dt^4}\langle (\Delta z)\rangle+
	\left(\kappa_{5tz}+\frac{\kappa_{4tz}\kappa_{tz}}{\kappa_z}\right)
\frac{d^5}{dt^5}\langle (\Delta z)\rangle+
	\cdots \nonumber\\
&&=\sum_{n=1}^{\infty}\left(\kappa_{(n+1)tz}
+\kappa_{ntz}\frac{\kappa_{tz}}{\kappa_z}\right)
\frac{d^{(n+1)}}{dt^{(n+1)}}\langle (\Delta z)\rangle.
	\label{Ttz}
\end{eqnarray}
Combining Equations (\ref{dzdt Rst of 1st PtE})
and (\ref{dz2dt Rst of 1st PtE})
we can obtain
\begin{eqnarray}
\frac{1}{2}\frac{d\sigma^2}{dt}=&&\kappa_{zz}
+\frac{\kappa_{tz}}{\kappa_z}
\left[\langle (\Delta z)\rangle
\frac{d^2}{dt^2}\langle (\Delta z)\rangle
-\frac{1}{2}\frac{d^2}{dt^2}
\langle (\Delta z)^2\rangle\right]-T_{tz}
\label{variance definition of Rst of 1st PtI operation}
\end{eqnarray}

In order to simplify the latter equation,
we need to investigate the term $T_{tz}$.
Applying the operator $\partial^n{}/\partial {t^n}$ with
$n=1,2,3,\cdots$ on
Equation (\ref{dzdt Rst of 1st PtE}) gives
\begin{eqnarray}
\frac{d^{n+1}}{dt^{n+1}}\langle (\Delta z)\rangle=
-\frac{\kappa_{tz}}{\kappa_z}
\frac{d^{n+2}}{dt^{n+2}}\langle (\Delta z)\rangle.
\label{dzdt Rst with ddtn}
\end{eqnarray}
Employing the above equation, we can find that Equation
(\ref{Ttz}) can be simplified as
\begin{eqnarray}
T_{tz}=
\lim_{n\rightarrow \infty}
\left[\left(\frac{\kappa_{tz}}{\kappa_z}
\right)^n\kappa_{tz}+\kappa_{(n+1)tz}\right]
\frac{d^{n+1}}{dt^{n+1}}\langle (\Delta z)\rangle.
\end{eqnarray}
Here, the subscript $(n+1)tz$ of $\kappa_{(n+1)tz}$
in the latter formula denotes
there are $(n+1)$ letters $t$. That is, $\kappa_{2tz}=
\kappa_{ttz}$, $\kappa_{3tz}=\kappa_{tttz}$,
$\kappa_{4tz}=\kappa_{ttttz}$, and so on.
Using Equation (\ref{kntz-k2tz}) in Appendix
\ref{The formulas of kntz with for 
all natural number except of n=0 and n=1}, 
we can rewrite the
latter equation as
\begin{eqnarray}
T_{tz}\approx\lim_{n\rightarrow \infty}
\left[\left(\frac{\kappa_{tz}}{\kappa_z}\right)^n
\kappa_{tz}+\left(\frac{1}{2D}\right)^{n-1}\kappa_{2tz}\right]
\frac{d^{n+1}}{dt^{n+1}}\langle (\Delta z)\rangle.
\end{eqnarray}
Inserting $\kappa_{tz}$, $\kappa_z$, and 
$\kappa_{2tz}$ (see Equation (\ref{ktz for first order xi}),
(\ref{kz for first order xi}), and 
(\ref{kttz-value}) in Appendix, respectively)
into the above equation yields
\begin{eqnarray}
\frac{1}{v}T_{tz}\approx\lim_{n\rightarrow \infty}
\left[\left(\frac{2}{3}\right)^n
\frac{2}{9}\xi
+\left(\frac{1}{2}\right)^{n-1}\left(-\frac{13}{108}
\xi\right)\right]
\frac{d^{n+1}}{d(Dt)^{n+1}}\langle (\Delta z)\rangle.
\label{nondimensionlized limit of Ttz}
\end{eqnarray}
In order to evaluate the latter equation, we have to
compute $d^{n+1}\langle (\Delta z)\rangle/dt^{n+1}$
in which the quantity $\langle (\Delta z)\rangle$ 
is the solutions 
of Equations (\ref{dzdt Rst of 1st PtE}) as
\begin{eqnarray} 
\langle (\Delta z) \rangle=c_1+c_2e^{-\kappa_zt/\kappa_{tz}}+\kappa_zt.
\label{delta z for Rtz of 1st PtI operation}
\end{eqnarray}
Here, $c_1$ and $c_2$ are the undetermined coefficients. 
From Equation (\ref{delta z for Rtz of 1st PtI operation})
we can obtain
\begin{eqnarray}
\frac{d^{n+1}}{dt^{n+1}}\langle (\Delta z)\rangle
= c_2\left(-1\right)^{n+1}
\left(\frac{\kappa_z}{\kappa_{tz}}\right)^{n+1}
e^{-\kappa_zt/\kappa_{tz}}.
\end{eqnarray}
Using Equations (\ref{ktz for first order xi}) and 
(\ref{kz for first order xi}) in Appendix gives
\begin{eqnarray}
\frac{d^{n+1}}{dt^{n+1}}\langle 
(\Delta z)\rangle \approx c_2
\left(-1\right)^{n+1}\left(\frac{3}{2}\right)^{n+1}D^{n+1}
e^{-3Dt/2}.
\end{eqnarray}
Considering the latter equation and 
the limit
$t\rightarrow t_{\infty}$, 
we can find that Equation
(\ref{nondimensionlized limit of Ttz}) becomes 
\begin{eqnarray}
\frac{1}{v}\lim_{t\rightarrow t_{\infty}}T_{tz}
\approx\lim_{n\rightarrow \infty}
\lim_{t\rightarrow t_{\infty}}\left[\frac{1}{3}\xi
+\frac{9}{4}\left(\frac{3}{4}\right)^{n-1}
\left(-\frac{13}{108}\xi\right)\right]
\left(-1\right)^{n+2}c_1e^{-3Dt/2}=0,
\end{eqnarray}
where $\xi\ll 1$, and $c_1$, $v$, $D$ are constants.
The above equation can be rewritten as
\begin{eqnarray}
\lim_{t\rightarrow t_{\infty}}T_{tz}=0.
\label{Ttz=0}
\end{eqnarray}

For the limit $t\rightarrow t_{\infty}$,
Equation (\ref{variance definition of Rst
	of 1st PtI operation}) becomes
\begin{eqnarray}
\frac{1}{2}\lim_{t\rightarrow t_{\infty}}
\frac{d\sigma^2}{dt}&=&\kappa_{zz}
+\frac{\kappa_{tz}}{\kappa_z}
\left[\lim_{t\rightarrow t_{\infty}}\langle (\Delta z)\rangle
\frac{d^2}{dt^2}\langle (\Delta z)\rangle
-\frac{1}{2}\lim_{t\rightarrow t_{\infty}}\frac{d^2}{dt^2}
\langle (\Delta z)^2\rangle\right],
\label{variance definition of Rst of 1st PtI
	operation without Ttz}
\end{eqnarray}
where 
Equation (\ref{Ttz=0}) is used. 
From Equations (\ref{dz2dt Rst of 1st PtE}), 
(\ref{Ttz}) and
(\ref{dzdt Rst with ddtn})
yields
\begin{eqnarray}
\frac{d}{dt}\langle (\Delta z)^2\rangle&
=&2\kappa_z\langle (\Delta z)
\rangle+2\kappa_{zz}-\frac{\kappa_{tz}}{\kappa_z}
\frac{d^2}{dt^2}\langle (\Delta z)^2\rangle-T_{tz}
\label{dz2dt Rst of 1st PtE without Ttz}
\end{eqnarray}
with
\begin{eqnarray}
T_{tz}=\lim_{n\rightarrow \infty}(-1)^{n-1}c_2 
\frac{\kappa_z}{\kappa_{tz}}e^{-\kappa_zt/\kappa_{tz}}.
\end{eqnarray}
The solution of Equation (\ref{dz2dt Rst of 1st PtE without Ttz}) 
can be obtained as
\begin{eqnarray}
\langle (\Delta z)^2 \rangle&=&c'_1+c'_2e^{-\kappa_zt/\kappa_{tz}}
+2(\kappa_{zz}+\kappa_z c_1)t+\kappa_z^2 t^2-2\kappa_z\kappa_{tz}t
-2\kappa_z c_2 t e^{-\kappa_zt/\kappa_{tz}}
+2c_2 \frac{\kappa_z}{\kappa_{tz}}
\lim_{n\rightarrow \infty}(-1)^{n-1}te^{-\kappa_zt/\kappa_{tz}}
\label{delta z2}
\end{eqnarray}
with the undetermined constants $c'_1$ and $c'_2$.
Inserting Equations (\ref{delta z for Rtz of 1st PtI operation})
and (\ref{delta z2}) into Equation
(\ref{variance definition of Rst of 1st PtI operation without Ttz}),
we can obtain
\begin{eqnarray}
\frac{1}{2}\frac{d\sigma^2}{dt}=&&\kappa_{zz}
+\left(c_1+c_2e^{-\kappa_zt/\kappa_{tz}}+\kappa_zt\right)
\frac{\kappa_{z}}{\kappa_{tz}}c_2 e^{-\kappa_zt/\kappa_{tz}}%
-\frac{1}{2}\left(\frac{\kappa_{z}}{\kappa_{tz}}c'_2
e^{-\kappa_zt/\kappa_{tz}}+2\kappa_{tz}\kappa_{z}
+4\kappa_{z}c_2e^{-\kappa_zt/\kappa_{tz}}
-2 \frac{\kappa_{z}^2}{\kappa_{tz}} c_2 t e^{-\kappa_zt/\kappa_{tz}}
\right)\nonumber\\
&&+2c_2\lim_{n\rightarrow \infty}(-1)^{n-1}\left[2
\frac{\kappa_z}{\kappa_{tz}}
-\left(\frac{\kappa_z}{\kappa_{tz}}\right)^2t 
\right]e^{-\kappa_zt/\kappa_{tz}}.
\label{variance definition for Rtz of 1st-order 
	PtI operation without Ttz}
\end{eqnarray}
Employing Equations (\ref{ktz for first order xi}) 
and (\ref{kz for first order xi}) in Appendix gives
\begin{eqnarray}
\kappa_z/\kappa_{tz}=3D/2>0.
\end{eqnarray}
Using the latter equation yields
\begin{eqnarray}
\lim_{t\rightarrow t_{\infty}} e^{-\kappa_zt/\kappa_{tz}}\approx 0.
\label{exponent function tends to zero for limit}
\end{eqnarray}
Inserting formula (\ref{exponent function tends to zero for limit}) into
Equation (\ref{variance definition for Rtz of 
1st-order PtI operation without Ttz}),
for the limit $t\rightarrow t_{\infty}$  we can find 
\begin{eqnarray}
\kappa_{zz}^{DV}=\frac{1}{2}\lim_{t\rightarrow
t_{\infty}}\frac{d\sigma^2}{dt}&=&\kappa_{zz}-\kappa_{tz}\kappa_{z},
\end{eqnarray}
which is identical with Equation (\ref{kzzvd0}).
Thus, we find that the displacement variance
definition $\kappa_{zz}^{DV}$ is invariant quantity for $R_{tz}$
of the first-order PtI operation.

\subsubsection{Displacement Variance definition
$\kappa_{zz}^{DV}$ for $R_{tz}$ 
of the 1st-order PtI operation under
the special condition}
\label{Displacement Variance definition
kzzDV for Rtz of the 1st PtI operation under
the strong restrictive conditions}

The derivation process in the latter subsection 
is very lengthy and complicate.  
In fact, employing some special condition we can give 
the same result through a simpler derivation.

According to Equations 
(\ref{exponent function tends to zero for limit}), 
we can find that the terms containing exponent function 
$e^{-\kappa_zt/\kappa_{tz}}$ in  
Equations (\ref{delta z for Rtz of 1st PtI operation})
and (\ref{delta z2})
tend to zero for the limit 
$t\rightarrow t_{\infty}$. 
However, if using the special condition 
\begin{eqnarray}
c_2=c'_2=0,
\label{c2=c2=0}
\end{eqnarray}
we can also find that 
the same terms are all equal to zero. 
In what follows, we provide the derivation process 
of the displacement variance definition for 
$R_{tz}$ of the first-order PtI operation
under the special condition.

For the special condition Equation (\ref{c2=c2=0}),
Equation (\ref{delta z for Rtz of 1st PtI operation})
becomes 
\begin{eqnarray}
\langle (\Delta z) \rangle=c_1+\kappa_zt,
\label{delta z for Rtz of 1st PtI operation for
strong condition}
\end{eqnarray}
employing which, we can find 
\begin{eqnarray}
\lim_{n\rightarrow \infty}
T_{tz}=0.
\label{T}
\end{eqnarray}
Thus,
Equation (\ref{dz2dt Rst of 1st PtE}) becomes
\begin{eqnarray}
\frac{d}{dt}\langle (\Delta z)^2\rangle&
=&2\kappa_z\langle (\Delta z)
\rangle+2\kappa_{zz}-\frac{\kappa_{tz}}{\kappa_z}
\frac{d^2}{dt^2}\langle (\Delta z)^2\rangle.
\label{dz2dt Rst of 1st PtE with strong condition}
\end{eqnarray}
Similarily, for the special condition Equation (\ref{c2=c2=0}), 
Equation (\ref{delta z2}) becomes
\begin{eqnarray}
\langle (\Delta z)^2 \rangle=c'_1+\kappa_z^2t^2
+2\left(\kappa_zc_1+\kappa_{zz}-\kappa_z\kappa_{tz}\right)t.
\label{delta z2 for Rtz of 1st PtI operation for strong condition}
\end{eqnarray}
Thereafter, inserting Equations (\ref{delta z for
Rtz of 1st PtI operation for strong condition}),
(\ref{T}),
and (\ref{delta z2 for Rtz of 1st PtI operation
for strong condition})
into Equation (\ref{variance definition of Rst
of 1st PtI operation}) yields, for the limit $t\rightarrow t_{\infty}$,
\begin{equation}
\kappa_{zz}^{DV}=\frac{1}{2}\lim_{t\rightarrow
t_{\infty}}\frac{d\sigma^2}{dt}
=\kappa_{zz}-\kappa_z\kappa_{tz}.
\end{equation}
The latter equation is identical with Equation (\ref{kzzvd0}).
Although the special condition used in this subsection is not necessary,
it can simplify the derivation process significantly. 

\subsubsection{$R_{tzz}$ of the 1st-order PtI operation}
\label{Displacement Variance definition and Fick's
law definition for Rtzz of the 1st PtI operation}

In this subsection, we introduce another DIO, i.e., 
$R_{tzz}$ of the first-order PtI
operation. 
Equation (\ref{first PtE}) can be rewritten as
\begin{eqnarray}
\frac{\partial^3{F}}{\partial{t}\partial{z^2}}
=&&\frac{1}{\kappa_{zz}}\Bigg[\frac{\partial^2{F}}
{\partial{t^2}}
-\left(-\kappa_z\frac{\partial^2{F}}
{\partial{t}\partial{z}}
+\kappa_{zzz}
\frac{\partial^{4}{F}}{\partial{t}\partial{z^3}}
+\kappa_{zzzz}
\frac{\partial^{5}{F}}{\partial{t}\partial{z^4}}
+\cdots\right)\nonumber\\
&&-\left( \kappa_{tz}
\frac{\partial^3{F}}{\partial{t^2}
	\partial{z}}
+ \kappa_{ttz}\frac{\partial^4{F}}
{\partial{t^3}\partial{z}}
+ \kappa_{tttz}\frac{\partial^{5}{F}}
{\partial{t^{4}}\partial{z}}
+\cdots\right)\nonumber\\
&&-\left( \kappa_{tzz}
\frac{\partial^{4}{F}}{\partial{t^{2}}
	\partial{z^2}}
+ \kappa_{ttzz}\frac{\partial^{5}{F}}
{\partial{t^{3}}\partial{z^2}}
+ \kappa_{tttzz}\frac{\partial^{6}{F}}
{\partial{t^{4}}\partial{z^2}}
+\cdots\right)
-\cdots\Bigg].
\end{eqnarray}
Inserting the latter equation into
Equation (\ref{equation of F with
constant coefficient}) yields
\begin{eqnarray}
\frac{\partial{F}}{\partial{t}}=
&&\left(-\kappa_z\frac{\partial{F}}
{\partial{z}}+\kappa_{zz}
\frac{\partial^2{F}}{\partial{z^2}}
+\kappa_{zzz}
\frac{\partial^3{F}}{\partial{z^3}}
+\kappa_{zzzz}
\frac{\partial^4{F}}{\partial{z^4}}
+\cdots\right)
+\frac{\kappa_{tzz}}{\kappa_{zz}}\frac{\partial^2{F}}
{\partial{t^2}}\nonumber\\
&&+\left[\left(\kappa_{tz}+\frac{\kappa_{tzz}}{\kappa_{zz}}
\kappa_z\right)
\frac{\partial^2{F}}{\partial{t}\partial{z}}
+\left(\kappa_{ttz}-
\frac{\kappa_{tzz}}{\kappa_{zz}}
\kappa_{tz}\right)\frac{\partial^3{F}}
{\partial{t^2}\partial{z}}
+ \left(\kappa_{tttz}-\frac{\kappa_{tzz}}{\kappa_{zz}}\kappa_{ttz}
\right)\frac{\partial^4{F}}
{\partial{t^3}\partial{z}}
+\cdots\right]\nonumber\\
&&+ \left[\left(\kappa_{ttzz}
-\frac{\kappa_{tzz}}{\kappa_{zz}}\kappa_{tzz}\right)
\frac{\partial^4{F}}
{\partial{t^2}\partial{z^2}}
+ \left(\kappa_{tttzz}
-\frac{\kappa_{tzz}}{\kappa_{zz}}
\kappa_{ttzz}\right)
\frac{\partial^5{F}}
{\partial{t^3}\partial{z^2}}
+\cdots\right]
+\cdots.
\label{Rtzz equation of the first-order PtI operation}
\end{eqnarray}

From Equation 
(\ref{Rtzz equation of the first-order PtI operation}), 
we can find that the Fick's
law definition $\kappa_{zz}^{FL}$ is equal to $\kappa_{zz}$,
$\kappa_{zz}^{FL}=\kappa_{zz}$, which 
is identical with Equation (\ref{kzzfl0}). Thus, $\kappa_{zz}^{FL}$
is invariant for $R_{tzz}$ of the 1st-order PtI operation.
In order to explore the displacement 
variance definition $\kappa_{zz}^{DV}$ 
for $R_{tzz}$ of the 1st-order PtI
operation,
the following formulas have to be derived
from Equation (\ref{Rtzz equation of the first-order PtI operation})
\begin{eqnarray}
\frac{d}{dt}\langle (\Delta z)
\rangle&=&\kappa_z+\frac{\kappa_{tzz}}{\kappa_{zz}}
\frac{d^2}{dt^2}\langle (\Delta z)\rangle,
\label{dzdt Rtzz of 1st PtE}\\
\frac{d}{dt}\langle (\Delta z)^2\rangle&
=&2\kappa_z\langle (\Delta z)
\rangle+2\kappa_{zz}+\frac{\kappa_{tzz}}{\kappa_{zz}}
\frac{d^2}{dt^2}\langle (\Delta z)^2\rangle
-2\left(\kappa_{tz}+\frac{\kappa_{tzz}}{\kappa_{zz}}
\kappa_z\right)\frac{d}{dt}
\langle (\Delta z)\rangle
-2T_{tzz}
\label{dz2dt Rtzz of 1st PtE}
\end{eqnarray}
with
\begin{eqnarray}
T_{tzz}=&&\left(\kappa_{ttz}-\frac{\kappa_{tzz}}{\kappa_{zz}}
\kappa_{tz}\right)\frac{d^2}{dt^2}\langle (\Delta z)\rangle
+
\left(\kappa_{tttz}-\frac{\kappa_{tzz}}{\kappa_{zz}}\kappa_{ttz}
\right)\frac{d^3}{dt^3}\langle (\Delta z)\rangle+
\cdots.
\label{Ttz for Rtz of the first-order PtI operation}
\end{eqnarray}
Combining Equation (\ref{dzdt Rtzz of 1st PtE})
and (\ref{dz2dt Rtzz of 1st PtE})
gives
\begin{eqnarray}
\frac{1}{2}\frac{d\sigma^2}{dt}=&&\kappa_{zz}
+\frac{1}{2}\frac{\kappa_{tzz}}{\kappa_{zz}}
\left[\frac{d^2}{dt^2}
\langle (\Delta z)^2\rangle-2\langle (\Delta z)\rangle
\frac{d^2}{dt^2}\langle (\Delta z)\rangle
\right]
-\left(\kappa_{tz}+\frac{\kappa_{tzz}}{\kappa_{zz}}\kappa_z
\right)\frac{d}{dt}
\langle (\Delta z)\rangle
-T_{tzz}.
\label{variance definition of Rtzz of 1st PtI operation}
\end{eqnarray}
For the limit $t\rightarrow t_{\infty}$, 
the latter equation can be written as
\begin{eqnarray}
\kappa_{zz}^{DV}=\frac{1}{2}\lim_{t\rightarrow t_{\infty}}
\frac{d\sigma^2}{dt}
=&&\lim_{t\rightarrow t_{\infty}}\left[\kappa_{zz}
+\frac{1}{2}\frac{\kappa_{tzz}}{\kappa_{zz}}
\left[\frac{d^2}{dt^2}
\langle (\Delta z)^2\rangle-2\langle (\Delta z)\rangle
\frac{d^2}{dt^2}\langle (\Delta z)\rangle
\right]
-\left(\kappa_{tz}+\frac{\kappa_{tzz}}{\kappa_{zz}}
\kappa_z\right)\frac{d}{dt}
\langle (\Delta z)\rangle\right]
-\lim_{t\rightarrow t_{\infty}}T_{tzz}.\nonumber\\
\end{eqnarray}

In what follows, we firstly deal with 
Equation (\ref{Ttz for Rtz of the first-order PtI operation}) 
which is the function of the first-order moment 
$\langle (\Delta z)\rangle$.
From Equation (\ref{dzdt Rtzz of 1st PtE}) we can obtain
\begin{eqnarray}
\langle (\Delta z) \rangle&=&c_1
+c_2e^{\kappa_{zz}t/\kappa_{tzz}}+\kappa_zt,
\label{delta z for Rtzz}
\end{eqnarray}
where $c_1$ and $c_2$ are undetermined constants.
To proceed,
taking $n$th order derivative of Equation 
(\ref{dzdt Rtzz of 1st PtE})
over time $t$ 
gives
\begin{eqnarray}
\frac{d^n}{dt^n}\langle (\Delta z)\rangle=
\frac{\kappa_{tzz}}{\kappa_{zz}}\frac{d^{n+1}}
{dt^{n+1}}\langle (\Delta z)\rangle,
\end{eqnarray}
and the latter equation can be rewritten as
\begin{eqnarray}
\frac{d^{n+1}}{dt^{n+1}}\langle (\Delta z)\rangle=
\frac{\kappa_{zz}}{\kappa_{tzz}}\frac{d^n}{dt^n}
\langle (\Delta z)\rangle.
\label{Rtzz of the 1st PtI operation}
\end{eqnarray}
Combining the above equation and 
Equation (\ref{Ttz for Rtz of the first-order PtI operation}) 
we can find
\begin{eqnarray}
T_{tzz}=
\lim_{n\rightarrow \infty}
\left[-\kappa_{tz}\frac{\kappa_{tzz}}{\kappa_{zz}}
+\kappa_{ntz}\left(\frac{\kappa_{zz}}{\kappa_{tzz}}\right)^{n-2}
\right]\frac{d^2}{dt^2}
\langle\left(\Delta z  \right)\rangle.
\end{eqnarray}
Inserting Equation (\ref{delta z for Rtzz}) into
the latter equation yields
\begin{eqnarray}
T_{tzz}=c_2
\lim_{n\rightarrow \infty}
\left[-\kappa_{tz}\frac{\kappa_{zz}}{\kappa_{tzz}}
+\kappa_{ntz}\left(\frac{\kappa_{zz}}
{\kappa_{tzz}}\right)^{n}\right]
e^{\frac{\kappa_{zz}}{\kappa_{tzz}}t}.
\label{Ttzz for Rtzz of the first-order PtI operation}
\end{eqnarray}
The value of $T_{tzz}$ in Equation 
(\ref{Ttzz for Rtzz of the first-order PtI operation}) 
is determined 
by the constant $c_2$, the exponent function
$e^{\kappa_{zz}t/\kappa_{tzz}}$,  
$\kappa_{tz} \kappa_{zz}/\kappa_{tzz}$, and  
$\kappa_{ntz}(\kappa_{zz}/\kappa_{tzz})^n$. 

Since $\kappa_{tzz}<0$ as shown 
in Appendix \ref{Determine the sign of ktzz}, we
can find $\lim_{t\rightarrow t_{\infty}} 
e^{\kappa_{zz}t/\kappa_{tzz}}\approx 0$.
However, the value of
$\lim_{n\rightarrow \infty}
\left(\kappa_{zz}/\kappa_{tzz}\right)^{n}$
is determined by $\kappa_{tzz}$,
the form of which is too complicated and hard to be evaluated.
For the mathematical tractability, 
for Equation (\ref{delta z for Rtzz}) we directly
employ the special condition
$c_2=0$ 
and obtain
\begin{eqnarray}
\langle (\Delta z) \rangle&=&c_1+\kappa_zt.
\label{delta z for Rtzz-2}
\end{eqnarray}
Accordingly, we can find
\begin{eqnarray}
\frac{d^n}{dt^n}\langle (\Delta z)\rangle=0
\hspace{1cm} n=2,3,4,\cdots,
\end{eqnarray}
so
Equation (\ref{Ttzz for Rtzz of the first-order PtI operation}) 
becomes
\begin{eqnarray} 
T_{tzz}=0.
\label{Ttzz for Rtzz of 1st PtI operation}
\end{eqnarray}
Inserting the latter equation into
Equation (\ref{dz2dt Rtzz of 1st PtE})
gives
\begin{eqnarray}
\frac{d}{dt}\langle 
(\Delta z)^2\rangle&=&2\kappa_z\langle (\Delta z)
\rangle+2\kappa_{zz}+\frac{\kappa_{tzz}}{\kappa_{zz}}
\frac{d^2}{dt^2}\langle (\Delta z)^2\rangle
-2\left(\kappa_{tz}+\frac{\kappa_{tzz}}{\kappa_{zz}}
\kappa_z\right)\frac{d}{dt}
\langle (\Delta z)\rangle.
\label{d2zdt2 for Rtzz of the 1st PtI operation}
\end{eqnarray}
Combining Equations (\ref{variance definition
of Rtzz of 1st PtI operation})
and (\ref{Ttzz for Rtzz of 1st PtI operation}) we can find
\begin{eqnarray}
\frac{1}{2}\frac{d\sigma^2}{dt}=&&\kappa_{zz}
+\frac{1}{2}\frac{\kappa_{tzz}}{\kappa_{zz}}
\left[\frac{d^2}{dt^2}
\langle (\Delta z)^2\rangle-2\langle (\Delta z)\rangle
\frac{d^2}{dt^2}\langle (\Delta z)\rangle
\right]
-\left(\kappa_{tz}+\frac{\kappa_{tzz}}{\kappa_{zz}}
\kappa_z\right)\frac{d}{dt}
\langle (\Delta z)\rangle.
\label{dz2dt Rtzz of 1st PtE with Ttzz=0}
\end{eqnarray}
Because of Equation 
(\ref{delta z for Rtzz-2}),
we can obtain the following formula
from Equation 
(\ref{d2zdt2 for Rtzz of the 1st PtI operation})
\begin{eqnarray}
\langle (\Delta z)^2 \rangle&=&c'_1
+c'_2 e^{\kappa_{zz}t/\kappa_{tzz}}
+\left(2\kappa_z\kappa_{tz}-2\kappa_z c_1- 2\kappa_{zz}+2
\kappa_z^2 \frac{\kappa_{tzz}}{\kappa_{zz}} \right)t
+2 \kappa_z^2 \frac{\kappa_{tzz}}{\kappa_{zz}}
+2\kappa_{z}^2t^2,
\end{eqnarray}
where $c'_1$ and $c'_2$ are all 
the undetermined constants.
For the special condition $c'_2=0$,
the latter equation becomes
\begin{eqnarray}
\langle (\Delta z)^2 \rangle&=&c'_1
+\left(2\kappa_z\kappa_{tz}
-2\kappa_z c_1- 2\kappa_{zz}+2
\kappa_z^2 \frac{\kappa_{tzz}}{\kappa_{zz}} \right)t
+2 \kappa_z^2 \frac{\kappa_{tzz}}{\kappa_{zz}}
+2\kappa_{z}^2t^2,
\label{delta z2 for Rtzz of 1st PtI operation}
\end{eqnarray}
Inserting Equations (\ref{delta z for Rtzz-2})
and (\ref{delta z2 for Rtzz of 1st PtI operation})
into (\ref{dz2dt Rtzz of 1st PtE with Ttzz=0})
we can easily obtain  
\begin{eqnarray}
\kappa_{zz}^{DV}=\frac{1}{2}\lim_{t\rightarrow
t_{\infty}}\frac{d\sigma^2}{dt}&=&\kappa_{zz}
-\kappa_{tz}\kappa_{z},
\end{eqnarray}
which is identical with Equation (\ref{kzzvd0}).
From the above deduction, we can find that at least for the
special condition $c_2=c'_2=0$ the displacement
variance definition $\kappa_{zz}^{DV}$ is an invariance
for $R_{tzz}$ of the
first-order PtI operation.
Of course, it is possible that 
the special condition is unnecessary.

\subsubsection{$R_{ntmz}$ of the 1st-order PtI operation}
\label{Two definitions 
for Rntmz of the 1st-order PtI operation}

To combine Equation
(\ref{equation of F with constant coefficient}) and
the deformation of Equation (\ref{first PtE}),
one can produce a lot of the EIDFs
by the DIOs 
for the first-order PtI operation, 
i.e., $R_{ntmz}$ with $n=2,3,4,\cdots,
 m=1,2,3,\cdots$.
Because Equation (\ref{first PtE}) 
does not contain the second-order spatial derivative 
terms, 
any replacement manipulation, i.e., the DIO,  of 
the first-order PtI operation, cannot 
influence the Fick's law definition. 
In the following, we explore the 
displacement variance definition for $R_{ntmz}$ of 
the first-order PtI operation.

The equation of $\langle (\Delta z) \rangle$
can be derived as follows
\begin{eqnarray}
\frac{d}{dt}\langle (\Delta z)
\rangle&=&\kappa_z+\frac{\kappa_{ntmz}}{\kappa_{(n-1)tmz}}
\frac{d^2}{dt^2}\langle (\Delta z)\rangle.
\label{dzdt for Rntmz of 1st PtI operation}
\end{eqnarray}
The solution of the latter equation can be found
\begin{eqnarray}
\langle (\Delta z) \rangle&=&c_1
+c_2e^{t\kappa_{(n-1)tmz}/\kappa_{ntmz}}+\kappa_zt.
\label{delta z for Rntmz}
\end{eqnarray}
Similarily, 
from the EIDF for $R_{ntmz}$ of the 1st-order PtI operation
the equation of $d\langle (\Delta z)^2 \rangle/dt$
can be found
\begin{eqnarray}
\frac{d}{dt}\langle (\Delta z)^2\rangle&
=&2\kappa_z\langle (\Delta z)
\rangle+2\kappa_{zz}+\frac{\kappa_{ntmz}}{\kappa_{(n-1)tmz}}
\frac{d^2}{dt^2}\langle (\Delta z)\rangle
-2\left(\kappa_{tz}+\frac{\kappa_{ntmz}}{\kappa_{(n-1)tmz}}
\kappa_z\right)\frac{d}{dt}
\langle (\Delta z)\rangle
-2T_{ntmz}
\label{dz2dt Rntmz of 1st PtE}
\end{eqnarray}
with
\begin{eqnarray}
T_{ntmz}=&&\left(\kappa_{ttz}-\frac{\kappa_{ntmz}}{\kappa_{(n-1)tmz}}
\kappa_{tz}\right)\frac{d^2}{dt^2}\langle (\Delta z)\rangle
+\left(\kappa_{tttz}-\frac{\kappa_{ntmz}}{\kappa_{(n-1)tmz}}
\kappa_{ttz}\right)\frac{d^3}{dt^3}\langle (\Delta z)\rangle+
\cdots.
\label{Tntmz for the first-order PtI operation}
\end{eqnarray}
From Equation (\ref{dzdt for Rntmz of 1st PtI operation})
we can obtain 
\begin{eqnarray}
\frac{d^n}{dt^n}\langle (\Delta z)
\rangle&=&\frac{\kappa_{ntmz}}{\kappa_{(n-1)tmz}}
\frac{d^{n+1}}{dt^{n+1}}\langle (\Delta z)\rangle,
\end{eqnarray}
using which we can find that 
Equation (\ref{Tntmz for the first-order PtI operation})
becomes
\begin{eqnarray}
T_{ntmz}=c_2
\lim_{n\rightarrow \infty}
\left[-\kappa_{tz}   \frac{\kappa_{(n-1)tmz}}{\kappa_{ntmz}}
+\kappa_{ntz}\left(\frac{\kappa_{(n-1)tmz}}{\kappa_{ntmz}}
 \right)^{n}\right]
e^{t\kappa_{(n-1)tmz}/\kappa_{ntmz}}.
\label{Tntmz by sum}
\end{eqnarray}

Because the forms of the parameters 
$\kappa_{(n-1)tmz}$ and $\kappa_{ntmz}$
are very complicated and hard to evaluate, for the simplification 
we only consider the case satisfying the special condition
$c_2=0$ in this subsection. 
Thus, Equation (\ref{delta z for Rntmz}) can be simplified as
\begin{eqnarray}
\langle (\Delta z) \rangle&=&c_1+\kappa_zt,
\label{delta z for Rntmz with condition}
\end{eqnarray}
consequently, Equation (\ref{Tntmz by sum}) becomes
\begin{eqnarray}
T_{ntmz}= 0.
\end{eqnarray}
Inserting the latter equation into Equation 
(\ref{dz2dt Rntmz of 1st PtE})
gives
\begin{eqnarray}
\frac{d}{dt}\langle (\Delta z)^2\rangle&
=&2\kappa_z\langle (\Delta z)
\rangle+2\kappa_{zz}+\frac{\kappa_{ntmz}}{\kappa_{(n-1)tmz}}
\frac{d^2}{dt^2}\langle (\Delta z)\rangle
-2\left(\kappa_{tz}+\frac{\kappa_{ntmz}}{\kappa_{(n-1)tmz}}
\kappa_z\right)\frac{d}{dt}
\langle (\Delta z)\rangle,
\label{dz2dt Rntmz of 1st PtE with condition}
\end{eqnarray}
the solution of which can be
obtained by employing
Equation (\ref{delta z for Rntmz})
\begin{eqnarray}
\langle (\Delta z)^2 \rangle&=&c'_1+c'_2
e^{t\kappa_{(n-1)tmz}/\kappa_{ntmz}}
+\left(2\kappa_z\kappa_{tz}-2\kappa_z c_1- 2\kappa_{zz}
+2 \kappa_z^2 \frac{\kappa_{ntmz}}{\kappa_{(n-1)tmz}} \right)t
+2 \kappa_z^2 \frac{\kappa_{ntmz}}{\kappa_{(n-1)tmz}}
+2\kappa_{z}^2t^2.
\label{delta z2 for Rntmz of 1st PtI operation}
\end{eqnarray}
For the special condition $c'_2=0$ the latter equation becomes
\begin{eqnarray}
\langle (\Delta z)^2 \rangle&=&c'_1
+\left(2\kappa_z\kappa_{tz}-2\kappa_z c_1- 2\kappa_{zz}
+2 \kappa_z^2 \frac{\kappa_{ntmz}}{\kappa_{(n-1)tmz}} \right)t
+2 \kappa_z^2 \frac{\kappa_{ntmz}}{\kappa_{(n-1)tmz}}
+2\kappa_{z}^2t^2.
\label{delta z2 for Rntmz of 1st PtI operation with condition}
\end{eqnarray}
Considering the Equations (\ref{dzdt for Rntmz of 1st PtI operation})
and (\ref{dz2dt Rntmz of 1st PtE with condition}),
we can find
\begin{eqnarray}
\frac{1}{2}\frac{d\sigma^2}{dt}
=\kappa_{zz}+\frac{1}{2}\frac{\kappa_{ntmz}}{\kappa_{(n-1)tmz}}
\left[\frac{d^2}{dt^2}\langle (\Delta z)^2\rangle
-2\langle (\Delta z)\rangle
\frac{d^2}{dt^2}\langle (\Delta z)\rangle\right]
-\left(\kappa_{tz}+\frac{\kappa_{ntmz}}{\kappa_{(n-1)tmz}}
\kappa_z\right)\frac{d}{dt}
\langle (\Delta z)\rangle.
\end{eqnarray}
Inserting Equations (\ref{delta z for Rntmz with condition})
and (\ref{delta z2 for Rntmz of 1st PtI operation with condition})
into the latter equation yields
\begin{eqnarray}
\kappa_{zz}^{DV}=\frac{1}{2}\lim_{t\rightarrow t_{\infty}}
\frac{d\sigma^2}{dt}&=&\kappa_{zz}-\kappa_{tz}\kappa_{z},
\end{eqnarray}
which is identical with Equation (\ref{kzzvd0}). 

From the investigation in this subsection,
we can find that for $R_{ntmz}$ of
the 1st-order PtI operation, 
the Fick's law definition $\kappa_{zz}^{FL}$
is invariant, and the displacement variance definition 
$\kappa_{zz}^{DV}$ is also 
an invariance
at least for the special condition. 

\subsection{The 2nd-order PtI operation}
\label{The 2nd PtI operation}

For $n=2$, i.e., the second-order 
PtI operation, Equation (\ref{n PtE}) becomes
\begin{eqnarray}
	\frac{\partial^{3}{F}}{\partial{t^{3}}}=
	&&\left(-\kappa_z\frac{\partial^{3}{F}}
	{\partial{t^2}\partial{z}}+\kappa_{zz}
	\frac{\partial^{4}{F}}{\partial{t^2}\partial{z^2}}
	+\kappa_{zzz}
	\frac{\partial^{5}{F}}{\partial{t^2}\partial{z^3}}
	+\kappa_{zzzz}
	\frac{\partial^{6}{F}}{\partial{t^2}\partial{z^4}}
	+\cdots\right)
	+\left( \kappa_{tz}
	\frac{\partial^{4}{F}}{\partial{t^{3}}
		\partial{z}}
	+ \kappa_{ttz}\frac{\partial^{5}{F}}
	{\partial{t^{4}}\partial{z}}
	+ \kappa_{tttz}\frac{\partial^{6}{F}}
	{\partial{t^{5}}\partial{z}}
	+\cdots\right)\nonumber\\
	&&+\left( \kappa_{tzz}
	\frac{\partial^{5}{F}}{\partial{t^{3}}
		\partial{z^2}}
	+ \kappa_{ttzz}\frac{\partial^{6}{F}}
	{\partial{t^{4}}\partial{z^2}}
	+ \kappa_{tttzz}\frac{\partial^{7}{F}}
	{\partial{t^{5}}\partial{z^2}}
	+\cdots\right)
	+\cdots.
	\label{2nd PtE}
\end{eqnarray}
As shown in subsection \ref{The 1st PtI operation},
rewritting the latter equation and 
inserting the results into Equation 
(\ref{equation of F with constant coefficient}),
we can obtain the new EIDFs. 

\subsubsection{$R_{ttz}$ of the 2nd-order PtI operation}
\label{Displacement Variance definition and Fick's
law definition for Rttz of the 2nd PtI operation}

In this part, by pulling out the term
$\partial^3{F}/(\partial{t^2}\partial{z})$
on the right hand side of Equation (\ref{2nd PtE}),
we rewrite this equation
and obtain
\begin{eqnarray}
\frac{\partial^{3}{F}}{\partial{t^2}\partial{z}}=&&
-\frac{1}{\kappa_z}\frac{\partial^{3}{F}}{\partial{t^{3}}}
+\left(\frac{1}{\kappa_z}\kappa_{zz}\frac{\partial^{4}{F}}
{\partial{t^2}\partial{z^2}}
+\frac{1}{\kappa_z}\kappa_{zzz}\frac{\partial^{5}{F}}
{\partial{t^2}\partial{z^3}}
+\frac{1}{\kappa_z}\kappa_{zzzz}\frac{\partial^{6}{F}}
{\partial{t^2}\partial{z^4}}
+\cdots\right)\nonumber\\
&&+\left(\frac{1}{\kappa_z}\kappa_{tz}\frac{\partial^{4}{F}}
{\partial{t^{3}}\partial{z}}
+\frac{1}{\kappa_z}\kappa_{ttz}\frac{\partial^{5}{F}}
{\partial{t^{4}}\partial{z}}
+\frac{1}{\kappa_z}\kappa_{tttz}\frac{\partial^{6}{F}}
{\partial{t^{5}}\partial{z}}
+\cdots\right)\nonumber\\
&&+\left(\frac{1}{\kappa_z}\kappa_{tzz}\frac{\partial^{5}{F}}
{\partial{t^{3}}\partial{z^2}}
+\frac{1}{\kappa_z}\kappa_{ttzz}\frac{\partial^{6}{F}}
{\partial{t^{4}}\partial{z^2}}
+\frac{1}{\kappa_z}\kappa_{tttzz}\frac{\partial^{7}{F}}
{\partial{t^{5}}\partial{z^2}}
+\cdots\right)
+\cdots.
\label{d3Fdt2dz}
\end{eqnarray}
After multiplying the latter equation by $\kappa_{ttz}$,
we can find
\begin{eqnarray}
\kappa_{ttz}\frac{\partial^{3}{F}}{\partial{t^2}\partial{z}}=&&
-\frac{\kappa_{ttz}}{\kappa_z}\frac{\partial^{3}{F}}{\partial{t^{3}}}
+\left(\frac{\kappa_{ttz}}{\kappa_z}\kappa_{zz}
\frac{\partial^{4}{F}}{\partial{t^2}\partial{z^2}}
+\frac{\kappa_{ttz}}{\kappa_z}\kappa_{zzz}
\frac{\partial^{5}{F}}{\partial{t^2}\partial{z^3}}
+\frac{\kappa_{ttz}}{\kappa_z}\kappa_{zzzz}
\frac{\partial^{6}{F}}{\partial{t^2}\partial{z^4}}
+\cdots\right)\nonumber\\
&&+\left(\frac{\kappa_{ttz}}{\kappa_z}\kappa_{tz}
\frac{\partial^{4}{F}}{\partial{t^{3}}\partial{z}}
+\frac{\kappa_{ttz}}{\kappa_z}\kappa_{ttz}\frac{\partial^{5}{F}}
{\partial{t^{4}}\partial{z}}
+\frac{\kappa_{ttz}}{\kappa_z}\kappa_{tttz}\frac{\partial^{6}{F}}
{\partial{t^{5}}\partial{z}}
+\cdots\right)\nonumber\\
&&+\left(\frac{\kappa_{ttz}}{\kappa_z}\kappa_{tzz}\frac{\partial^{5}{F}}
{\partial{t^{3}}\partial{z^2}}
+\frac{\kappa_{ttz}}{\kappa_z}\kappa_{ttzz}\frac{\partial^{6}{F}}
{\partial{t^{4}}\partial{z^2}}
+\frac{\kappa_{ttz}}{\kappa_z}\kappa_{tttzz}\frac{\partial^{7}{F}}
{\partial{t^{5}}\partial{z^2}}
+\cdots\right)
+\cdots.
\label{kttz-d3Fdt2dz}
\end{eqnarray}
Replacing the term $\partial^3{F}/(\partial{t^2}\partial{z})$ in Equation
(\ref{equation of F with
constant coefficient}) by the terms on the right-hand
side of the latter equation
(\ref{equation of F with constant coefficient}) gives
\begin{eqnarray}
\frac{\partial{F}}{\partial{t}}=
&&\left(-\kappa_z\frac{\partial{F}}
{\partial{z}}+\kappa_{zz}
\frac{\partial^2{F}}{\partial{z^2}}
+\kappa_{zzz}
\frac{\partial^3{F}}{\partial{z^3}}
+\kappa_{zzzz}
\frac{\partial^4{F}}{\partial{z^4}}
+\cdots\right)-\frac{\kappa_{ttz}}{\kappa_z}\frac{\partial^{3}{F}}
{\partial{t^{3}}}\nonumber\\
&&+\kappa_{tz}
\frac{\partial^2{F}}{\partial{t}\partial{z}}
+\kappa_{tzz}
\frac{\partial^3{F}}{\partial{t}\partial{z^2}}
+\kappa_{tzzz}
\frac{\partial^4{F}}{\partial{t}\partial{z^3}}
+\kappa_{tzzzz}
\frac{\partial^5{F}}{\partial{t}\partial{z^4}}
+\kappa_{tzzzzz}
\frac{\partial^6{F}}{\partial{t}\partial{z^5}}
+\cdots
\nonumber\\
&&+\left(\frac{\kappa_{ttz}}{\kappa_z}\kappa_{zz}+\kappa_{ttzz}\right)
\frac{\partial^{4}{F}}{\partial{t^2}\partial{z^2}}
+\left(\frac{\kappa_{ttz}}{\kappa_z}\kappa_{zzz}+\kappa_{ttzzz}\right)
\frac{\partial^{5}{F}}{\partial{t^2}\partial{z^3}}+\cdots
\nonumber\\
&&+\left(\frac{\kappa_{ttz}}{\kappa_z}\kappa_{tz}+\kappa_{tttz}\right)
\frac{\partial^{4}{F}}{\partial{t^{3}}\partial{z}}
+\left(\frac{\kappa_{ttz}}{\kappa_z}\kappa_{ttz}+\kappa_{ttttz}\right)
\frac{\partial^{5}{F}}{\partial{t^{4}}\partial{z}}
+\cdots\nonumber\\
&&+\left(\frac{\kappa_{ttz}}{\kappa_z}\kappa_{tzz}+\kappa_{tttzz}\right)
\frac{\partial^{5}{F}}{\partial{t^{3}}\partial{z^2}}
+\left(\frac{\kappa_{ttz}}{\kappa_z}\kappa_{ttzz}+\kappa_{ttttzz}\right)
\frac{\partial^{6}{F}}{\partial{t^{4}}\partial{z^2}}
+\cdots\nonumber\\
&&+\left(\frac{\kappa_{ttz}}{\kappa_z}\kappa_{tzzz}+\kappa_{tttzzz}\right)
\frac{\partial^{6}{F}}{\partial{t^{3}}\partial{z^3}}
+\cdots,
\label{Rttz equation}
\end{eqnarray}
The manipulation in this part, which is one of the DIOs, 
is called the $R_{ttz}$ of the 2nd-order PtI
operation.

As shown in Equation (\ref{Rttz equation}), 
the Fick's law definition
$\kappa_{zz}^{FL}$ is equal to $\kappa_{zz}$, 
$\kappa_{zz}^{FL}=\kappa_{zz}$,
which is identical with Equation (\ref{kzzfl0}). 
Thus, the Fick's law definition is 
not changed by $R_{ttz}$ of the 2nd-order PtI operation.
In the following, we explore the displacement variance 
definition $\kappa_{zz}^{DV}$ for the new manipulation.

From Equation (\ref{Rttz equation}) 
we can derive the following formulas
\begin{eqnarray}
\frac{d}{dt}\langle (\Delta z)
\rangle&=&\kappa_z-\frac{\kappa_{ttz}}{\kappa_{z}}
\frac{d^3}{dt^3}\langle (\Delta z)\rangle,
\label{dzdt for Rttz of 2nd PtE}\\
\frac{d}{dt}\langle (\Delta z)^2\rangle&
=&2\kappa_z\langle (\Delta z)
\rangle+2\kappa_{zz}-\frac{\kappa_{ttz}}{\kappa_{z}}
\frac{d^3}{dt^3}\langle (\Delta z)^2\rangle
-2\kappa_{tz}\frac{d}{dt}
\langle (\Delta z)\rangle
-2T_{ttz}
\label{dz2dt Rttz of 2nd PtE}
\end{eqnarray}
with
\begin{eqnarray}
T_{ttz}=&&\left(\frac{\kappa_{ttz}}{\kappa_{z}}\kappa_{tz}
+\kappa_{tttz}\right)\frac{d^3}{dt^3}\langle (\Delta z)\rangle
+
\left(\frac{\kappa_{ttz}}{\kappa_{z}}\kappa_{ttz}+\kappa_{ttttz}\right)
\frac{d^4}{dt^4}\langle (\Delta z)\rangle+
\cdots.
\label{T_ttz}
\end{eqnarray}
Taking the $(n-1)$th-order derivative of Equation
(\ref{dzdt for Rttz of 2nd PtE})
yields
\begin{eqnarray}
\frac{d^{n+2}}{dt^{n+2}}\langle (\Delta z)\rangle
=-\frac{\kappa_{z}}{\kappa_{ttz}}
\frac{d^n}{dt^n}\langle (\Delta z)
\rangle.
\label{the n-order derivative of dz2dt Rttz of 2nd PtE}
\end{eqnarray}
By inserting the latter equation into Equation (\ref{T_ttz})
we can obtain
\begin{eqnarray}
T_{ttz}=&&\lim_{n\rightarrow\infty}\left(\frac{\kappa_{ttz}}
{\kappa_{z}}\kappa_{tz}
+\kappa_{ntz}\left(-\frac{\kappa_{z}}
{\kappa_{ttz}}\right)^{\lceil n/2-2\rceil}\right)
\frac{d^3}{dt^3}\langle (\Delta z)\rangle
+\left(\frac{\kappa_{ttz}}{\kappa_{z}}\kappa_{ttz}
+\kappa_{ntz}\left(-\frac{\kappa_{z}}
{\kappa_{ttz}}\right)^{\lceil n/2-2\rceil}\right)
\frac{d^4}{dt^4}\langle (\Delta z)\rangle,
\label{simplified T_ttz}
\end{eqnarray}
where the subscript $ntz$ of $\kappa_{ntz}$ denotes 
$n$ letters $t$ and one letter $z$, e.g.,
$\kappa_{ttz}$ can be rewritten as $\kappa_{2tz}$,
$\kappa_{tttz}$ can be rewritten as $\kappa_{3tz}$, and so on.
The symbol $\lceil n/2-2\rceil$ indicates
the smallest integer not smaller than $(n/2-2)$.

If considering the parameter $\kappa_{ntz}$
(see Equation (\ref{kntz-k2tz}) in Appendix), we can find
\begin{eqnarray}
\kappa_{ntz}\left(-\frac{\kappa_{z}}{\kappa_{ttz}}\right)^{\{n/2-2\}}
\approx
\left(\frac{1}{2D}\right)^{n-2}\kappa_{2tz}\left(\sqrt{-\frac{\kappa_{z}}
{\kappa_{ttz}}}\right)^{n-4}
=\frac{1}{4}\left(\frac{1}{D}\right)^{n-2}\kappa_{2tz}\left(\frac{1}{2}
\sqrt{-\frac{\kappa_{z}}{\kappa_{ttz}}}\right)^{n-4}.
\label{T-term}
\end{eqnarray}
Combining the parameters $\kappa_{tz}$  and 
$\kappa_{ttz}$ (see Equation (\ref{ktz for first order xi}) 
and  (\ref{kttz-value}) in Appendix), we find that 
Equation (\ref{T-term}) becomes 
\begin{eqnarray}
\kappa_{ntz}\left(-\frac{\kappa_{z}}{\kappa_{ttz}}
\right)^{\{n/2-2\}}\approx
\frac{1}{4D^2}\frac{\left(0.8320502943\right)^{n}}{0.4792899408}
\left(-\frac{13}{108}\frac{v}{D^2}\xi\right).
\label{T-term-2}
\end{eqnarray}
Since $\lim_{n\rightarrow\infty}\left(0.8320502943\right)^{n}=0$,
Equation (\ref{T-term-2}) becomes
\begin{eqnarray}
\lim_{n\rightarrow\infty}\kappa_{ntz}\left(-\frac{\kappa_{z}}
{\kappa_{ttz}}\right)^{\{n/2-2\}}\approx
\lim_{n\rightarrow\infty}\frac{1}{4D^2}\frac{\left(0.8320502943\right)^{n}}
{0.4792899408}
\left(-\frac{13}{108}\frac{v}{D^2}\xi\right)=0,
\label{limit of T-term}
\end{eqnarray}
where $D$, $\xi$ and $v$ are all constants.
Inserting the latter result into Equation (\ref{simplified T_ttz}) gives
\begin{eqnarray}
T_{ttz}=&&\frac{\kappa_{ttz}}{\kappa_{z}}\kappa_{tz}
\frac{d^3}{dt^3}\langle (\Delta z)\rangle
+\frac{\kappa_{ttz}}{\kappa_{z}}\kappa_{ttz}
\frac{d^4}{dt^4}\langle (\Delta z)\rangle.
\label{Tttz-2}
\end{eqnarray}
Actually, Equation (\ref{Tttz-2}) can be simplified again.
From Equation (\ref{the n-order derivative of dz2dt Rttz of 2nd PtE})
we can find
\begin{eqnarray}
\frac{d^{3}}{dt^{3}}\langle (\Delta z)\rangle
&=&\kappa_{z}\frac{\kappa_{z}}{\kappa_{ttz}}
-\frac{\kappa_{z}}{\kappa_{ttz}}\frac{d}{dt}
\langle (\Delta z)
\rangle,\\
\frac{d^{4}}{dt^{4}}\langle (\Delta z)\rangle
&=&-\frac{\kappa_{z}}{\kappa_{ttz}}
\frac{d^2}{dt^2}\langle (\Delta z)\rangle,
\end{eqnarray}
inserting which into Equation (\ref{Tttz-2}), one obtains
\begin{eqnarray}
T_{ttz}=\kappa_{tz}\kappa_z-\kappa_{ttz}\frac{d^2}{dt^2}\langle
(\Delta z)\rangle
-\kappa_z\frac{d}{dt}\langle (\Delta z)\rangle.
\label{simplified limit T}
\end{eqnarray}

Combining Equations (\ref{dz2dt Rttz of 2nd PtE}) and
(\ref{simplified limit T}) gives
\begin{eqnarray}
\frac{d}{dt}\langle (\Delta z)^2\rangle&
=&2\kappa_{zz}-2\kappa_{tz}\kappa_z+2\kappa_z\langle
(\Delta z)
\rangle-\frac{\kappa_{ttz}}{\kappa_{z}}
\frac{d^3}{dt^3}\langle (\Delta z)^2\rangle
+2\kappa_{ttz}\frac{d^2}{dt^2}\langle (\Delta z)\rangle,
\label{dz2dt Rttz of 2nd PtE-2}
\end{eqnarray}
which can be rewritten as
\begin{eqnarray}
\frac{d^3}{dt^3}\langle (\Delta z)^2\rangle+\frac{\kappa_z}
{\kappa_{ttz}}\frac{d}{dt}\langle (\Delta z)^2\rangle
=2\kappa_{zz}\frac{\kappa_z}{\kappa_{ttz}}-2\kappa_{tz}
\kappa_z\frac{\kappa_z}{\kappa_{ttz}}
+2\kappa_z\frac{\kappa_z}{\kappa_{ttz}}\langle (\Delta z)\rangle
+2\kappa_{z}\frac{d^2}{dt^2}\langle (\Delta z)\rangle.
\label{simplified dz2dt Rttz of 2nd PtE-2}
\end{eqnarray}
In order to derive the displacement variance
\begin{equation}
\kappa_{zz}^{DV}=\frac{1}{2}\lim_{t\rightarrow\infty}
\frac{d\sigma^2}{dt}=\frac{1}{2}\lim_{t\rightarrow\infty}
\frac{d}{dt}\left(\langle\left(\Delta z\right)^2 \rangle-\langle
\left(\Delta z\right)\rangle^2\right),
\label{1}
\end{equation}
we have to obtain the formulas of the first- 
and second-order moment
$\langle\left(\Delta z\right)\rangle$ and
$\langle\left(\Delta z\right)^2 \rangle$,
which are the solution of Equations (\ref{dzdt for Rttz of 2nd PtE})
and (\ref{simplified dz2dt Rttz of 2nd PtE-2}), respectively.
$\langle\left(\Delta z\right)\rangle$ can be obtained as
\begin{eqnarray}
\langle (\Delta z)\rangle&=&c_1+c_2e^{-t\sqrt{-\kappa_{z}/\kappa_{ttz}}}
+c_3e^{t\sqrt{-\kappa_{z}/\kappa_{ttz}}}+\kappa_zt,
\label{delta z for Rttz}
\end{eqnarray}
where $c_1$, $c_2$ and $c_3$ are the undetermined constants.
After inserting Equation (\ref{delta z for Rttz}) 
into Equation
(\ref{simplified dz2dt Rttz of 2nd PtE-2}),
we can find 
\begin{eqnarray}
\langle (\Delta z)^2\rangle&=&c'_1
+c'_2e^{-t\sqrt{-\kappa_{z}/\kappa_{ttz}}}
+c'_3e^{t\sqrt{-\kappa_{z}/\kappa_{ttz}}}+2\left(\kappa_{zz}
-\kappa_{tz}\kappa_{z}+\kappa_{z}c_1\right)t
+\kappa_{z}^2 t^2
\label{delta2 z for Rttz}
\end{eqnarray}
with the undetermined constants $c'_1$, $c'_2$ and $c'_3$.

Equation (\ref{dzdt for Rttz of 2nd PtE}) can be rewritten as
\begin{eqnarray}
\kappa_z=\frac{d}{dt}\langle (\Delta z)
\rangle+\frac{\kappa_{ttz}}{\kappa_{z}}
\frac{d^3}{dt^3}\langle (\Delta z)\rangle.
\label{185}
\end{eqnarray}
Combining Equations (\ref{dz2dt Rttz of 2nd PtE-2}) 
and (\ref{185}) gives
\begin{eqnarray}
\frac{1}{2}\lim_{t\rightarrow t_{\infty}}\frac{d\sigma^2}{dt}
=\kappa_{zz}-\kappa_{tz}\kappa_z
+\frac{\kappa_{ttz}}{\kappa_z}\langle (\Delta z)\rangle
\frac{d^3}{dt^3}\langle (\Delta z)\rangle
+\kappa_{ttz}\frac{d^2}{dt^2}\langle (\Delta z)\rangle
-\frac{1}{2}\frac{\kappa_{ttz}}{\kappa_{z}}
\frac{d^3}{dt^3}\langle (\Delta z)^2\rangle,
\label{kzzvd for Rttz of 2nd PtE}
\end{eqnarray}
where the formula $\sigma^2=\langle
\left(\Delta z\right)^2 \rangle
-\langle\left(\Delta z\right)\rangle^2$ is used.
Inserting Equations (\ref{delta z for Rttz}) and
(\ref{delta2 z for Rttz}) into Equation (\ref{kzzvd for Rttz of 2nd PtE}) gives
\begin{eqnarray}
\frac{1}{2}\lim_{t\rightarrow t_{\infty}}\frac{d\sigma^2}{dt}
=&&\kappa_{zz}-\kappa_{tz}\kappa_z
+c_2^2\sqrt{-\frac{\kappa_z}{\kappa_{tz}}}
\lim_{t\rightarrow t_{\infty}}
e^{-2t\sqrt{-\kappa_{z}/\kappa_{ttz}}}
-c_3^2\sqrt{-\frac{\kappa_{z}}{\kappa_{ttz}}}\lim_{t\rightarrow t_{\infty}}e^{2t\sqrt{-\kappa_{z}/\kappa_{ttz}}}\nonumber\\
&&+\lim_{t\rightarrow t_{\infty}}
\left(c_1c_2\sqrt{-\frac{\kappa_{z}}
{\kappa_{ttz}}}-c_2\kappa_z-\frac{1}{2}c'_2 
\sqrt{-\frac{\kappa_{z}}{\kappa_{ttz}}}
+c_2\kappa_z t \sqrt{-\frac{\kappa_{z}}{\kappa_{ttz}}}\right) 
e^{-t\sqrt{-\kappa_{z}/\kappa_{ttz}}}\nonumber\\
&&+\lim_{t\rightarrow t_{\infty}}\left(-c_1c_3\sqrt{-\frac{\kappa_{z}}
{\kappa_{ttz}}}-c_3\kappa_z-\frac{1}{2}c'_3 
\sqrt{-\frac{\kappa_{z}}{\kappa_{ttz}}}
-c_3\kappa_z t \sqrt{-\frac{\kappa_{z}}{\kappa_{ttz}}}\right) 
e^{-t\sqrt{-\kappa_{z}/\kappa_{ttz}}}.
\label{kzzVD for Rttz of 2nd PtE-2}
\end{eqnarray}
From Equations
(\ref{kz for first order xi}) and (\ref{kttz-value}) in Appendix
the relation $\kappa_{z}/\kappa_{ttz}<0$ can be found, therewith 
one have
\begin{eqnarray}
\lim_{t\rightarrow t_{\infty}}e^{-t\sqrt{-\kappa_{z}/\kappa_{ttz}}}=0.
\label{limt exponent function}
\end{eqnarray}
Inserting Equation  (\ref{limt exponent function})
into Equation (\ref{kzzVD for Rttz of 2nd PtE-2})
gives
\begin{eqnarray}
\frac{1}{2}\lim_{t\rightarrow t_{\infty}}\frac{d\sigma^2}{dt}
=&&\kappa_{zz}-\kappa_{tz}\kappa_z
-c_3^2\sqrt{-\frac{\kappa_{z}}{\kappa_{ttz}}}
\lim_{t\rightarrow t_{\infty}}e^{2t\sqrt{-\kappa_{z}/\kappa_{ttz}}}.
\label{kzzVD for Rttz of 2nd PtE-3}
\end{eqnarray}
Because $\sigma^2$ is nonnegative, 
the third term on the right-hand side of the 
latter equation has to be zero. Thus, the condition 
$c_3=0$ need to be satisfied, and Equation 
(\ref{kzzVD for Rttz of 2nd PtE-3}) becomes 
\begin{eqnarray}
\frac{1}{2}\lim_{t\rightarrow t_{\infty}}\frac{d\sigma^2}{dt}
=&&\kappa_{zz}-\kappa_{tz}\kappa_z,
\label{kzzVD for Rttz of 2nd PtE-4}
\end{eqnarray}
which is identical with Equation (\ref{kzzvd0}). 
Therefore, the displacement variance definition 
is an invariance for $R_{ttz}$ of the 2nd-order
PtI operation.

Actually, if adopting the special condition
$c_2=c_3=c'_2=c'_3=0$, 
we can  obtain Equation 
(\ref{kzzVD for Rttz of 2nd PtE-4}) easily.  
For the special condition, Equations 
(\ref{delta z for Rttz})
and (\ref{delta2 z for Rttz}) become
 \begin{eqnarray}
\langle (\Delta z)\rangle&=&c_1+\kappa_zt,
\label{simplified delta z for Rttz}\\
\langle (\Delta z)^2\rangle&=&c'_1
+2\left(\kappa_{zz}
-\kappa_{tz}\kappa_{z}+\kappa_{z}c_1\right)t
+\kappa_{z}^2 t^2.
\label{simplified delta2 z for Rttz}
 \end{eqnarray}
Combining Equations (\ref{dzdt for Rttz of 2nd PtE})
and (\ref{dz2dt Rttz of 2nd PtE})
gives
\begin{eqnarray}
\frac{1}{2}\frac{d\sigma^2}{dt}
=\kappa_{zz}
+\frac{\kappa_{ttz}}{\kappa_z}\langle (\Delta z)\rangle
\frac{d^3}{dt^3}\langle (\Delta z)\rangle
-\frac{1}{2}\frac{\kappa_{ttz}}{\kappa_{z}}
\frac{d^3}{dt^3}\langle (\Delta z)^2\rangle
-\kappa_{tz}\frac{d}{dt}\langle (\Delta z)\rangle
-T_{ttz},
\label{simplified kzzvd for Rttz of 2nd PtE}
\end{eqnarray}
and inserting Equation (\ref{simplified delta z for Rttz})
into Equation (\ref{T_ttz}) gives
\begin{eqnarray}
T_{ttz}=0.
\label{Tttz on the restrictive condition}
\end{eqnarray}
Thereafter, combining Equations 
(\ref{simplified delta z for Rttz})-(\ref{Tttz on 
the restrictive condition})
yields
\begin{eqnarray}
\frac{1}{2}\frac{d\sigma^2}{dt}
=&&\kappa_{zz}-\kappa_{tz}\kappa_z,
\end{eqnarray}
which also
holds for the limit $t\rightarrow t_{\infty}$
\begin{eqnarray}
\kappa_{zz}^{DV}=\frac{1}{2}
\lim_{t\rightarrow t_{\infty}}\frac{d\sigma^2}{dt}
=&&\kappa_{zz}-\kappa_{tz}\kappa_z.
\label{kzzVD for Rttz of 2nd PtE-5}
\end{eqnarray}

\subsubsection{$R_{ttzz}$ of the 2nd-order PtI operation}
\label{Displacement Variance definition and Fick's law
definition for Rttzz of the 2nd PtI operation}

In this subsection, we introduce another DIO, i.e., 
$R_{ttzz}$ of the 2nd-order PtI operation. 
As the above subsection,
we can obtain the governing equation of $R_{ttzz}$ as
\begin{eqnarray}
\frac{\partial{F}}{\partial{t}}=
&&\left(-\kappa_z\frac{\partial{F}}
{\partial{z}}+\kappa_{zz}
\frac{\partial^2{F}}{\partial{z^2}}
+\kappa_{zzz}
\frac{\partial^3{F}}{\partial{z^3}}
+\kappa_{zzzz}
\frac{\partial^4{F}}{\partial{z^4}}
+\cdots\right)+\frac{\kappa_{ttzz}}{\kappa_{zz}}
\frac{\partial^{3}{F}}{\partial{t^{3}}}\nonumber\\
&&+\kappa_{tz}
\frac{\partial^2{F}}{\partial{t}\partial{z}}
+\kappa_{tzz}
\frac{\partial^3{F}}{\partial{t}\partial{z^2}}
+\kappa_{t3z}
\frac{\partial^4{F}}{\partial{t}\partial{z^3}}
+\kappa_{t4z}
\frac{\partial^5{F}}{\partial{t}\partial{z^4}}
+\kappa_{t5z}
\frac{\partial^6{F}}{\partial{t}\partial{z^5}}
+\cdots
\nonumber\\
&&+\left(\kappa_{ttz}+\kappa_z\frac{\kappa_{ttzz}}{\kappa_{zz}}\right)
\frac{\partial^{3}{F}}{\partial{t^2}\partial{z}}
+\left(\kappa_{3tz}-\kappa_{tz}\frac{\kappa_{ttzz}}{\kappa_{zz}}\right)
\frac{\partial^{4}{F}}{\partial{t^3}\partial{z}}
+\left(\kappa_{4tz}-\kappa_{ttz}\frac{\kappa_{ttzz}}{\kappa_{zz}}\right)
\frac{\partial^{5}{F}}{\partial{t^4}\partial{z}}
+\cdots\nonumber\\
&&+\left(\kappa_{3t2z}-\kappa_{tzz}\frac{\kappa_{ttzz}}{\kappa_{zz}}\right)
\frac{\partial^{5}{F}}{\partial{t^3}\partial{z^2}}
+\left(\kappa_{4t2z}-\kappa_{ttzz}\frac{\kappa_{ttzz}}{\kappa_{zz}}\right)
\frac{\partial^{6}{F}}{\partial{t^4}\partial{z^2}}+\cdots\nonumber\\
&&+\left(\kappa_{tt3z}-\kappa_{3z}\frac{\kappa_{ttzz}}{\kappa_{zz}}\right)
\frac{\partial^{5}{F}}{\partial{t^2}\partial{z^3}}
+\left(\kappa_{3t3z}-\kappa_{t3z}\frac{\kappa_{ttzz}}{\kappa_{zz}}\right)
\frac{\partial^{6}{F}}{\partial{t^3}\partial{z^3}}\nonumber\\
&&+\left(\kappa_{tt4z}-\kappa_{4z}\frac{\kappa_{ttzz}}{\kappa_{zz}}\right)
\frac{\partial^{6}{F}}{\partial{t^2}\partial{z^4}}+\cdots.
\label{Rttzz equation}
\end{eqnarray}

It is obvious that in 
Equation (\ref{Rttzz equation}) 
the Fick's law definition $\kappa_{zz}^{FL}$ is equal to 
$\kappa_{zz}$, i.e., $\kappa_{zz}^{FL}=\kappa_{zz}$,
which is identical with Eequation (\ref{kzzfl0}).

From Equation (\ref{Rttzz equation})
we can obtain the following equations
\begin{eqnarray}
\frac{d}{dt}\langle (\Delta z)
\rangle&=&\kappa_z+\frac{\kappa_{ttzz}}{\kappa_{zz}}
\frac{d^3}{dt^3}\langle (\Delta z)\rangle,
\label{dzdt Rttzz of 2nd PtE}\\
\frac{d}{dt}\langle (\Delta z)^2\rangle&
=&2\kappa_z\langle (\Delta z)
\rangle+2\kappa_{zz}+\frac{\kappa_{ttzz}}{\kappa_{zz}}
\frac{d^3}{dt^3}\langle (\Delta z)^2\rangle
-2\kappa_{tz}\frac{d}{dt}
\langle (\Delta z)\rangle-2T_{ttzz}
\label{dz2dt Rttzz of 2nd PtE}
\end{eqnarray}
with
\begin{eqnarray}
T_{ttzz}=&&\left(\kappa_{ttz}+\kappa_{z}
\frac{\kappa_{ttzz}}{\kappa_{zz}}\right)
\frac{d^2}{dt^2}\langle (\Delta z)\rangle
+\left(\kappa_{tttz}-\kappa_{tz}
\frac{\kappa_{ttzz}}{\kappa_{zz}}\right)
\frac{d^3}{dt^3}\langle (\Delta z)\rangle
+\left(\kappa_{4tz}-\kappa_{ttz}
\frac{\kappa_{ttzz}}{\kappa_{zz}}\right)
\frac{d^4}{dt^4}\langle (\Delta z)\rangle
+\cdots.
\label{Tttzz}
\end{eqnarray}
The solution of Equation (\ref{dzdt Rttzz of 2nd PtE}) 
can be obtained easily as
\begin{eqnarray}
\langle (\Delta z)\rangle&=&c_1
+c_2e^{-t\sqrt{\kappa_{zz}/\kappa_{ttzz}}}
+c_3e^{t\sqrt{\kappa_{zz}/\kappa_{ttzz}}}
+\kappa_{z} t.
\label{delta2 z for Rttzz of the 2nd PtI operation}
\end{eqnarray}
Because the formula of $\kappa_{ttzz}$ 
is very complicated and hard to be evaluated, 
in this subsection we only explore the  
displacement variance definition $\kappa_{zz}^{DV}$ for 
$R_{ttzz}$ of the 2nd-order PtI operation
under the special condition, which requires
the coefficients of the exponent functions in 
$\langle (\Delta z)\rangle$ and 
$\langle (\Delta z)^2\rangle$ as zero. Thus, 
Equation (\ref{delta2 z for Rttzz of the 2nd PtI operation}) becomes 
\begin{eqnarray}
\langle (\Delta z)\rangle
&=&c_1+\kappa_{z} t,
\label{delta2 z for Rttzz of the 2nd PtI 
operation under restrictive condition}
\end{eqnarray}
inserting which into Equation 
(\ref{Tttzz}), one can find 
\begin{eqnarray}
T_{ttzz}=0.
\label{Tttzz for Rttzz of the 2nd PtI operation 
under restrictive condition}
\end{eqnarray}
Inserting Equations 
(\ref{delta2 z for Rttzz of the 2nd PtI operation 
under restrictive condition})
and (\ref{Tttzz for Rttzz of the 2nd PtI 
operation under restrictive condition})
into Equation (\ref{dz2dt Rttzz of 2nd PtE}) 
yields
\begin{eqnarray}
\frac{d}{dt}\langle (\Delta z)^2\rangle&
=&2\kappa_z\left(c_1+\kappa_{z} t\right)+2\kappa_{zz}
+\frac{\kappa_{ttzz}}{\kappa_{zz}}
\frac{d^3}{dt^3}\langle (\Delta z)^2\rangle
-2\kappa_{tz}\kappa_z.
\end{eqnarray}
The solution of the latter equation can been 
easily obtained as follows
\begin{eqnarray}
\langle (\Delta z)^2\rangle&=&c'_1
+c'_2e^{-t\sqrt{\kappa_{zz}/\kappa_{ttzz}}}
+c'_3e^{t\sqrt{\kappa_{zz}/\kappa_{ttzz}}}
+2\left(\kappa_{tz}\kappa_z-\kappa_{zz}
-\kappa_zc_1\right)t+\kappa_{z}^2 t^2.
\end{eqnarray}
For the special condition $c'_2=c'_3=0$, 
the latter equation becomes
\begin{eqnarray}
\langle (\Delta z)^2\rangle&=&c'_1
+2\left(\kappa_{tz}\kappa_z-\kappa_{zz}
-\kappa_zc_1\right)t+\kappa_{z}^2 t^2.
\label{delta2 z for Rttzz for strong condition}
\end{eqnarray}

By combining Equations (\ref{dzdt Rttzz of 2nd PtE}) and
(\ref{dz2dt Rttzz of 2nd PtE}) we can find
\begin{eqnarray}
\frac{1}{2}\frac{d\sigma^2}{dt}&=&
-\frac{\kappa_{ttzz}}{\kappa_{zz}}
\langle (\Delta z)\rangle
\frac{d^3}{dt^3}\langle (\Delta z)\rangle
+\kappa_{zz}
+\frac{1}{2}\frac{\kappa_{ttzz}}{\kappa_{zz}}
\frac{d^3}{dt^3}\langle (\Delta z)^2\rangle
-\kappa_{tz}\frac{d}{dt}
\langle (\Delta z)\rangle-T_{ttzz}.
\end{eqnarray}
Inserting Equations 
(\ref{delta2 z for Rttzz of the 2nd PtI 
operation under restrictive condition}), 
(\ref{delta2 z for Rttzz for strong condition}),
and (\ref{Tttzz for Rttzz of the 2nd PtI 
operation under restrictive condition})
into the latter equation yields
\begin{eqnarray}
\frac{1}{2}\frac{d\sigma^2}{dt}
=\kappa_{zz}-\kappa_{tz}\kappa_z.
\label{vc for Rttzz of 2nd PtI}
\end{eqnarray}
For the limit $t\rightarrow t_{\infty}$,
the latter equation also holds
\begin{eqnarray}
\kappa_{zz}^{VD}=\frac{1}{2}\lim_{t\rightarrow t_{\infty}}
\frac{d\sigma^2}{dt}
=\kappa_{zz}-\kappa_{tz}\kappa_z,
\end{eqnarray}
which is identical with Equation (\ref{kzzvd0}). 
From the above investigation
we can find that at least for the special condition
the displacement variance definition is an invariant for the
$R_{ttzz}$ of the 2nd-order PtI operation.

\subsubsection{$R_{ntmz}$ for the 2nd PtI operation}
\label{Displacement Variance definition and Fick's
law definition for the other case for the 2nd PtI operation}

Here, we give the general DIO, i.e., $R_{ntmz}$ of the 
second-order PtI operation. As done in subsection 
\ref{Displacement Variance definition and Fick's
law definition for Rtzz of the 1st PtI operation},
the equation corresponding to $R_{ntmz}$ of the 
second-order PtI operation can be obtained from
Equation (\ref{first PtE}) as follows
\begin{eqnarray}
\frac{\partial{F}}{\partial{t}}=
&&\left(-\kappa_z\frac{\partial{F}}
{\partial{z}}+\kappa_{zz}
\frac{\partial^2{F}}{\partial{z^2}}
+\kappa_{zzz}
\frac{\partial^3{F}}{\partial{z^3}}
+\kappa_{zzzz}
\frac{\partial^4{F}}{\partial{z^4}}
+\cdots\right)+\frac{\kappa_{ntmz}}{\kappa_{(n-2)tmz}}
\frac{\partial^{3}{F}}{\partial{t^{3}}}\nonumber\\
&&+\kappa_{tz}
\frac{\partial^2{F}}{\partial{t}\partial{z}}
+\kappa_{tzz}
\frac{\partial^3{F}}{\partial{t}\partial{z^2}}
+\kappa_{t3z}
\frac{\partial^4{F}}{\partial{t}\partial{z^3}}
+\kappa_{t4z}
\frac{\partial^5{F}}{\partial{t}\partial{z^4}}
+\kappa_{t5z}
\frac{\partial^6{F}}{\partial{t}\partial{z^5}}
+\cdots
\nonumber\\
&&+\left(\kappa_{ttz}
+\kappa_z\frac{\kappa_{ntmz}}{\kappa_{(n-2)tmz}}\right)
\frac{\partial^{3}{F}}{\partial{t^2}\partial{z}}
+\left(\kappa_{3tz}-\kappa_{tz}\frac{\kappa_{ntmz}}{\kappa_{(n-2)tmz}}\right)
\frac{\partial^{4}{F}}{\partial{t^3}\partial{z}}
+\left(\kappa_{4tz}-\kappa_{ttz}\frac{\kappa_{ntmz}}{\kappa_{(n-2)tmz}}\right)
\frac{\partial^{5}{F}}{\partial{t^4}\partial{z}}
+\cdots\nonumber\\
&&+\left(\kappa_{2t2z}-\kappa_{zz}\frac{\kappa_{ntmz}}{\kappa_{(n-2)tmz}}\right)
\frac{\partial^{4}{F}}{\partial{t^2}\partial{z^2}}
+\left(\kappa_{3t2z}-\kappa_{tzz}\frac{\kappa_{ntmz}}{\kappa_{(n-2)tmz}}\right)
\frac{\partial^{5}{F}}{\partial{t^3}\partial{z^2}}
+\left(\kappa_{4t2z}-\kappa_{ttzz}\frac{\kappa_{ntmz}}{\kappa_{(n-2)tmz}}\right)
\frac{\partial^{6}{F}}{\partial{t^4}\partial{z^2}}\cdots\nonumber\\
&&+\left(\kappa_{ttzzz}-\kappa_{3z}\frac{\kappa_{ntmz}}{\kappa_{(n-2)tmz}}\right)
\frac{\partial^{5}{F}}{\partial{t^2}\partial{z^3}}
+\left(\kappa_{3t3z}-\kappa_{t3z}\frac{\kappa_{ntmz}}{\kappa_{(n-2)tmz}}\right)
\frac{\partial^{6}{F}}{\partial{t^3}\partial{z^3}}\nonumber\\
&&+\left(\kappa_{tt4z}-\kappa_{4z}\frac{\kappa_{ntmz}}{\kappa_{(n-2)tmz}}\right)
\frac{\partial^{6}{F}}{\partial{t^2}\partial{z^4}}+\cdots.
\label{Rntmz equation for second-order PtI operation}
\end{eqnarray}

From Equation (\ref{Rntmz equation for second-order PtI operation}) 
we can find easily that the Fick's law definition 
$\kappa_{zz}^{FL}$ is equal to $\kappa_{zz}$, i.e., 
$\kappa_{zz}^{FL}=\kappa_{zz}$, which 
is same as Equation (\ref{kzzfl0}). 
In what follows, we investigate the 
displacement variance definition
for $R_{ntmz}$ for the 2nd-order PtI operation.

As done in subsection \ref{Two definitions 
for Rntmz of the 1st-order PtI operation},
from Equation (\ref{Rntmz equation for second-order PtI operation}) 
we can obtain
\begin{eqnarray}
\frac{d}{dt}\langle (\Delta z)
\rangle&=&\kappa_z+\frac{\kappa_{ntmz}}{\kappa_{(n-2)tmz}}
\frac{d^2}{dt^2}\langle (\Delta z)\rangle.
\label{dzdt for Rntmz of 2nd PtI operation}\\
\frac{d}{dt}\langle (\Delta z)^2\rangle&
=&2\kappa_z\langle (\Delta z)
\rangle+2\kappa_{zz}+\frac{\kappa_{ntmz}}{\kappa_{(n-2)tmz}}
\frac{d^3}{dt^3}\langle (\Delta z)\rangle
-2\kappa_{tz}\frac{d}{dt}
\langle (\Delta z)\rangle
-2T_{ntmz}
\label{dz2dt Rntmz of 2nd PtE}
\end{eqnarray}
with
\begin{eqnarray}
T_{ntmz}=&&\left(\kappa_{ttz}-\frac{\kappa_{ntmz}}{\kappa_{(n-2)tmz}}
\kappa_{tz}\right)\frac{d^2}{dt^2}\langle (\Delta z)\rangle
+\left(\kappa_{tttz}-\frac{\kappa_{ntmz}}{\kappa_{(n-2)tmz}}
\kappa_{ttz}\right)\frac{d^3}{dt^3}\langle (\Delta z)\rangle+
\cdots.
\label{Tntmz for the second-order PtI operation}
\end{eqnarray}
Solving Equation (\ref{dzdt for Rntmz of 1st PtI operation}) gives
\begin{eqnarray}
\langle (\Delta z)\rangle&=&c_1
+c_2e^{-t\sqrt{\kappa_{(n-2)tmz}/\kappa_{ntmz}}}
+c_3e^{t\sqrt{\kappa_{(n-2)tmz}/\kappa_{ntmz}}}
+\kappa_{z} t,
\label{delta z for Rntmz of the 2nd PtI operation}
\end{eqnarray}
which, for the special condition $c_2=c_3=0$,  becomes
\begin{eqnarray}
\langle (\Delta z)\rangle&=&c_1
+\kappa_{z} t.
\label{delta z for Rntmz of the 2nd PtI operation under restrictive condition}
\end{eqnarray}
Thus, Equation (\ref{Tntmz for the second-order PtI operation}) 
is simplified as 
\begin{eqnarray}
T_{ntmz}=0.
\label{Tntmz for the second-order PtI 
operation under restrictive condition}
\end{eqnarray}
Inserting Equations (\ref{delta z for Rntmz of 
the 2nd PtI operation under restrictive condition})
and (\ref{Tntmz for the second-order PtI 
operation under restrictive condition}) into 
Equation (\ref{dz2dt Rntmz of 2nd PtE}) yields
\begin{eqnarray}
\frac{d}{dt}\langle (\Delta z)^2\rangle&
=&2\kappa_z(c_1
+\kappa_{z} t)+2\kappa_{zz}
-2\kappa_{tz}\kappa_z.
\label{dz2dt Rntmz of 2nd PtE-2}
\end{eqnarray}
The solution of the above equation can be found easily
\begin{eqnarray}
\langle (\Delta z)\rangle&=&c'_1
+c'_2e^{-t\sqrt{\kappa_{(n-2)tmz}/\kappa_{ntmz}}}
+c'_3e^{t\sqrt{\kappa_{(n-2)tmz}/\kappa_{ntmz}}}
+2(\kappa_{zz}-\kappa_{tz}\kappa_z+\kappa_zc_1)t
+\kappa_{z}^2 t^2.
\end{eqnarray}
We only explore the case for the special condition,
so the latter equation becomes 
\begin{eqnarray}
\langle (\Delta z)^2\rangle&=&c'_1
+2(\kappa_{zz}-\kappa_{tz}\kappa_z+\kappa_zc_1)t
+\kappa_{z}^2 t^2.
\label{delta z2 for Rntmz of the 2nd PtI operation}
\end{eqnarray}
Combining Equations (\ref{dzdt for Rntmz of 2nd PtI operation}), and 
(\ref{dz2dt Rntmz of 2nd PtE}) we can obtain 
the equation of the displacement variance as follows
\begin{eqnarray}
\frac{1}{2}\frac{d\sigma^2}{dt}&=&
-\frac{\kappa_{ntmz}}{\kappa_{(n-2)tmz}}
\langle (\Delta z)\rangle
\frac{d^3}{dt^3}\langle (\Delta z)\rangle
+\kappa_{zz}
+\frac{1}{2}\frac{\kappa_{ntmz}}{\kappa_{(n-2)tmz}}
\frac{d^3}{dt^3}\langle (\Delta z)^2\rangle
-\kappa_{tz}\frac{d}{dt}
\langle (\Delta z)\rangle.
\end{eqnarray}
Inserting Equations 
(\ref{delta z for Rntmz of the 2nd PtI 
operation under restrictive condition}),
(\ref{Tntmz for the second-order PtI operation 
under restrictive condition}), and 
(\ref{delta z2 for Rntmz of the 2nd PtI operation})
into the latter equation we can find, for the limit 
$t\rightarrow t_{\infty}$,
\begin{eqnarray}
\kappa_{zz}^{VD}=\frac{1}{2}\lim_{t\rightarrow t_{\infty}}
\frac{d\sigma^2}{dt}
=\kappa_{zz}-\kappa_{tz}\kappa_z,
\end{eqnarray}
which is identical with Equation (\ref{kzzvd0}).
Therefore, the displacement variance definition 
$\kappa_{zz}^{VD}$ is invariant for 
$R_{ntmz}$ for the 2nd-order PtI operation
at least for the special condition.

\subsection{The third-order PtI operation}
\label{The third-th PtI operation}

For $n=3$, Equation (\ref{n PtE}) becomes
\begin{eqnarray}
\frac{\partial^{4}{F}}{\partial{t^{4}}}=
&&\left(-\kappa_z\frac{\partial^{4}{F}}
{\partial{t^3}\partial{z}}+\kappa_{zz}
\frac{\partial^{5}{F}}{\partial{t^3}\partial{z^2}}
+\kappa_{zzz}
\frac{\partial^{6}{F}}{\partial{t^3}\partial{z^3}}
+\kappa_{zzzz}
\frac{\partial^{7}{F}}{\partial{t^3}\partial{z^4}}
+\cdots\right)
+\left( \kappa_{tz}
\frac{\partial^{5}{F}}{\partial{t^{4}}
	\partial{z}}
+ \kappa_{ttz}\frac{\partial^{6}{F}}
{\partial{t^{5}}\partial{z}}
+ \kappa_{tttz}\frac{\partial^{7}{F}}
{\partial{t^{6}}\partial{z}}
+\cdots\right)\nonumber\\
&&+\left( \kappa_{tzz}
\frac{\partial^{6}{F}}{\partial{t^{4}}
	\partial{z^2}}
+ \kappa_{ttzz}\frac{\partial^{7}{F}}
{\partial{t^{5}}\partial{z^2}}
+ \kappa_{tttzz}\frac{\partial^{8}{F}}
{\partial{t^{6}}\partial{z^2}}
+\cdots\right)
+\cdots,
\label{third PtE}
\end{eqnarray}
which is the governing equation of the third-order PtI operation. 

\subsubsection{$R_{tttz}$ of the third PtI operation}

Combining Equations (\ref{equation of F with constant coefficient}) 
and (\ref{third PtE}) we can find 
\begin{eqnarray}
\frac{\partial{F}}{\partial{t}}=
&&\left(-\kappa_z\frac{\partial{F}}
{\partial{z}}+\kappa_{zz}
\frac{\partial^2{F}}{\partial{z^2}}
+\kappa_{zzz}
\frac{\partial^3{F}}{\partial{z^3}}
+\kappa_{zzzz}
\frac{\partial^4{F}}{\partial{z^4}}
+\cdots\right)-\frac{\kappa_{3tz}}{\kappa_z}\frac{\partial^{4}{F}}
{\partial{t^{4}}}\nonumber\\
&&+\kappa_{tz}
\frac{\partial^2{F}}{\partial{t}\partial{z}}
+\kappa_{tzz}
\frac{\partial^3{F}}{\partial{t}\partial{z^2}}
+\kappa_{tzzz}
\frac{\partial^4{F}}{\partial{t}\partial{z^3}}
+\kappa_{tzzzz}
\frac{\partial^5{F}}{\partial{t}\partial{z^4}}
+\kappa_{tzzzzz}
\frac{\partial^6{F}}{\partial{t}\partial{z^5}}
+\cdots
\nonumber\\
&&+\kappa_{ttz}
\frac{\partial^3{F}}{\partial{t^2}\partial{z}}
+\kappa_{ttzz}
\frac{\partial^4{F}}{\partial{t^2}\partial{z^2}}
+\kappa_{tzzz}
\frac{\partial^5{F}}{\partial{t^2}\partial{z^3}}
+\kappa_{tzzzz}
\frac{\partial^6{F}}{\partial{t^2}\partial{z^4}}
+\kappa_{tzzzzz}
\frac{\partial^7{F}}{\partial{t^2}\partial{z^5}}
+\cdots
\nonumber\\
&&+\left(\frac{\kappa_{tttz}}{\kappa_z}\kappa_{zz}+\kappa_{ttzz}\right)
\frac{\partial^{5}{F}}{\partial{t^3}\partial{z^2}}
+\left(\frac{\kappa_{tttz}}{\kappa_z}\kappa_{zzz}+\kappa_{ttzzz}\right)
\frac{\partial^{6}{F}}{\partial{t^3}\partial{z^3}}+\cdots
\nonumber\\
&&+\left(\frac{\kappa_{tttz}}{\kappa_z}\kappa_{tz}+\kappa_{4tz}\right)
\frac{\partial^{5}{F}}{\partial{t^{4}}\partial{z}}
+\left(\frac{\kappa_{tttz}}{\kappa_z}\kappa_{ttz}+\kappa_{5tz}\right)
\frac{\partial^{6}{F}}{\partial{t^{5}}\partial{z}}
+\cdots\nonumber\\
&&+\left(\frac{\kappa_{tttz}}{\kappa_z}\kappa_{tzz}+\kappa_{4tzz}\right)
\frac{\partial^{6}{F}}{\partial{t^{4}}\partial{z^2}}
+\left(\frac{\kappa_{3tz}}{\kappa_z}\kappa_{tzzz}+\kappa_{4t3z}\right)
\frac{\partial^{7}{F}}{\partial{t^{5}}\partial{z^2}}
+\cdots,
\label{Rtttz equation for third-order PtI operation}
\end{eqnarray}
which is the EIDF corresponding to the 
$R_{tttz}$, one of the DIOs, of the third PtI operation.  

From Equation 
(\ref{Rtttz equation for third-order PtI operation}) 
we can easily find that 
the Fick's law definition $\kappa_{zz}^{FL}$ is equal
to $\kappa_{zz}$, which is identical with Equation
(\ref{kzzfl0}).  

From Equation 
(\ref{Rtttz equation for third-order PtI operation})
We can find 
\begin{eqnarray}
\frac{d}{dt}\langle (\Delta z)
\rangle&=&\kappa_z-\frac{\kappa_{3tz}}{\kappa_{z}}
\frac{d^4}{dt^4}\langle (\Delta z)\rangle,
\label{dzdt for Rntmz of third PtI operation}\\
\frac{d}{dt}\langle (\Delta z)^2\rangle&
=&2\kappa_z\langle (\Delta z)
\rangle+2\kappa_{zz}-\frac{\kappa_{3tz}}{\kappa_{z}}
\frac{d^4}{dt^4}\langle (\Delta z)\rangle
-2\kappa_{tz}\frac{d}{dt}
\langle (\Delta z)\rangle
-2\kappa_{ttz}\frac{d^2}{dt^2}
\langle (\Delta z)\rangle
-2T_{3tz}
\label{dz2dt Rntmz of third PtE}
\end{eqnarray}
with
\begin{eqnarray}
T_{3tz}=&&\left(\frac{\kappa_{3tz}}{\kappa_{z}}
\kappa_{tz}+\kappa_{4tz}\right)\frac{d^4}{dt^4}\langle (\Delta z)\rangle
+\left(\frac{\kappa_{3tz}}{\kappa_{z}}
\kappa_{ttz}+\kappa_{5tz}\right)\frac{d^5}{dt^5}\langle (\Delta z)\rangle+
\cdots.
\label{T3tz for the third-order PtI operation}
\end{eqnarray}
The solution of 
Equation (\ref{dzdt for Rntmz of third PtI operation})
can be obtained as follows
\begin{eqnarray}
\langle (\Delta z) \rangle=c_0+c_1 e^{r_1t} +c_2 e^{r_2t} 
+c_3 e^{r_3t} +\kappa_z t
\label{Delta z for R3tz of third-order PtI operation-1}
\end{eqnarray}
with 
\begin{eqnarray}
r_1&=&\left(\frac{\kappa_z}{\kappa_{3tz}}\right)^{1/3} e^{i\pi /3},
\label{r1}\\
r_2&=&-\left(\frac{\kappa_z}{\kappa_{3tz}}\right)^{1/3},
\label{r2}\\
r_1&=&\left(\frac{\kappa_z}{\kappa_{3tz}}\right)^{1/3} e^{i5\pi/3},
\label{r3} 
\end{eqnarray}
where $r_1$, $r_2$, and $r_3$ satisfy the following formulas
\begin{eqnarray}
&&r_1+r_2+r_3=0,
\label{3r=0}\\
&&r_1^3=r_2^3=r_3^3=-\frac{\kappa_z}{\kappa_{3tz}}.
\label{r1=r2=r3=}
\end{eqnarray}
From Equation (\ref{dzdt for Rntmz of 2nd PtI operation})
we can find
\begin{eqnarray}
\frac{d^{n+3}}{dt^{n+3}}\langle (\Delta z)
\rangle&=&-\frac{\kappa_{z}}{\kappa_{3tz}}
\frac{d^n}{dt^n}\langle (\Delta z)\rangle,
\end{eqnarray}
inserting which into Equation 
(\ref{T3tz for the third-order PtI operation}), one obtains
\begin{eqnarray}
T_{3tz}=-\kappa_{3tz}\frac{d^3}{dt^3}\langle (\Delta z)\rangle
-\kappa_{ttz}\frac{d^2}{dt^2}\langle (\Delta z)\rangle
+\kappa_{tz}\kappa_z
-\kappa_{tz}\frac{d}{dt}\langle (\Delta z)\rangle.
\label{T3tz for the third-order PtI operation-2}
\end{eqnarray}
Replacing term $T_{3tz}$ in Equation 
(\ref{dz2dt Rntmz of 2nd PtE}) by the latter equation
gives
\begin{eqnarray}	
\frac{1}{2}\frac{d}{dt}\langle (\Delta z)^2\rangle&
=&\kappa_z\langle (\Delta z)
\rangle+\kappa_{zz}-\frac{1}{2}\frac{\kappa_{3tz}}{\kappa_{z}}
\frac{d^4}{dt^4}\langle (\Delta z)\rangle
-\kappa_{tz}\frac{d}{dt}
\langle (\Delta z)\rangle
-\kappa_{ttz}\frac{d^2}{dt^2}
\langle (\Delta z)\rangle\nonumber\\
&&+\kappa_{3tz}\frac{d^3}{dt^3}\langle (\Delta z)\rangle
+\kappa_{ttz}\frac{d^2}{dt^2}\langle (\Delta z)\rangle
-\kappa_{tz}\kappa_z
+\kappa_{tz}\frac{d}{dt}\langle (\Delta z)\rangle.
\label{dz2dt Rntmz of 2nd PtE-3}
\end{eqnarray}
Considering Equation 
(\ref{Delta z for R3tz of third-order PtI operation-1}),
from Equation (\ref{dz2dt Rntmz of 2nd PtE-3})
we can find
\begin{eqnarray}
\langle (\Delta z)^2 \rangle=c'_0+c'_1 e^{r_1t} +c'_2 e^{r_2t} 
+c'_3 e^{r_3t} 
+2\left(\kappa_{zz}-\kappa_{tz}\kappa_z+\kappa_zc_0 \right)
+\kappa_z^2 t^2,
\label{Delta z for R3tz of third-order PtI operation}
\end{eqnarray}
where $r_1$, $r_2$, and $r_3$ are shown from Equations (\ref{r1})-(\ref{r3}). 

Combining Equations (\ref{dzdt for Rntmz of third PtI operation}), 
(\ref{dz2dt Rntmz of third PtE}),
and (\ref{T3tz for the third-order PtI operation-2})
we can obtain
\begin{eqnarray}
\frac{1}{2}\frac{d\sigma^2}{dt}=&&\kappa_{zz}-\kappa_{tz}\kappa_{z}
+\frac{\kappa_{3tz}}{\kappa_z}c_0c_1r_1^4 e^{r_1t}
+\frac{\kappa_{3tz}}{\kappa_z}c_0c_2r_2^4 e^{r_2t}
+\frac{\kappa_{3tz}}{\kappa_z}c_0c_3r_3^4 e^{r_3t}\nonumber\\
&&+\frac{\kappa_{3tz}}{\kappa_z}c_1^2r_1^4 e^{2r_1t}
+\frac{\kappa_{3tz}}{\kappa_z}c_2^2r_2^4 e^{2r_2t}
+\frac{\kappa_{3tz}}{\kappa_z}c_3^2r_3^4 e^{2r_3t}\nonumber\\
&&+\frac{\kappa_{3tz}}{\kappa_z}c_1c_2r_1^4 e^{(r_1+r_2)t}
+\frac{\kappa_{3tz}}{\kappa_z}c_1c_2r_2^4 e^{(r_1+r_2)t}
+\frac{\kappa_{3tz}}{\kappa_z}c_1c_3r_3^4 e^{(r_1+r_3)t}
+\frac{\kappa_{3tz}}{\kappa_z}c_1c_3r_1^4 e^{(r_1+r_3)t}\nonumber\\
&&+\frac{\kappa_{3tz}}{\kappa_z}c_2c_3r_3^4 e^{(r_2+r_3)t}
+\frac{\kappa_{3tz}}{\kappa_z}c_2c_3r_2^4 e^{(r_2+r_3)t}\nonumber\\
&&+r_1^4c_1e^{r_1t}\frac{\kappa_{3tz}}{\kappa_z}\kappa_zt
+r_2^4c_2e^{r_2t}\frac{\kappa_{3tz}}{\kappa_z}\kappa_zt
+r_3^4c_3e^{r_3t}\frac{\kappa_{3tz}}{\kappa_z}\kappa_zt\nonumber\\
&&+\kappa_{3tz}r_1^3c_1e^{r_1t}
+\kappa_{3tz}r_2^3c_2e^{r_2t}+\kappa_{3tz}r_3^3c_3e^{r_3t}\nonumber\\
&&-\frac{1}{2}\frac{\kappa_{3tz}}{\kappa_z}r_1^4c'_1e^{r_1t}
-\frac{1}{2}\frac{\kappa_{3tz}}{\kappa_z}r_2^4c'_2e^{r_2t}
-\frac{1}{2}\frac{\kappa_{3tz}}{\kappa_z}r_3^4c'_3e^{r_3t},
\end{eqnarray}
which, because of Equations (\ref{3r=0}) 
and (\ref{r1=r2=r3=}), 
can be rewritten as 
\begin{eqnarray}
\frac{1}{2}\frac{d\sigma^2}{dt}
=&&\kappa_{zz}-\kappa_{tz}\kappa_{z}-c_0c_1r_1 e^{r_1t}
-\kappa_{z}c_1 e^{r_1t}-c_0c_2r_2 e^{r_2t}
-\kappa_{z}c_2 e^{r_2t}
-c_0c_3r_3 e^{r_3t}-\kappa_{z}c_3 e^{r_3t}\nonumber\\
&&-r_1c_1^2 e^{2r_1t}-r_2c_2^2 e^{2r_2t}-r_3c_3^2 e^{2r_3t}
+c_1c_2 r_3 e^{-r_3t}+c_1c_3r_2 e^{-r_2t}+c_2c_3r_1 e^{-r_1t}\nonumber\\
&&-\kappa_z r_1c_1 e^{r_1t}t-\kappa_z r_2c_2 e^{r_2t}t
-\kappa_z r_3c_3 e^{r_3t}t
+\frac{1}{2}r_1c'_1e^{r_1t}
+\frac{1}{2}r_2c'_2e^{r_2t}
+\frac{1}{2}r_3c'_3e^{r_3t}.
\label{kzzdv for R3tz of third-order PtI operation-1}
\end{eqnarray}
In order to ensure that $d\sigma^2/(2dt)$ is real,
we have to neglect the terms containing $c_1$ and $c_3$
in the above equation, i.e., 
to employ the condition 
$c_1=c'_1=c_3=c'_3=0$, 
so Equation (\ref{kzzdv for R3tz of third-order PtI operation-1}) becomes
\begin{eqnarray}
\frac{1}{2}\frac{d\sigma^2}{dt}
=&&\kappa_{zz}-\kappa_{tz}\kappa_{z}-c_0c_2r_2 e^{r_2t}
-\kappa_{z}c_2 e^{r_2t}
-r_2c_2^2 e^{2r_2t}
-\kappa_z r_2c_2 e^{r_2t}t
+\frac{1}{2}r_2c'_2e^{r_2t}.
\end{eqnarray}
Since $r_2<0$, for the limit $t\rightarrow t_{\infty}$ 
the latter equation becomes
\begin{eqnarray}
\kappa_{zz}^{VD}=\frac{1}{2}\lim_{t\rightarrow t_{\infty}}
\frac{d\sigma^2}{dt}
=\kappa_{zz}-\kappa_{tz}\kappa_z,
\label{kzzdv for R3tz of third-order PtI}
\end{eqnarray}
which is identical with Equation (\ref{kzzvd0}).
Therefore, the displacement variance definition $\kappa_{zz}^{DV}$ 
is an invariant quantity for
$R_{tttz}$ of the third-order PtI operation.  
From the above part, we find that the special condition 
$c_1=c'_1=c_2=c'_2=c_3=c'_3=0$
is an inference of the derivation process.

As shown in subsections 
\ref{Two definitions  for Rntmz of the 1st-order PtI operation}
and 
\ref{Displacement Variance definition and Fick's 
law definition for the other case for the 2nd PtI operation},
if using the special condition, 
we  can also find that for $R_{ntmz}$ of the third-order PtI operation both
the Fick's law definition $\kappa_{zz}^{FL}$ 
and the displacement variance definition $\kappa_{zz}^{FL}$
are the invariant quantities. For the  simplification, 
in this subsection we do not give the detailed deduction process. 

\subsection{The $i$th-order PtI operation}
\label{The i-th PtI operation}

Setting $n=i$ 
in Equation (\ref{n PtE}), yields
\begin{eqnarray}
\frac{\partial^{i+1}{F}}{\partial{t^{i+1}}}=
&&\left(-\kappa_z\frac{\partial^{i+1}{F}}
	{\partial{t^i}\partial{z}}+\kappa_{zz}
	\frac{\partial^{i+2}{F}}{\partial{t^i}\partial{z^2}}
	+\kappa_{zzz}
	\frac{\partial^{i+3}{F}}{\partial{t^i}\partial{z^3}}
	+\kappa_{zzzz}
	\frac{\partial^{i+4}{F}}{\partial{t^i}\partial{z^4}}
	+\cdots\right)
	+\left( \kappa_{tz}
	\frac{\partial^{i+2}{F}}{\partial{t^{i+1}}
		\partial{z}}
	+ \kappa_{ttz}\frac{\partial^{i+3}{F}}
	{\partial{t^{i+2}}\partial{z}}
	+ \kappa_{tttz}\frac{\partial^{i+4}{F}}
	{\partial{t^{i+3}}\partial{z}}
	+\cdots\right)\nonumber\\
	&&+\left( \kappa_{tzz}
	\frac{\partial^{i+3}{F}}{\partial{t^{i+1}}
		\partial{z^2}}
	+ \kappa_{ttzz}\frac{\partial^{i+4}{F}}
	{\partial{t^{i+2}}\partial{z^2}}
	+ \kappa_{tttzz}\frac{\partial^{i+5}{F}}
	{\partial{t^{i+3}}\partial{z^2}}
	+\cdots\right)
	+\cdots,
	\label{i PtE}
\end{eqnarray}
which is the equation of the $i$th-order PtI operation.

\subsubsection{$R_{itz}$ of the i-th PtI operation}
\label{Displacement Variance definition and Fick's law definition
for Ritz of the i-th PtI operation}

Equation (\ref{i PtE}) can be rewritten as
\begin{eqnarray}
\kappa_{itz}\frac{\partial^{i+1}{F}}{\partial{t^i}\partial{z}}=&&
-\frac{\kappa_{itz}}{\kappa_z}\frac{\partial^{i+1}{F}}{\partial{t^{i+1}}}
+\left(\frac{\kappa_{itz}}{\kappa_z}\kappa_{zz}
\frac{\partial^{i+2}{F}}{\partial{t^i}\partial{z^2}}
+\frac{\kappa_{itz}}{\kappa_z}\kappa_{zzz}
\frac{\partial^{i+3}{F}}{\partial{t^i}\partial{z^3}}
+\frac{\kappa_{itz}}{\kappa_z}\kappa_{zzzz}
\frac{\partial^{i+4}{F}}{\partial{t^i}\partial{z^4}}
+\cdots\right)\nonumber\\
&&+\left(\frac{\kappa_{itz}}{\kappa_z}\kappa_{tz}
\frac{\partial^{i+2}{F}}{\partial{t^{i+1}}\partial{z}}
+\frac{\kappa_{itz}}{\kappa_z}\kappa_{ttz}
\frac{\partial^{i+3}{F}}{\partial{t^{i+2}}\partial{z}}
+\frac{\kappa_{itz}}{\kappa_z}\kappa_{tttz}
\frac{\partial^{i+4}{F}}{\partial{t^{i+3}}\partial{z}}
+\cdots\right)\nonumber\\
&&+\left(\frac{\kappa_{itz}}{\kappa_z}
\kappa_{tzz}\frac{\partial^{i+3}{F}}{\partial{t^{i+1}}\partial{z^2}}
+\frac{\kappa_{itz}}{\kappa_z}\kappa_{ttzz}
\frac{\partial^{i+4}{F}}{\partial{t^{i+2}}\partial{z^2}}
+\frac{\kappa_{itz}}{\kappa_z}\kappa_{tttzz}
\frac{\partial^{i+5}{F}}{\partial{t^{i+3}}\partial{z^2}}
+\cdots\right)
+\cdots,
\label{di+1ditdz of i PtE}
\end{eqnarray}
from which, considering Equation
(\ref{equation of F with
constant coefficient}), we find
\begin{eqnarray}
\frac{\partial{F}}{\partial{t}}=
&&\left(-\kappa_z\frac{\partial{F}}
{\partial{z}}+\kappa_{zz}
\frac{\partial^2{F}}{\partial{z^2}}
+\kappa_{zzz}
\frac{\partial^3{F}}{\partial{z^3}}
+\kappa_{zzzz}
\frac{\partial^4{F}}{\partial{z^4}}
+\cdots\right)-\frac{\kappa_{itz}}{\kappa_{z}}
\frac{\partial^{i+1}{F}}{\partial{t^{i+1}}}\nonumber\\
&&+\kappa_{tz}\frac{\partial^2{F}}{\partial{t}\partial{z}}
+\kappa_{ttz}\frac{\partial^3{F}}{\partial{t^2}\partial{z}}
+\kappa_{3tz}\frac{\partial^4{F}}{\partial{t^3}\partial{z}}
+\kappa_{4tz}\frac{\partial^5{F}}{\partial{t^4}\partial{z}}
+\kappa_{5tz}\frac{\partial^6{F}}{\partial{t^5}\partial{z}}
+\cdots\nonumber\\
&&+\kappa_{itz}\frac{\partial^{i+1}{F}}{\partial{t^i}\partial{z}}
+\left(\kappa_{itz}+\kappa_{tz}\frac{\kappa_{itz}}
{\kappa_{z}}\right)\frac{\partial^{i+2}{F}}{\partial{t^{i+1}}\partial{z}}
+\left(\kappa_{(i+1)tz}+\kappa_{ttz}\frac{\kappa_{itz}}
{\kappa_{z}}\right)\frac{\partial^{i+3}{F}}{\partial{t^{i+2}}\partial{z}}
+\cdots
\nonumber\\
&&+\kappa_{tzz}\frac{\partial^3 F}{\partial{t}\partial{z^2}}
+\kappa_{ttzz}\frac{\partial^4 F}{\partial{t^2}\partial{z^2}}
+\kappa_{3tzz}\frac{\partial^5 F}{\partial{t^3}\partial{z^2}}
+\cdots
+\left(\kappa_{itzz}+\kappa_{zz}\frac{\kappa_{itz}}
{\kappa_{z}}\right)\frac{\partial^{i+2}{F}}{\partial{t^i}\partial{z^2}}
+\cdots
\nonumber\\
&&+\kappa_{tzzz}\frac{\partial^4 F}{\partial{t}\partial{z^3}}
+\kappa_{ttzzz}\frac{\partial^5 F}{\partial{t^2}\partial{z^3}}
+\kappa_{tt4z}\frac{\partial^6 F}{\partial{t^3}\partial{z^3}}
+\cdots
+\left(\kappa_{it3z}+\kappa_{zzz}\frac{\kappa_{itz}}
{\kappa_{z}}\right)\frac{\partial^{i+3}{F}}{\partial{t^i}\partial{z^3}}
\nonumber\\
&&+\cdots.
\label{Ritz equation of ith operation}
\end{eqnarray}
The latter manipulation is a new DIO,  
i.e., $R_{itz}$ of the
i-th order operation. 
It is obvious that the Fick's law definition $\kappa_{zz}^{FL}$
is equal to $\kappa_{zz}$, $\kappa_{zz}^{FL}=\kappa_{zz}$,
which is identical with Equation (\ref{kzzfl0}).
Thus, $\kappa_{zz}^{FL}$ is an invairant quantity for 
$R_{itz}$ of the
i-th order operation.

From Equation (\ref{Ritz equation of ith operation})
we can obtain the first and second moment equations
of the displacement as
\begin{eqnarray}
\frac{d}{dt}\langle (\Delta z)
\rangle&=&\kappa_z-\frac{\kappa_{itz}}{\kappa_{z}}
\frac{d^{i+1}}{dt^{i+1}}\langle (\Delta z)\rangle,
\label{dzdt Ritz of ith PtE}\\
\frac{d}{dt}\langle (\Delta z)^2\rangle&
=&2\kappa_z\langle (\Delta z)
\rangle+2\kappa_{zz}-\frac{\kappa_{itz}}{\kappa_{z}}
\frac{d^{i+1}}{dt^{i+1}}\langle (\Delta z)^2\rangle
-2\kappa_{tz}\frac{d}{dt}
\langle (\Delta z)\rangle-2T_{itz}
\label{dz2dt Ritz of ith PtE}
\end{eqnarray}
with
\begin{eqnarray}
T_{itz}=&&\kappa_{ttz}\frac{d^2}{dt^2}\langle (\Delta z)\rangle
+\kappa_{3tz}\frac{d^3}{dt^3}\langle (\Delta z)\rangle
+\kappa_{4tz}\frac{d^4}{dt^4}\langle (\Delta z)\rangle
+\cdots
+\kappa_{itz}\frac{d^i}{dt^i}\langle (\Delta z)\rangle\nonumber\\
&&+\left(\kappa_{itz}+\kappa_{tz}\frac{\kappa_{itz}}
{\kappa_z}\right)\frac{d^{i+1}}{dt^{i+1}}\langle (\Delta z)\rangle
+\left(\kappa_{(+1)itz}+\kappa_{ttz}\frac{\kappa_{itz}}
{\kappa_z}\right)\frac{d^{i+2}}{dt^{i+2}}\langle (\Delta z)\rangle
+\cdots.
\label{Titz for Ritz of ith PtI operation}
\end{eqnarray}
The solution of Equation (\ref{dzdt Ritz of ith PtE})
 can be easily found
\begin{eqnarray}
\langle (\Delta z)\rangle&=&c_1+c_2e^{r_1t}+c_3e^{r_2t}
+c_4e^{r_3t}+\cdots+c_{i+1}e^{r_{i}t}+\kappa_zt,
\label{solution of delta z for Ritz}
\end{eqnarray}
where $r_1$, $r_2$, $r_3$, $\cdots$, and $r_i$ are
the solution of the corresponding characteristic equation,
and $c_1$, $c_2$, $c_3$, $\cdots$, and $c_{i+1}$ are
undetermined constants.
For the special condition
$c_2=c_3=\cdots=c_{i+1}=0$,
Equation (\ref{solution of delta z for Ritz}) becomes
\begin{eqnarray}
\langle (\Delta z)\rangle&=&c_1+\kappa_zt.
\label{solution of delta z for Ritz under strong condition}
\end{eqnarray}
Inserting the latter formula into Equation
(\ref{dz2dt Ritz of ith PtE})
gives
\begin{eqnarray}
\frac{\kappa_{itz}}{\kappa_{z}}
\frac{d^{i+1}}{dt^{i+1}}\langle (\Delta z)^2
\rangle+\frac{d}{dt}\langle (\Delta z)^2\rangle&
=&2\left(\kappa_zc_1+\kappa_{zz}-\kappa_{tz}
\kappa_z\right)+2\kappa_z^2 t,
\label{dz2dt Ritz of ith PtE-2}
\end{eqnarray}
the solution of which can be found
\begin{eqnarray}
\langle (\Delta z)^2\rangle&=&c'_1+c'_2e^{r'_1t}
+c'_3e^{r'_2t}+c'_4e^{r'_3t}+\cdots+c'_ie^{r'_{i-1}t}
+2\left(\kappa_zc'_1+\kappa_{zz}-\kappa_{tz}
\kappa_z\right)+\kappa_z^2 t^2.\label{solution of delta z2 for Ritz}
\end{eqnarray}
Here, $r'_1$, $r'_2$, $r'_3$, $\cdots$, and $r'_i$
are the solution of the characteristic equation corresponding to
Equation (\ref{dz2dt Ritz of ith PtE-2}),
and $c'_1$, $c'_2$, $c'_3$, $\cdots$, and $c'_i$
are undetermined constants.
With the special condition
$c'_2=c'_3=\cdots=c'_i=0$, 
Equation (\ref{solution of delta z2 for Ritz}) becomes
\begin{eqnarray}
\langle (\Delta z)^2\rangle&=&c'_1
+2\left(\kappa_zc'_1+\kappa_{zz}-\kappa_{tz}
\kappa_z\right)t+\kappa_z^2 t^2.
\label{solution of delta z2 for Ritz under
strong condition}
\end{eqnarray}
Combining Equations (\ref{solution of delta
z for Ritz under strong condition})
and (\ref{solution of delta z2 for Ritz under
strong condition}) yields, for the limit $t\rightarrow t_{\infty}$,
\begin{eqnarray}
\kappa_{zz}^{VD}=\frac{1}{2}\lim_{t\rightarrow t_{\infty}}
\frac{d\sigma^2}{dt}
=\kappa_{zz}-\kappa_{tz}\kappa_z,
\end{eqnarray}
which is identical with equation (\ref{kzzvd0}). 
The above investigation shows that at least for the
special condition the displacement
variance definition $\kappa_{zz}^{DV}$
is invariant for $R_{itz}$ of the third-order PtI operation. 

\subsubsection{$R_{itz}$ of the ith-order
PtI operation and $R_{tz}$ of the 1st-order PtI operation}
\label{Displacement Variance definition and
Fick's law definition for Ritz of the i-th PtI
operation and Rtz of the 1st one}

If considering Equations (\ref{Rst of 1st PtE})
and (\ref{Ritz equation of ith operation}),
that is, combining the $R_{itz}$ of the ith PtI
operation and $R_{tz}$ of the 1st-order  PtI operation,
which is another new DIO,
we can obtain the following equation
\begin{eqnarray}
\frac{\partial{F}}{\partial{t}}=
&&\left(-\kappa_z\frac{\partial{F}}{\partial{z}}+\kappa_{zz}
\frac{\partial^2{F}}{\partial{z^2}}
+\kappa_{zzz}\frac{\partial^3{F}}{\partial{z^3}}
+\kappa_{zzzz}\frac{\partial^4{F}}{\partial{z^4}}
+\cdots\right)-\frac{\kappa_{tz}}{\kappa_z}
\frac{\partial^2{F}}{\partial{t^2}}
-\frac{\kappa_{itz}}{\kappa_z}\frac{\partial^{i+1}{F}}
{\partial{t^{i+1}}}
\nonumber\\
&&+\left(\kappa_{tzz}+\frac{\kappa_{tz}\kappa_{zz}}{\kappa_z}\right)
\frac{\partial^3{F}}{\partial{t}\partial{z^2}}
+\left(\kappa_{tzzz}+\frac{\kappa_{tz}\kappa_{zzz}}{\kappa_z}\right)
\frac{\partial^4{F}}{\partial{t}\partial{z^3}}
+\cdots
\nonumber\\
&&+\left(\kappa_{ttz}+\frac{\kappa_{tz}^2}{\kappa_z}\right)
\frac{\partial^3{F}}{\partial{t^2}\partial{z}}
+\left(\kappa_{tttz}+\frac{\kappa_{tz}\kappa_{ttz}}{\kappa_z}
\right)\frac{\partial^4{F}}{\partial{t^3}\partial{z}}
+\left(\kappa_{ttttz}+\frac{\kappa_{tz}\kappa_{tttz}}{\kappa_z}
\right)\frac{\partial^5{F}}{\partial{t^4}\partial{z}}+\cdots\nonumber\\
&&+\left(\kappa_{ttzz}+\frac{\kappa_{tz}\kappa_{tzz}}{\kappa_z}\right)
\frac{\partial^4{F}}{\partial{t^2}\partial{z^2}}
+\left(\kappa_{tttzz}+\frac{\kappa_{tz}\kappa_{ttzz}}{\kappa_z}\right)
\frac{\partial^5{F}}{\partial{t^3}\partial{z^2}}
+\left(\kappa_{ttttzz}+\frac{\kappa_{tz}\kappa_{tttzz}}{\kappa_z}\right)
\frac{\partial^6{F}}{\partial{t^4}\partial{z^2}}+\cdots\nonumber\\
&&+\left(\kappa_{ttzzz}+\frac{\kappa_{tz}\kappa_{tzzz}}{\kappa_z}\right)
\frac{\partial^5{F}}{\partial{t^2}\partial{z^3}}
+\left(\kappa_{tttzzz}+\frac{\kappa_{tz}\kappa_{ttzzz}}{\kappa_z}\right)
\frac{\partial^6{F}}{\partial{t^3}\partial{z^3}}+\cdots\nonumber\\
&&+\left(\kappa_{it3z}+\kappa_{zzz}\frac{\kappa_{itz}}{\kappa_z}
+\frac{\kappa_{tz}}{\kappa_z}\kappa_{(i-1)t3z}\right)
\frac{\partial^{i+3}{F}}{\partial{t^i}\partial{z^3}}+\cdots
\nonumber\\
&&+\left(\kappa_{itz}+\kappa_{(i-1)tz}\frac{\kappa_{tz}}{\kappa_z}
\right)
\frac{\partial^{i+1}{F}}{\partial{t^i}\partial{z}}
+\left(\kappa_{(i+1)tz}+\kappa_{tz}\frac{\kappa_{itz}}{\kappa_z}
+\frac{\kappa_{tz}}{\kappa_z}\kappa_{itz}\right)
\frac{\partial^{i+2}{F}}{\partial{t^{(i+1)}}\partial{z}}\nonumber\\
&&+\left(\kappa_{(i+2)tz}+\kappa_{ttz}\frac{\kappa_{itz}}{\kappa_z}
+\frac{\kappa_{tz}}{\kappa_z}\kappa_{(i+1)tz}\right)
\frac{\partial^{i+3}{F}}{\partial{t^{(i+2)}}\partial{z}}+\cdots
\nonumber\\
&&+\left(\kappa_{itz}+\kappa_{zz}\frac{\kappa_{itz}}{\kappa_z}
+\frac{\kappa_{tz}}{\kappa_z}\kappa_{(i-1)tzz}\right)
\frac{\partial^{i+2}{F}}{\partial{t^{i}}\partial{z^2}}+\cdots. 
\end{eqnarray}
From the latter equation we can find
\begin{eqnarray}
\frac{d}{dt}\langle (\Delta z)
\rangle&=&\kappa_z
-\frac{\kappa_{tz}}{\kappa_z}\frac{d^2}{dt^2}
\langle \left(\Delta z\right) \rangle
-\frac{\kappa_{itz}}{\kappa_{z}}
\frac{d^{i+1}}{dt^{i+1}}\langle (\Delta z)\rangle,
\label{dzdt Ritz of ith PtE and Rtz of 1st PtE}\\
\frac{d}{dt}\langle (\Delta z)^2\rangle&
=&2\kappa_z\langle (\Delta z)
\rangle+2\kappa_{zz}
-\frac{\kappa_{tz}}{\kappa_z}\frac{d^2}{dt^2}
\langle (\Delta z)^2\rangle
-\frac{\kappa_{itz}}{\kappa_{z}}
\frac{d^{i+1}}{dt^{i+1}}\langle (\Delta z)^2\rangle+T_{itz+tz}
\label{dz2dt Ritz of ith PtE  and Rtz of 1st PtE}
\end{eqnarray}
with
\begin{eqnarray}
T_{itz+tz}=&&\left(\kappa_{ttz}+\frac{\kappa_{tz}^2}{\kappa_z}\right)
\frac{d^2}{dt^2}\langle (\Delta z)\rangle+
\left(\kappa_{tttz}+\frac{\kappa_{ttz}\kappa_{tz}}{\kappa_z}\right)
\frac{d^3}{dt^3}\langle (\Delta z)\rangle
+\left(\kappa_{4tz}+\frac{\kappa_{tttz}\kappa_{tz}}{\kappa_z}\right)
\frac{d^4}{dt^4}\langle (\Delta z)\rangle
+\cdots\nonumber\\
&&+\left(\kappa_{itz}+\kappa_{(i-1)tz}\frac{\kappa_{tz}}{\kappa_z}\right)
\frac{d^{i}}{dt^{i}}\langle (\Delta z)\rangle
+\left(\kappa_{(i+1)tz}+\kappa_{tz}\frac{\kappa_{itz}}{\kappa_z}
+\kappa_{itz}\frac{\kappa_{tz}}{\kappa_z}\right)
\frac{d^{(i+1)}}{dt^{(i+1)}}\langle (\Delta z)\rangle\nonumber\\
&&+\left(\kappa_{(i+2)tz}+\kappa_{ttz}\frac{\kappa_{itz}}{\kappa_z}
+\kappa_{(i+1)tz}\frac{\kappa_{tz}}{\kappa_z}\right)
\frac{d^{(i+2)}}{dt^{(i+2)}}\langle (\Delta z)\rangle+\cdots.
\label{Titz+tz}
\end{eqnarray}
The characteristic equation corresponding to 
Equation (\ref{dzdt Ritz of ith PtE and Rtz of 1st PtE})
can be found
\begin{eqnarray}
r^{i+1}+\frac{\kappa_{tz}}{\kappa_{itz}}r^2+
\frac{\kappa_{z}}{\kappa_{itz}}r=0,
\end{eqnarray}
which has $(i+1)$ solutions. 
As the above subsection, for the special 
condition we can obtain the solution
of Equation (\ref{dzdt Ritz of ith PtE and Rtz of 1st PtE}) as
\begin{eqnarray}
\langle (\Delta z)\rangle&=&c_1+\kappa_zt,
\label{solution of delta z for Ritz+Rtz for strong condition}
\end{eqnarray}
inserting which into Equations
(\ref{dz2dt Ritz of ith PtE  and Rtz of 1st PtE})
and (\ref{Titz+tz}), one can find 
\begin{eqnarray}
\frac{d}{dt}\langle (\Delta z)^2\rangle&=&2\kappa_z\left(c_1
+\kappa_zt\right)
+2\kappa_{zz}
-\frac{\kappa_{tz}}{\kappa_z}\frac{d^2}{dt^2}\langle
(\Delta z)^2\rangle
-\frac{\kappa_{itz}}{\kappa_{z}}
\frac{d^{i+1}}{dt^{i+1}}\langle (\Delta z)^2\rangle.
\label{dz2dt Ritz of ith PtE  and Rtz of 1st PtE-2}
\end{eqnarray}
For the special
condition, the solution of the latter equation  is shown as
\begin{eqnarray}
\langle (\Delta z)^2\rangle&=&c'_1
+2\left(\kappa_zc'_1+\kappa_{zz}-\kappa_{tz}\kappa_z\right)t
+\kappa_z^2 t^2.
\label{solution of delta z2 for Ritz+Rtz for strong condition}
\end{eqnarray}

Combining Equations (\ref{dzdt Ritz of ith PtE and Rtz of 1st PtE})
and (\ref{dz2dt Ritz of ith PtE  and Rtz of 1st PtE-2}) gives
\begin{eqnarray}
\frac{1}{2}\frac{d\sigma^2}{dt}
&=&\kappa_{zz}+\frac{\kappa_{tz}}{\kappa_z}\langle
(\Delta z)\rangle\frac{d^2}{dt^2}\langle (\Delta z)\rangle
+\frac{\kappa_{itz}}{\kappa_{z}}\langle (\Delta z
)\rangle\frac{d^{i+1}}{dt^{i+1}}\langle (\Delta z)\rangle
-\frac{1}{2}\frac{\kappa_{tz}}{\kappa_{z}}
\frac{d^2}{dt^2}\langle (\Delta z)^2\rangle
-\frac{1}{2}\frac{\kappa_{itz}}{\kappa_{z}}
\frac{d^{i+1}}{dt^{i+1}}\langle (\Delta z)^2\rangle.
\label{dz2dt Ritz of ith PtE-2}
\end{eqnarray}
Considering Equations (\ref{solution of delta z for Ritz+Rtz
for strong condition}),
(\ref{solution of delta z2 for Ritz+Rtz for strong condition}), 
and (\ref{dz2dt Ritz of ith PtE-2}), for the limit $t\rightarrow t_{\infty}$
we can find 
\begin{eqnarray}
\kappa_{zz}^{VD}=\frac{1}{2}\lim_{t\rightarrow t_{\infty}}
\frac{d\sigma^2}{dt}
=\kappa_{zz}-\kappa_{tz}\kappa_z,
\end{eqnarray}
which is identical with Equation (\ref{kzzvd0}) and shows, 
at least for the special condition, that the displacement 
variance definition is an invariant quantity 
for the $i$th-order
PtI operation and $R_{tz}$ of the 1st-order PtI operation. 

Because there is not second-order spatial derivative term in 
Equation (\ref{n PtE}) with $n=1,2,3,\cdots$, 
any manipulation of inserting  
a deformation of Equation (\ref{n PtE}) 
into Equation
(\ref{equation of F with constant coefficient})
cannot change the Fick's law  definition of the SPDC.
Analogous to the deduction in the previous part 
in this subsection, 
at least for the special condition,
the displacement vairance definition is invariant 
for the more complicated combination 
of the DIOs of the PtI operations. 

\section{TGK definition FOR FOCUSING FIELD}
\label{TGK definition FOR FOCUSING FIELD}

The Taylor-Green-Kubo (TGK) formulation
is a useful tool to calculate diffusion coefficients
\citep{Taylor1922,Green1951, Kubo1962},
for the SPDC, it is shown as
\begin{eqnarray}
\kappa_{zz}^{TGK}=\int_0^{\infty}dt
\langle v_z(t)v_z(0)\rangle.
\label{TGK formulation}
\end{eqnarray}
Here, the z-component of energetic charged particle velocity 
$v_z$ is equal to $v\mu$ with the particle velocity 
$v$ and the pitch-angle cosine $\mu$, so 
Equation (\ref{TGK formulation}) can be rewritten as
\begin{eqnarray}
\kappa_{zz}^{TGK}=v^2\int_0^{\infty}dt 
\langle \mu(t)\mu(0)\rangle.
\label{TGK formulation with mu}
\end{eqnarray}
The integral in the latter equation 
is shown 
\begin{eqnarray}
\int_0^{\infty}dt \langle \mu(t)\mu(0)\rangle=\frac{1}{4}
\int_{-\infty}^{\infty}dz\int_0^{\infty}dt\int_{-1}^1d\mu_0
\mu_0\int_{-1}^1d\mu\mu f(z,\mu,t),
\label{integral in TGK formulation with mu}
\end{eqnarray}
where $f(z,\mu,t)$ is the distribution function
 and satisfies the Fokker-Planck equation,
$\mu_0$ is the initial pitch angle cosine of energetic particle.
Thus, TGK formulation becomes
\begin{eqnarray}
\kappa_{zz}^{TGK}=\frac{v^2}{4}\int_{-\infty}^{\infty}dz
\int_0^{\infty}dt\int_{-1}^1d\mu_0\mu_0\int_{-1}^1d\mu\mu f(z,\mu,t),
\end{eqnarray}
which, 
with Equations (\ref{f=F+g})-(\ref{integrate g over mu=0}),
can be expressed as 
\begin{eqnarray}
\kappa_{zz}^{TGK}=\frac{v^2}{4}\int_{-\infty}^{\infty}dz
\int_0^{\infty}dt\int_{-1}^1d\mu_0\mu_0\int_{-1}^1d\mu\mu g(z,\mu,t).
\label{TGK formulation with g}
\end{eqnarray}
Here, $\int_{-1}^{1}d\mu\mu F(z,t)=0$ is used.

\label{From TGK formulation to derive the parallel 
diffusion coefficient formula of Shalchi and Danos (2013)}

With the anisotropic distribution function 
$g(z,\mu,t)$ (see Equation (\ref{g})),
Equation (\ref{TGK formulation with g}) becomes
\begin{eqnarray}
\kappa_{zz}^{TGK}&=&\frac{v^2}{4}\int_{-\infty}^{\infty}
dz\int_0^{\infty}dt\int_{-1}^1d\mu_0\mu_0\int_{-1}^1d\mu\mu
\Bigg\{L\left(\frac{\partial{F}}
{\partial{z}}
-\frac{F}{L}\right)\left[1-
\frac{2e^{M(\mu)}}{\int_{-1}^{1}
	d\mu e^{M(\mu) }}\right]\nonumber\\
&&+e^{M(\mu)}\left[R(\mu)
-\frac{\int_{-1}^{1}d\mu
	e^{M(\mu)}R(\mu)}
{\int_{-1}^{1}d\mu
	e^{M(\mu) }}\right]\Bigg\}.
\label{combining TGK formulation and g}
\end{eqnarray}
Because isotropic distribution function $F(z,t)$ 
does not contain variable $\mu_0$, the following integral
can be obtained
\begin{eqnarray}
\kappa_{zz}^{TGK}=\frac{v^2}{4}\int_{-\infty}^{\infty}dz
\int_0^{\infty}dt\int_{-1}^1d\mu\mu\int_{-1}^1d\mu_0\mu_0
L\left(\frac{\partial{F}}
{\partial{z}}
-\frac{F}{L}\right)\left[1-
\frac{2e^{M(\mu)}}{\int_{-1}^{1}
	d\mu e^{M(\mu) }}\right]=0,
\label{integral=0}
\end{eqnarray}
from which, Equation (\ref{combining TGK formulation and g}) becomes
\begin{eqnarray}
\kappa_{zz}^{TGKF}&=&\frac{v^2}{4}(X_1-X_2)
\end{eqnarray}
with
\begin{eqnarray}
&&X_1=\int_{-\infty}^{\infty}dz\int_0^{\infty}dt
\int_{-1}^1d\mu_0\mu_0\int_{-1}^1d\mu\mu
e^{M(\mu)}R(\mu),
\label{X1}\\
&&X_2=\frac{\int_{-1}^1d\mu\mu e^{M(\mu)}}
{\int_{-1}^{1}d\mu e^{M(\mu) }}\int_{-\infty}^{\infty}dz\int_0^{\infty}dt
\int_{-1}^1d\mu_0\mu_0\int_{-1}^{1}d\mu e^{M(\mu)}R(\mu).
\label{X2}
\end{eqnarray}
Inserting Equation (\ref{R(mu)}) with Equation (\ref{Phi}) 
into Equations (\ref{X1}) and (\ref{X2})
yields
\begin{eqnarray}
&&X_1=Y_1+Y_2,
\label{X1=Y1+Y2}\\
&&X_2=Y_3+Y_4
\label{X2=Y3+Y4}
\end{eqnarray}
with
\begin{eqnarray}
&&Y_1=\int_{-\infty}^{\infty}dz\int_0^{\infty}dt
\int_{-1}^1d\mu_0\mu_0\int_{-1}^1d\mu\mu
e^{M(\mu)}\int_{-1}^{\mu} d\nu
\frac{e^{-M(\nu)}}{D_{\nu\nu}(\nu)}
\left(\frac{\partial{F}}
{\partial{t}}\nu
+\int_{-1}^{\nu}\frac{\partial{g}}{\partial{t}}d\rho\right),\\
&&Y_2=\frac{v}{2}\int_{-\infty}^{\infty}dz\int_0^{\infty}dt
\int_{-1}^1d\mu_0\mu_0\int_{-1}^1d\mu\mu
e^{M(\mu)}\int_{-1}^{\mu} d\nu
\frac{e^{-M(\nu)}}{D_{\nu\nu}(\nu)}
\left(2\int_{-1}^{\nu}d\rho \rho \frac{\partial{g}}{\partial{z}}-
\int_{-1}^{1}d\mu \mu \frac{\partial{g}}{\partial{z}}\right),\\
&&Y_3=\frac{\int_{-1}^1d\mu\mu e^{M(\mu)}}
{\int_{-1}^{1}d\mu e^{M(\mu) }}\int_{-\infty}^{\infty}dz\int_0^{\infty}dt
\int_{-1}^1d\mu_0\mu_0\int_{-1}^{1}d\mu e^{M(\mu)}\int_{-1}^{\mu} d\nu
\frac{e^{-M(\nu)}}{D_{\nu\nu}(\nu)}
\left(\frac{\partial{F}}
{\partial{t}}\nu
+\int_{-1}^{\nu}\frac{\partial{g}}{\partial{t}}d\rho\right) ,\\
&&Y_4=\frac{\int_{-1}^1d\mu\mu e^{M(\mu)}}
{\int_{-1}^{1}d\mu e^{M(\mu) }}\int_{-\infty}^{\infty}dz
\int_0^{\infty}dt\int_{-1}^1d\mu_0\mu_0
\int_{-1}^{1}d\mu e^{M(\mu)}\int_{-1}^{\mu} d\nu
\frac{e^{-M(\nu)}}{D_{\nu\nu}(\nu)}
\left(2\int_{-1}^{\nu}d\rho \rho \frac{\partial{g}}{\partial{z}}-
\int_{-1}^{1}d\mu \mu \frac{\partial{g}}{\partial{z}}\right).
\end{eqnarray}
By using the following formulas
\begin{eqnarray}
&&\int_{-1}^1 d\mu_0\mu_0 F=0,\\
&&\int_0^{\infty}dt \int_{-\infty}^{\infty}dz\frac{\partial{g(z,t)}}{\partial{t}}
=\left[\int_{-\infty}^{\infty}dzg(z)\right](t=\infty)
-\left[\int_{-\infty}^{\infty}dzg(z)\right](t=0),\\
&&\int_0^{\infty}dz \frac{\partial{g}}{\partial{z}}=g(z=\infty)-g(z=-\infty)
\label{1}
\end{eqnarray}
with
\begin{eqnarray}
&&\left[\int_{-\infty}^{\infty}dzg(z)\right](t=\infty)=0,\\
&&\left[\int_{-\infty}^{\infty}dzg(z)\right](t=0)=2\delta(\mu-\mu_0),\\
&&g(z=\infty)=g(z=-\infty)=0,
\label{2}
\end{eqnarray}
we can find
\begin{eqnarray}
&&Y_1=-2\int_{-1}^1d\mu_0\mu_0\int_{-1}^1d\mu\mu
e^{M(\mu)}\int_{-1}^{\mu} d\nu
\frac{e^{-M(\nu)}}{D_{\nu\nu}(\nu)}
\int_{-1}^{\nu}\delta (\nu-\nu_0)d\rho,
\label{Y1}\\
&&Y_2=0,
\label{Y2}\\
&&Y_3=-2\frac{\int_{-1}^1d\mu\mu e^{M(\mu)}}
{\int_{-1}^{1}d\mu e^{M(\mu) }}\int_{-1}^1d\mu_0\mu_0
\int_{-1}^{1}d\mu e^{M(\mu)}\int_{-1}^{\mu} d\nu
\frac{e^{-M(\nu)}}{D_{\nu\nu}(\nu)}
\int_{-1}^{\nu}\delta (\nu-\nu_0)d\rho,
\label{Y3}\\
&&Y_4=0.
\label{Y4}
\end{eqnarray}

Equations (\ref{Y1})-(\ref{Y4}) can be combined with 
Equations (\ref{X1=Y1+Y2}) and (\ref{X2=Y3+Y4}) becomes
\begin{eqnarray}
&&X_1=-2\int_{-1}^1d\mu_0\mu_0\int_{-1}^1d\mu\mu
e^{M(\mu)}\int_{-1}^{\mu} d\nu
\frac{e^{-M(\nu)}}{D_{\nu\nu}(\nu)}
\int_{-1}^{\nu}\delta (\nu-\nu_0)d\rho,
\label{X1=Y1+Y2-2}\\
&&X_2=-2\frac{\int_{-1}^1d\mu\mu e^{M(\mu)}}
{\int_{-1}^{1}d\mu e^{M(\mu) }}\int_{-1}^1d\mu_0\mu_0
\int_{-1}^{1}d\mu e^{M(\mu)}\int_{-1}^{\mu} d\nu
\frac{e^{-M(\nu)}}{D_{\nu\nu}(\nu)}
\int_{-1}^{\nu}\delta (\nu-\nu_0)d\rho.
\label{X2=Y3+Y4-2}
\end{eqnarray}
With the formula
\begin{eqnarray}
\int_{-1}^\mu d\nu\int_{-1}^1d\mu_0\mu_0 \delta (\nu-\mu_0)=\frac{\mu^2-1}{2},
\end{eqnarray}
we can derive
\begin{eqnarray}
&&X_1=\int_{-1}^1d\mu\mu
e^{M(\mu)}\int_{-1}^{\mu} d\nu e^{-M(\nu)}
\frac{(1-\nu^2)}{D_{\nu\nu}(\nu)},
\label{X1=Y1+Y2-3}\\
&&X_2=\frac{\int_{-1}^1d\mu\mu e^{M(\mu)}}
{\int_{-1}^{1}d\mu e^{M(\mu) }}\int_{-1}^1d\mu_0\mu_0
\int_{-1}^{1}d\mu e^{M(\mu)}\int_{-1}^{\mu} d\nu
e^{-M(\nu)}
\frac{(1-\nu^2)}{D_{\nu\nu}(\nu)}.
\label{X2=Y3+Y4-4}
\end{eqnarray}
Combining Equations (\ref{X1=Y1+Y2-3}), (\ref{X2=Y3+Y4-4}), 
and Equations (\ref{kzz-xi-x2}) yields
\begin{eqnarray}
\kappa_{zz}^{TGKF}&=&\frac{v^2}{4}\int_{-1}^1d\mu\mu
e^{M(\mu)}\int_{-1}^{\mu} d\nu e^{-M(\nu)}
\frac{(1-\nu^2)}{D_{\nu\nu}(\nu)}\nonumber\\
&&-\frac{v^2}{4}\frac{\int_{-1}^1d\mu\mu e^{M(\mu)}}
{\int_{-1}^{1}d\mu e^{M(\mu) }}\int_{-1}^{1}d\mu 
e^{M(\mu)}\int_{-1}^{\mu} d\nu
e^{-M(\nu)}
\frac{(1-\nu^2)}{D_{\nu\nu}(\nu)}.
\label{kzz-xi-x2}
\end{eqnarray}
The latter equation is identical with Equation (56) 
in the paper of \citet{ShalchiEA2013}.

From Equations (\ref{1}) and
(\ref{2})
we can find that the contribution 
of the terms containing $\partial{g}/\partial{z}$ 
to the SPDC is equal to zero, 
i.e., neglecting the higher-order spatial derivative terms, which is used in
the previous papers
\citep{BeeckEA1986, Litvinenko2012b, HeEA2014}.
However, \citet{wq2018} showed that for focusing field the influence of
the terms containing $\partial{g}/\partial{z}$ 
to the SPDC
cannot be ignored.
Therefore, $\kappa_{zz}^{TGK}$ only give the approximate result
in such condition. 
In order to prove this inference, 
in what follows, 
we evaluate Equation 
(\ref{kzz-xi-x2}) for the isotropic 
pitch angle scattering model
$D_{\mu\mu}(\mu)=D(1-\mu^2)$ with the constant $D$. 

For this simple model,
Equation (\ref{kzz-xi-x2}) is easily simplified as 
\begin{eqnarray}
\kappa_{zz}^{TGK}=\kappa_\parallel^0(1+S)
\label{kzz-TGKF}
\end{eqnarray}
with
\begin{eqnarray}
S=-\frac{1}{15}\xi^2
+\frac{2}{315}\xi^4+\cdots.
\label{S in the maintext}
\end{eqnarray}
Here, $\kappa_\parallel^0$ is the SPDC 
for the constant mean magnetic field, 
and 
\begin{equation}
\xi= \frac{v}{2DL}.
\label{xi}
\end{equation}
is the dimensionless parameter
describing the focusing field with the adiabatic 
focusing characteristic length $L$.
Equations (\ref{kzz-TGKF}) and (\ref{S in the maintext}) show that
the adiabatic focusing effect reduces the value of the SPDC
regardless of the sign of the focusing length.

In fact, Equation (\ref{kzz-TGKF}) is identical with
that derived by many researchers \citep
{BeeckEA1986, BieberEA1990, Kota2000, Litvinenko2012a, ShalchiEA2013,
HeEA2014} (hereafter, the result is denoted as 
$\kappa_{\parallel}^{BW}$), however, 
which is an approximate result
\citep{wq2018}.
In the above derivation, we find  
$\kappa_\parallel^{TGK}=\kappa_\parallel^{BW}$, 
therefore, the TGK formula definition $\kappa_\parallel^{TGK}$
is also an approximate result. 
Thus, we confirm the inference in the above part.

Considering the formula of $g(z,\mu,t)$ (see 
Equation (\ref{g2})) and Equation (\ref{TGK formulation with g}),
we can find that the TGK definition
of the parallel diffusion $\kappa_{zz}^{TGK}$
is determined by two factors, the initial condition
\begin{eqnarray}
\left[\int_{-\infty}^{\infty}dzg(z)\right](t=0)=2\delta(\mu-\mu_0)
\end{eqnarray}
and the coefficients of the terms containing $\partial {g(z,\mu,t)}/dt$ 
in the formula of $g(z,\mu,t)$. 
The above factors are not influenced by 
the manipulations of
the PzI, PtI, and PtzI operations,
so the TGK definition $\kappa_{zz}^{TGK}$ as well as
the displacement variance definition $\kappa_{zz}^{DV}$ are invariant for
the PzI, PtI, and PtzI operations. However, 
for focusing field, 
the TGK definition is only approximate result.
Consequently, the displacement variance definition
$\kappa_{zz}^{DV}$
is more appropriate than 
the other ones. 

\section{To evaluate THE displacement VARIANCE DEFINITION}
\label{EVALUATING THE displacement VARIANCE DEFINITION}

In this section, we evaluate the displacement variance definition
through the formula 
\begin{eqnarray}
\kappa_{zz}^{VD}
=\kappa_{zz}-\kappa_{tz}\kappa_z.
\label{kzzdv for evaluation}
\end{eqnarray}
For the isotropic pitch angle scattering model
$D_{\mu\mu}(\mu)=D(1-\mu^2)$ with
the constant $D$,
the coefficient $\kappa_{tz}$ becomes
\begin{equation}
\kappa_{tz}\approx \frac{2v}{9D}\xi,
\label{ktz-xi}
\end{equation}
where the focusing parameter is $\xi=v/(2DL)$.
The detailed derivation is shown 
in Appendix \ref{The accurate formula of ktz}.
The streaming coefficient $\kappa_z$ is also evaluated for the model
$D_{\mu\mu}(\mu)=D(1-\mu^2)$ as
\begin{eqnarray}
\kappa_{z}\approx \frac{v}{3}\xi,
\label{kz for first order xi-main paper}
\end{eqnarray}
the evaluation process of which 
can be found in Appendix \ref{Evaluate ktz for isotropic pitch
angle scattering}.

\citet{wq2019} found the formula $\kappa_{zz}^{DV}=\kappa_{zz}
-\kappa_z\kappa_{tz}$
from Equation (\ref{equation of F with constant coefficient}), 
where $\kappa_{zz}$
was obtained 
as
\begin{equation}
\kappa_{zz}=\kappa_{zz}^{BW}+\eta_{0,2,0}
\label{kzzWQ}
\end{equation}
with
\begin{eqnarray}
\eta_{0,2,0}\approx
\frac{1}{5}\xi^2
\kappa_{\parallel 0}.
\label{A1 with kappa0}
\end{eqnarray}
Here, the quantity $\kappa_{zz}^{BW}$ is shown as follows 
\begin{eqnarray}
\kappa_{zz}^{BW}=\kappa_{\parallel 0}(1+S)
\label{kzzBW}
\end{eqnarray}
with
\begin{eqnarray}
S=-\frac{1}{15}\xi^2
+\frac{2}{315}\xi^4+\cdots.
\label{S-main text}
\end{eqnarray}
Equation (\ref{kzzBW})
is an approximate formula
of the SPDC
\citep{BeeckEA1986, Litvinenko2012b, ShalchiEA2013, HeEA2014}. 
Combining Equations (\ref{kzzWQ})-(\ref{S-main text}), \citet{wq2018}
found
\begin{equation}
\kappa_{zz}
\approx \kappa_{\parallel 0}
\left(1+\frac{2}{15}\xi^2
\right). 
\label{kzzWQ-main text with xi}
\end{equation}
Inserting Equations (\ref{ktz-xi}), (\ref{kz for
first order xi-main paper}), and
(\ref{kzzWQ-main text with xi}) into formula 
(\ref{kzzdv for evaluation})
yields
\begin{equation}
\kappa_{zz}^{DV}=\kappa_{zz}-\kappa_z\kappa_{tz}
\approx\kappa_{\parallel 0}
\left(1-\frac{14}{45}\xi^2
\right),
\label{Wang and Qin formula 2019 with xi}
\end{equation}
which shows the corrective action
induced by
the adiabatic focusing effect reduces the
parallel diffusion coefficient
regardless of the sign of the focusing parameter $\xi$.

\section{SUMMARY AND CONCLUSION}
\label{SUMMARY AND CONCLUSION}

In the previous years, much progress has been achieved in the
theoretical description of energetic charged particle transport
in turbulent magnetic field which is superposed on
large scale field. The SPDC
is one of the key parameters 
for modeling particle transport and acceration in the Galaxy and
the solar system
\citep{Schlickeiser2002, Shalchi2009, Shalchi2020}.
In the past, people have found three different 
definitions of the SPDC,
i.e., the displacement variance definition
$\kappa_{zz}^{DV}=\lim_{t\rightarrow t_{\infty}}d\sigma^2/(2dt)$,
the Fick's Law definition
$\kappa_{zz}^{FL}=J/X$ with $X=\partial{F}/\partial{z}$,
and
the TGK definition $\kappa_{zz}^{TGK}=\int_0^{\infty}dt
\langle v_z(t)v_z(0) \rangle$.
For the constant background magnetic field,
the three different definitions of the SPDCs give the same result.
However, some researchers \citep{DanosEA2013,LitvinenkoEA2013,
ShalchiEA2013,LasuikEA2017} found that
the displacement variance definition $\kappa_{zz}^{DV}$
and the TGK definition $\kappa_{zz}^{TGK}$ give different value
for the spatially varying mean magnetic field, and for which, 
the Fick's law definition $\kappa_{zz}^{FL}$
is not equal the displacement variance definition 
$\kappa_{zz}^{DV}$ \citep{wq2019}. 
Thus, the three different definitions of the SPDC are not 
equal one another for focusing field.

In this paper, employing perturbation theory,
starting from the Fokker-Planck equation with the simple 
BGK collision operator we achieve
the EIDF which is different from the one derived 
through the Fourier expansion \citep{GombosiEA1993}.
In addition, with some DIOs, one can not only interconvert 
these EIDFs into each other, but also produces countless new EIDFs.  
Therefore, we get different equations to describe the same physical process.
However, different EIDFs describing the same transport process
should give the same SPDC. 
If one definition of the SPDC is invariant for the DIOs,
it is also an invariance for the different EIDFs, therewith
it is an invariant quantity for different
DME. Therefore, in the present paper 
we explore whether the EIDFs are invariant quantities
for the DIOs.

Using the method of \citet{wq2018}, through 
the DIOs belonging to PzI, PtzI, and PtI operations 
we obtain a limitless variety of the EIDFs 
from the modified Fokker-Planck equation with adiabatic focusing
effect. 
The Fick's law definition $\kappa_{zz}^{FL}$ is invariant
with the DIOs of the PtzI and PtI operations,
but it is not with the PzI operation. 
The displacement variance definition 
$\kappa_{zz}^{DV}$ is invariant not only 
for the PzI and PtzI operations, but also 
for the PtI operation
at least under the special condition.  
The TGK definition $\kappa_{zz}^{TGK}$ 
is the third kind of the SPDC which is invariant 
for the iteration operations. However, 
the TGK definition $\kappa_{zz}^{TGK}$
ignores the effect of the higher-order spatial derivative terms, 
which actually have influence on 
the parallel diffusion coefficient in focusing field.
Therefore, $\kappa_{zz}^{TGK}$ only give the approximate result
in such condition. 
Consequently, the displacement variance definition
$\kappa_{zz}^{DV}$
is more appropriate than 
the other ones. 
Therefore, for data analysis and simulation
we should use $\kappa_{zz}^{DV}$
rather than $\kappa_{zz}^{FL}$ and $\kappa_{zz}^{TGK}$
for focusing field.

\citet{wq2018} derived the formula of $\kappa_{zz}$
and evaluated as
$\kappa_{zz}\approx \kappa_{\parallel 0}
\left(1+2\xi^2/15\right)$.
In this paper, using the method of \citet{wq2018}
we obtain
$\kappa_{tz}\approx 2v\xi/(9D)$ and
$\kappa_{z}\approx v\xi/3$ for the isotropic pitch angle scattering model
$D_{\mu\mu}(\mu)=D(1-\mu^2)$.
Through the formula $\kappa_{zz}^{DV}=
\kappa_{zz}-\kappa_{tz}\kappa_z$,
we find that the displacement variance definition $\kappa_{zz}^{DV}$ 
is approximately equal to $\kappa_{\parallel 0}(1-14\xi^2/45)$, 
where the focusing parameter is $\xi=3\kappa_{\parallel 0}/(vL)$.
This result shows that the corrective factor $-14\xi^2/45$
induced by
the adiabatic focusing effect reduces the
parallel diffusion coefficient
regardless of the sign of $\xi$.

In this work, it is suggested that 
the displacement variance definition 
$\kappa_{zz}^{DV}$ is invariant 
for the PzI and PtzI operations
as well as
for the PtI operation
at least under the special condition. 
Here, the requirement of the special condition 
might not be necessary and will be explored in the future.
In addition, the momentum diffusion with adiabatic focusing 
effect will be also investigated. 

\acknowledgments

We are partly supported by
grant  NNSFC 41874206.

\renewcommand{\theequation}{\Alph{section}-\arabic{equation}}
\begin{appendices}

\section{The accurate formula of $\kappa_{tz}$}
\label{The accurate formula of ktz}

Equation (\ref{Equation of F with g2}) can be rewritten as
\begin{eqnarray}
\frac{\partial{F}}{\partial{t}}
&+&\kappa_z^0\frac{\partial{F}}{\partial{z}}
=\kappa_{zz}^0\frac{\partial^2{F}}{\partial{z^2}}
+\left(\alpha_1
+\alpha_2\right)\frac{\partial^2{F}}{\partial{t}\partial{z}}
-\frac{v}{2}\int_{-1}^{1}d\mu \mu e^{M(\mu)}\int_{-1}^{\mu}d\nu e^{-M(\nu)}
\frac{1}{D_{\nu\nu}(\nu)}\int_{-1}^{\nu}
\frac{\partial^2{g}}{\partial{t}\partial{z}}d\rho
\nonumber \\
&&+\frac{v}{2}\frac{\int_{-1}^{1}d\mu \mu e^{M(\mu)}}
{\int_{-1}^{1}d\mu
e^{M(\mu)}}\int_{-1}^{1}d\mu
e^{M(\mu)}\int_{-1}^{\mu} d\nu
e^{-M(\nu)}
\frac{1}{D_{\nu\nu}(\nu)}
\int_{-1}^{\nu}\frac{\partial^2{g}}{\partial{t}\partial{z}}
d\rho\nonumber\\
&&-\frac{v^2}{4}\int_{-1}^{1}d\mu \mu 
e^{M(\mu)}\int_{-1}^{\mu}d\nu e^{-M(\nu)}
\frac{1}{D_{\nu\nu}(\nu)}
\left(2\int_{-1}^{\nu}d\rho \rho
\frac{\partial^2{g}}{\partial{z^2}}-
\int_{-1}^{1}d\mu \mu
\frac{\partial^2{g}}{\partial{z^2}}\right)\nonumber\\
&&+\frac{v^2}{4}\frac{\int_{-1}^{1}d\mu \mu e^{M(\mu)}}
{\int_{-1}^{1}d\mu
e^{M(\mu) }}\int_{-1}^{1}d\mu
e^{M(\mu)}\int_{-1}^{\mu} d\nu
e^{-M(\nu)}
\frac{1}{D_{\nu\nu}(\nu)}
\left(2\int_{-1}^{\mu}
d\nu \nu \frac{\partial^2{g}}{\partial{z^2}}-
\int_{-1}^{1}d\mu \mu
\frac{\partial^2{g}}{\partial{z^2}}\right).
\label{Equation of F with g2-2}
\end{eqnarray}
with
\begin{eqnarray}
\kappa_z^0&=&v\frac{\int_{-1}^{1}d\mu \mu e^{M(\mu)}}{\int_{-1}^{1}
d\mu e^{M(\mu)}}\\
\kappa_{zz}^0&=&vL\frac{\int_{-1}^{1}d\mu \mu e^{M(\mu)}}{\int_{-1}^{1}
d\mu e^{M(\mu)}}\\
\alpha_1&=&-\frac{v}{2}\int_{-1}^{1}d\mu \mu 
e^{M(\mu)}\int_{-1}^{\mu}d\nu e^{-M(\nu)}
\frac{1}{D_{\nu\nu}(\nu)}\nu\\
\alpha_2&=&\frac{v}{2}\frac{\int_{-1}^{1}d\mu \mu e^{M(\mu)}}
{\int_{-1}^{1}d\mu
e^{M(\mu)}}\int_{-1}^{1}d\mu
e^{M(\mu)}\int_{-1}^{\mu} d\nu
e^{-M(\nu)}
\frac{1}{D_{\nu\nu}(\nu)}\nu
\label{a2}
\end{eqnarray}

It is obvious that in Equation (\ref{Equation of F with g2-2})
there is no $\partial^2 F/(\partial t \partial z)$
in the terms containing $\partial^2 g/\partial z^2$,
but there might be $\partial^2 F/(\partial t \partial z)$
in the terms containing $\partial^2 g/(\partial t \partial z)$.
Therefore, the correction to the coefficient of $\partial^2 F/(\partial t \partial z)$
could only come from the fourth term
\begin{equation}
-\frac{v}{2}\int_{-1}^{1}d\mu \mu e^{M(\mu)}\int_{-1}^{\mu}d\nu e^{-M(\nu)}
\frac{1}{D_{\nu\nu}(\nu)}\int_{-1}^{\nu}
\frac{\partial^2{g}}{\partial{t}\partial{z}}d\rho,
\label{4th term}
\end{equation}
and the fifth term
\begin{equation}
+\frac{v}{2}\frac{\int_{-1}^{1}d\mu \mu e^{M(\mu)}}
{\int_{-1}^{1}d\mu
e^{M(\mu)}}\int_{-1}^{1}d\mu
e^{M(\mu)}\int_{-1}^{\mu} d\nu
e^{-M(\nu)}
\frac{1}{D_{\nu\nu}(\nu)}
\int_{-1}^{\nu}\frac{\partial^2{g}}{\partial{t}\partial{z}}
d\rho
\label{fifth term}
\end{equation}
on the right hand side of Equation (\ref{Equation of F with g2-2}).
Operating $\partial^2 /(\partial t \partial z)$
on Equation (\ref{g2}), we can obtain
\begin{eqnarray}
\frac{\partial^2 g}{\partial t\partial z}
&=&\left(L\frac{\partial^3 F}{\partial t\partial z^2}
-\frac{\partial^2 F}{\partial t\partial z}\right)\left[1-
\frac{2e^{M(\mu)}}{\int_{-1}^{1}
d\mu e^{M(\mu) }}\right]
+e^{M(\mu)}\Bigg\{\int_{-1}^{\mu}
d\nu
e^{-M(\nu)}
\frac{1}{D_{\nu\nu}(\nu)}
\Bigg[\left(\frac{\partial^3 F}{\partial t^2\partial z}\nu
+\int_{-1}^{\nu}
\frac{\partial^3 g}{\partial t^2\partial z}d\rho\right)
\nonumber \\
&&+\frac{v}{2}
\left(2\int_{-1}^{\nu}d\rho \rho
\frac{\partial^3 g}{\partial t\partial z^2}-
\int_{-1}^{1}d\mu \mu
\frac{\partial^3 g}{\partial t\partial z^2}\right)
\Bigg]\nonumber\\
&&-\frac{1}
{\int_{-1}^{1}d\mu
e^{M(\mu) }}\int_{-1}^{1}d\mu
e^{M(\mu)}\int_{-1}^{\mu} d\nu
e^{-M(\nu)}
\frac{1}{D_{\nu\nu}(\nu)}
\Bigg[\left(\frac{\partial^3 F}{\partial t^2\partial z}\nu
+\int_{-1}^{\nu}\frac{\partial^3 g}{\partial t^2\partial z}
d\rho\right)\nonumber\\
&&+\frac{v}{2}
\left(2\int_{-1}^{\mu}
d\nu \nu \frac{\partial^3 g}{\partial t\partial z^2}-
\int_{-1}^{1}d\mu \mu
\frac{\partial^3 g}{\partial t\partial z^2}\right)\Bigg]\Bigg\}.
\label{d2g/dtdz}
\end{eqnarray}
The latter equation can be rewritten as
\begin{eqnarray}
\frac{\partial^2 g}{\partial t\partial z}
&=&\frac{\partial^2 F}{\partial t\partial z}
\left[\frac{2e^{M(\mu)}}{\int_{-1}^{1}d\mu e^{M(\mu)}}-1\right]
+\frac{\partial^3 F}{\partial t\partial z^2}L\left[1-
\frac{2e^{M(\mu)}}{\int_{-1}^{1}
d\mu e^{M(\mu) }}\right]
+e^{M(\mu)}\Bigg\{\int_{-1}^{\mu}
d\nu
e^{-M(\nu)}
\frac{1}{D_{\nu\nu}(\nu)}
\Bigg[\left(\frac{\partial^3 F}{\partial t^2\partial z}\nu
+\int_{-1}^{\nu}
\frac{\partial^3 g}{\partial t^2\partial z}d\rho\right)
\nonumber \\
&&+\frac{v}{2}
\left(2\int_{-1}^{\nu}d\rho \rho
\frac{\partial^3 g}{\partial t\partial z^2}-
\int_{-1}^{1}d\mu \mu
\frac{\partial^3 g}{\partial t\partial z^2}\right)
\Bigg]
-\frac{1}
{\int_{-1}^{1}d\mu
e^{M(\mu) }}\int_{-1}^{1}d\mu
e^{M(\mu)}\int_{-1}^{\mu} d\nu
e^{-M(\nu)}
\frac{1}{D_{\nu\nu}(\nu)}
\Bigg[\Bigg(\frac{\partial^3 F}{\partial t^2\partial z}\nu\nonumber\\
&&+\int_{-1}^{\nu}\frac{\partial^3 g}{\partial t^2\partial z}
d\rho\Bigg)
+\frac{v}{2}
\left(2\int_{-1}^{\mu}
d\nu \nu \frac{\partial^3 g}{\partial t\partial z^2}-
\int_{-1}^{1}d\mu \mu
\frac{\partial^3 g}{\partial t\partial z^2}\right)\Bigg]\Bigg\}.
\label{d2g/dtdz-2}
\end{eqnarray}
From Equation (\ref{d2g/dtdz-2}) we can find that
the term containing $\partial^2 F/(\partial t\partial z)$ is
\begin{equation}
\frac{\partial^2 F}{\partial t\partial z}
\left[\frac{2e^{M(\mu)}}{\int_{-1}^{1}d\mu e^{M(\mu)}}-1\right],
\label{d2g/dtdz-1}
\end{equation}
by substituting which
for $\partial^2 g/(\partial t\partial z)$ 
in formulas (\ref{4th term}) and (\ref{fifth term}), 
we can obtain
$\alpha_3\partial^2 F/(\partial t\partial z)$
with
\begin{equation}
\alpha_3=-\frac{v}{2}\int_{-1}^{1}d\mu \mu e^{M(\mu)}
\int_{-1}^{\mu}d\nu e^{-M(\nu)}
\frac{1}{D_{\nu\nu}(\nu)}\int_{-1}^{\nu}
\left[\frac{2e^{M(\mu)}}{\int_{-1}^{1}d\mu e^{M(\mu)}}-1\right]d\rho,
\label{a3}
\end{equation}
and fifth term $\alpha_4\partial^2 F/(\partial t\partial z)$
with
\begin{equation}
\alpha_4=\frac{v}{2}\frac{\int_{-1}^{1}d\mu \mu e^{M(\mu)}}
{\int_{-1}^{1}d\mu
e^{M(\mu)}}\int_{-1}^{1}d\mu
e^{M(\mu)}\int_{-1}^{\mu} d\nu
e^{-M(\nu)}
\frac{1}{D_{\nu\nu}(\nu)}
\int_{-1}^{\nu}
\left[\frac{2e^{M(\mu)}}{\int_{-1}^{1}d\mu e^{M(\mu)}}-1\right]
d\rho.
\label{a4}
\end{equation}
Combining Equations (\ref{Equation of F with g2-2})-(\ref{a2}) 
and expressions (\ref{a3})-(\ref{a4}),
we can obtain the following formula 
\begin{eqnarray}
\kappa_{tz}=&&\alpha_1+\alpha_2+\alpha_3+\alpha_4\nonumber\\
=&&\frac{v}{2}\frac{\int_{-1}^{1}d\mu \mu e^{M(\mu)}}
{\int_{-1}^{1}d\mu
e^{M(\mu)}}
\int_{-1}^{1}d\mu e^{M(\mu)}\int_{-1}^{\mu} d\nu 
\frac{e^{-M(\nu)}}{D_{\nu\nu}(\nu)}
\left(2\frac{\int_{-1}^{\nu}d\rho e^{M(\rho)}}{\int_{-1}^{1}d\mu e^{M(\mu)}}
-1\right)\nonumber\\
&&-\frac{v}{2}\int_{-1}^{1}d\mu \mu e^{M(\mu)}\int_{-1}^{\mu}d\nu
\frac{e^{-M(\nu)}}{D_{\nu\nu}(\nu)}
\left(2\frac{\int_{-1}^{\nu}d\rho e^{M(\rho)}}{\int_{-1}^{1}d\mu e^{M(\mu)}}
-1\right).
\label{ktz}
\end{eqnarray}
In addition, we can find that the coefficient of the convection term is
\begin{eqnarray}
\kappa_z=\kappa_z^0=v\frac{\int_{-1}^{1}d\mu \mu e^{M(\mu)}}{\int_{-1}^{1}.
d\mu e^{M(\mu)}}
\end{eqnarray}

Similarly, by using the above method
the coefficient of the term $\partial^3{F}/(\partial{t}\partial{z^2})$
can be obtained as
\begin{eqnarray}
\kappa_{tzz}
=&&\frac{v}{2}\frac{\int_{-1}^{1}d\mu \mu e^{M(\mu)}}
{\int_{-1}^{1}d\mu e^{M(\mu)}}
\int_{-1}^{1}d\mu e^{M(\mu)}A
-\frac{v}{2}\int_{-1}^{1}d\mu \mu e^{M(\mu)}A.
\label{ktzz}
\end{eqnarray}
Here, the parameter $A$ is
\begin{eqnarray}
A=\int_{-1}^{\mu}d\nu \frac{e^{-M(\nu)}}{D_{\nu\nu}}
\left(\int_{-1}^{\nu}d\rho B_1 +v\int_{-1}^{\nu}d\rho\rho B_2
-\frac{v}{2}\int_{-1}^{1}d\mu\mu B_2  \right)
\label{A}
\end{eqnarray}
with
\begin{eqnarray}
B_1=&&L\left(1-\frac{2e^{M(\mu)}}{\int_{-1}^{1}d\mu e^{M(\mu)}} \right)
+v e^{M(\mu)} \Bigg[\int_{-1}^{\mu}d\nu \frac{e^{-M(\nu)}}{D_{\nu\nu}}D_1
-J\int_{-1}^{\mu}d\nu \frac{e^{-M(\nu)}}{D_{\nu\nu}}
 \Bigg]\nonumber\\
&&-\frac{v e^{M(\mu)}}{\int_{-1}^1d\mu e^{M(\mu)}}\int_{-1}^1d\mu e^{M(\mu)}
 \Bigg[\int_{-1}^{\mu}d\nu \frac{e^{-M(\nu)}}{D_{\nu\nu}}D_1
-
J\int_{-1}^{\mu}d\nu \frac{e^{-M(\nu)}}{D_{\nu\nu}} \Bigg],
\label{B1}\\
B_2=&&e^{M(\mu)} \Bigg[\int_{-1}^{\mu}d\nu \frac{e^{-M(\nu)}}{D_{\nu\nu}}
D_2
-\frac{1}{\int_{-1}^1d\mu e^{M(\mu)}}\int_{-1}^1d\mu e^{M(\mu)}
\int_{-1}^{\mu}d\nu \frac{e^{-M(\nu)}}{D_{\nu\nu}}
D_2 \Bigg].
\label{B2}
\end{eqnarray}
Here,
\begin{eqnarray}
J&=& \frac{\int_{-1}^{1}d\mu\mu e^{M(\mu)}}{\int_{-1}^1d\mu e^{M(\mu)}},\\
D_1&=&2\frac{\int_{-1}^{\nu}d\rho\rho e^{M(\rho)}}{\int_{-1}^1d\mu 
	e^{M(\mu)}}-\frac{\nu^2-1}{2},\\
D_2&=& 2\frac{\int_{-1}^{\nu}d\rho e^{M(\rho)}}{\int_{-1}^1d\mu e^{M(\mu)}}-1.
\end{eqnarray}
The other cofficients of Equation 
(\ref{equation of F with constant coefficient})
can also be obtained through the same method. 
In general, the higher-order derivative terms have the more complicated coefficients.

\section{To Evaluate $\kappa_{tz}$ 
for the isotropic pitch angle scattering}
\label{Evaluate ktz for isotropic pitch angle scattering}

For the isotropic pitch-angle
scattering model
$D_{\mu\mu}=D(1-\mu^2)$ with constant $D$, \citet{HeEA2014}
showed that 
Equation (\ref{M(mu)})
can be simplified as
\begin{equation}
M(\mu)=\xi (\mu+1)
\label{M(mu) for isotropic model}
\end{equation}
with
\begin{equation}
\xi= \frac{v}{2DL}.
\end{equation}
By using Equation (\ref{M(mu) for isotropic model}) with (\ref{xi}),
we can rewrite Equation (\ref{ktz}) as
\begin{eqnarray}
\kappa_{tz}&=&\frac{v}{2}\frac{\int_{-1}^{1}d\mu \mu e^{\xi\mu}}
{\int_{-1}^{1}d\mu
e^{\xi\mu}}
V_1
-\frac{v}{2}V_2,
\label{ktz with V1 and V2}
\end{eqnarray}
where,
\begin{eqnarray}
&&V_1=\int_{-1}^{1}d\mu e^{\xi\mu}\int_{-1}^{\mu} d\nu W(\nu)
\label{V1},\\
&&V_2=\int_{-1}^{1}d\mu \mu e^{\xi\mu}
\int_{-1}^{\mu}d\nu W(\nu)
\label{V2}
\end{eqnarray}
with
\begin{eqnarray}
W(\mu)=\frac{e^{-\xi\mu}}{D_{\mu\mu}(\mu)}
\left(2\frac{\int_{-1}^{\mu}d\nu e^{\xi\nu}}{\int_{-1}^{1}d\mu e^{\xi\mu}}
-1\right).
\end{eqnarray}
Employing integration in parts for Equations (\ref{V1}) and (\ref{V2}),
we can obtain
\begin{eqnarray}
&&V_1=\frac{e^{\xi}}{\xi}\int_{-1}^{1}d\mu W(\mu)-
\frac{1}{\xi}\int_{-1}^{1}d\mu e^{\xi\mu}W(\mu)
\label{V1 integration in parts},\\
&&V_2=\frac{e^{\xi}}{\xi}\int_{-1}^{1}d\mu W(\mu)-
\frac{1}{\xi}\int_{-1}^{1}d\mu \mu e^{\xi\mu} W(\mu)
-\frac{V_1}{\xi},
\label{V2 integration in parts}\\
&&W(\mu)=\frac{e^{-\xi\mu}}{D_{\mu\mu}(\mu)}
\left(2\frac{e^{\xi\mu}-e^{-\xi}}{e^{\xi}-e^{-\xi}}-1\right).
\label{W in xi}
\end{eqnarray}

From the formula
\begin{eqnarray}
vL\frac{\int_{-1}^{1}d\mu \mu e^{M(\mu)}}
{\int_{-1}^{1}d\mu
e^{M(\mu)}}=\kappa_{\parallel 0} (1+S)
\end{eqnarray}
with
\begin{eqnarray}
S=-\frac{1}{15}\xi^2
+\frac{2}{315}\xi^4+\cdots,
\label{S}
\end{eqnarray}
the following formula can be obtained
\begin{eqnarray}
\frac{\int_{-1}^{1}d\mu \mu e^{M(\mu)}}
{\int_{-1}^{1}d\mu
e^{M(\mu)}}=\frac{\xi}{3} (1+S).
\label{frac with S}
\end{eqnarray}
Combining Equations (\ref{ktz with V1 and V2}) 
with (\ref{V1 integration in parts}), (\ref{V2 integration in parts}),
and (\ref{frac with S}) gives
\begin{eqnarray}
\kappa_{tz}=\frac{v}{2}\left[\left(\frac{1}{3}
+\frac{S}{3}+\frac{1}{\xi^2}-\frac{1}{\xi}\right)e^{\xi}\int_{-1}^1 d\mu W(\mu)
-\left(\frac{1}{3}+\frac{S}{3}+\frac{1}{\xi^2}\right)\int_{-1}^1 d\mu 
e^{\xi\mu}W(\mu)+\frac{1}{\xi}\int_{-1}^1 d\mu \mu e^{\xi\mu}W(\mu)\right].\nonumber\\
\label{ktz with integration in parts}
\end{eqnarray}

For $\xi\ll 1$,  by employing the following formulas
\begin{eqnarray}
e^\xi&=&1+\xi+\frac{1}{2}\xi^2
+\frac{1}{6}
\xi^3+\frac{1}{24}\xi^4+\cdots,
\label{e1}\\
e^{\mu\xi}&=&1+\mu\xi
+\frac{1}{2}(\mu\xi)^2
+\frac{1}{6}(\mu\xi)^3
+\frac{1}{24}(\mu\xi)^4
+\cdots,
\label{e2}
\end{eqnarray}
we find that Equation (\ref{ktz with integration in parts})
becomes
\begin{eqnarray}
\kappa_{tz}\approx\frac{v}{2}\Bigg[&&\frac{1}{2}
\int_{-1}^1 d\mu \left(\mu^2-1\right)W(\mu)+
\frac{\xi}{3}
\int_{-1}^1 d\mu\mu \left(\mu^2-1\right)W(\mu)
+\frac{\xi^2}{24}
\int_{-1}^1 d\mu \left(3\mu^2-1\right)\left(\mu^2-1\right)W(\mu)\nonumber\\
&&+\frac{\xi^3}{90}
\int_{-1}^1 d\mu\mu \left(3\mu^2-2\right)\left(\mu^2-1\right)W(\mu)\Bigg].
\end{eqnarray}
Inserting
Equations (\ref{e1}) and (\ref{e2}) into Equation (\ref{W in xi}), 
we can obtain
\begin{eqnarray}
\kappa_{tz}\approx \frac{2v}{9D}\xi.
\label{ktz for first order xi}
\end{eqnarray}
Similarly, we can also obtain
\begin{eqnarray}
\kappa_{z}\approx \frac{v}{3}\xi. 
\label{kz for first order xi}
\end{eqnarray}

\section{To Determine the sign of $\kappa_{tzz}$}
\label{Determine the sign of ktzz}

In this paper, we only consider very 
weak adiabatic focusing effect, for which
the mean free path of the charged particles
is much less than the characteristic length
of the adiabatic focusing field, i.e.,
$\xi=\lambda/L\ll 1$ with $\lambda=v/(2D)$.
Therefore, the adiabatic focusing effect just has an very little
correction function on the coefficients of 
Equation (\ref{equation of F with constant coefficient}). 
Thus, if certain one coefficient is not equal zero, 
the sign of it cannot be changed by the very weak 
adiabatic focusing effect since which is too weak.
In order to judge the sign of one coefficient, 
one only need to explore it is negtive or positive
for the limit $\xi\rightarrow 0$. 
 In this section,
we only explore the sign of coefficient $\kappa_{tzz}$
(see Equation (\ref{ktzz})).

For the isotropic model
$D_{\mu\mu}=D(1-\mu^2)$ with the positive constant $D$ 
and the limit $\xi\rightarrow 0$, from
Equation (\ref{M(mu) for isotropic model})
we can obtain $M(\mu)\rightarrow 0$ and
therewith
the following results
\begin{eqnarray}
&&e^{M(\mu)}\rightarrow 1,\\
&&e^{-M(\mu)}\rightarrow 1.
\end{eqnarray}
To proceed, using the latter relations, we can find
$J=0$, $D_1=0$, and $D_2= \mu$,
inserting which into Equations (\ref{A}), (\ref{B1}) 
and (\ref{B2}), one can ontain
\begin{eqnarray}
A&\rightarrow&\int_{-1}^{\mu}d\nu \frac{1}{D_{\nu\nu}}
\left(\int_{-1}^{\nu}d\rho B_1 +v\int_{-1}^{\nu}d\rho\rho B_2
-\frac{v}{2}\int_{-1}^{1}d\mu\mu B_2  \right),
\label{simplified A}\\
B_1&\rightarrow&-\frac{v}{2D}\mu,
\label{simplified B1}\\
B_2&\rightarrow&\int_{-1}^{\mu}d\nu \frac{1}{D_{\nu\nu}}\nu
-\frac{1}{2}\int_{-1}^1d\mu \frac{1-\mu}{D_{\mu\mu}}\mu.
\label{simplified B2}
\end{eqnarray}
Combining Equations (\ref{ktzz})
and (\ref{simplified A})
yields
\begin{eqnarray}
\kappa_{tzz}\rightarrow-\frac{v}{2}\int_{-1}^1d\mu\mu\int_{-1}^{\mu}d\nu
\frac{1}{D_{\nu\nu}}
\left(\int_{-1}^{\nu}d\rho B_1+v\int_{-1}^{\nu}d\rho\rho B_2
-\frac{v}{2}\int_{-1}^1d\mu\mu B_2 \right).
\end{eqnarray}
Thereafter, using the integral in parts and employing the formula
$D_{\mu\mu}(\mu)=D(1-\mu^2)$, we can obtain
\begin{eqnarray}
\kappa_{tzz}\rightarrow-\frac{v}{4D}\int_{-1}^1d\mu
\left(\int_{-1}^{\mu}d\nu B_1+v\int_{-1}^{\mu}d\nu\nu B_2
-\frac{v}{2}\int_{-1}^1d\mu\mu B_2 \right),
\label{simplified ktzz-1}
\end{eqnarray}
considering which and 
Equations (\ref{simplified B1}) and
(\ref{simplified B2}), we can find 
\begin{eqnarray}
\kappa_{tzz}\rightarrow&&
-\frac{v}{4D}\int_{-1}^1d\mu\int_{-1}^{\mu}d\nu \left(-\frac{v}{2D}\nu\right)
-\frac{v^2}{4D}\int_{-1}^1d\mu\int_{-1}^{\mu}d\nu\nu
\left(\int_{-1}^{\nu}d\rho \frac{1}{D_{\rho\rho}}\rho
-\frac{1}{2}\int_{-1}^1d\mu \frac{1-\mu}{D_{\mu\mu}}\mu \right)\nonumber\\
&&+\frac{v^2}{8D}\int_{-1}^1d\mu\int_{-1}^1d\mu\mu
\left(\int_{-1}^{\mu}d\nu \frac{1}{D_{\nu\nu}}\nu
-\frac{1}{2}\int_{-1}^1d\mu \frac{1-\mu}{D_{\mu\mu}}\mu \right).
\label{simplified ktzz-2}
\end{eqnarray}

In order to simplify the latter equation, 
one can use the integration by parts for the first terms to derive
\begin{eqnarray}
\kappa_{tzz}\rightarrow&&
-\frac{v^2}{12D^2}
-\frac{v^2}{4D}\int_{-1}^1d\mu\left(\mu-\mu^2\right)
\left(\int_{-1}^{\nu}d\rho \frac{1}{D_{\rho\rho}}\rho
-\frac{1}{2}\int_{-1}^1d\mu \frac{1-\mu}{D_{\mu\mu}}\mu \right)\nonumber\\
&&+\frac{v^2}{8D}\int_{-1}^1d\mu\int_{-1}^1d\mu\mu
\left(\int_{-1}^{\mu}d\nu \frac{1}{D_{\nu\nu}}\nu
-\frac{1}{2}\int_{-1}^1d\mu \frac{1-\mu}{D_{\mu\mu}}\mu \right).
\label{simplified ktzz-3}
\end{eqnarray}
Integrating by parts again for the second term gives
\begin{eqnarray}
\kappa_{tzz}\rightarrow&&
-\frac{v^2}{12D^2}
-\frac{v^2}{12D}\int_{-1}^1d\mu\frac{\mu^4}{D_{\mu\mu}}
-\frac{v^2}{12D}\int_{-1}^1d\mu\frac{1-\mu}{D_{\mu\mu}}\mu
+\frac{v^2}{12D}\int_{-1}^1d\mu 
\frac{1-\mu^2}{D_{\mu\mu}}\mu,
\label{simplified ktzz-3}
\end{eqnarray}
therewith we can obtain at last
\begin{eqnarray}
\kappa_{tzz}\rightarrow-\frac{v^2}{36D^2}.
\label{simplified ktzz-4}
\end{eqnarray}
Because $v^2>0$ and $D^2>0$ the latter equation satisfies
\begin{eqnarray}
\kappa_{tzz}\rightarrow-\frac{v^2}{36D^2}<0,
\label{simplified ktzz-5}
\end{eqnarray}
from which, we can find that
$\kappa_{tzz}$ is negative for the limit $\xi\rightarrow 0$.

\section{The formulas of $\kappa_{ntz}$ with $n=2,3,4,\cdots$}
\label{The formulas of kntz with for all natural number except of n=0 and n=1}

Using the method in \citet{wq2018} we can obtain
the coefficients of the governing equation of $F(z,t)$,
among which, 
the formulas of $\kappa_{ntz}$ with $n=2,3,\cdots$ is shown as
\begin{eqnarray}
\kappa_{ntz}=&&\frac{v}{2}\frac{\int_{-1}^{1}d\mu \mu e^{M(\mu)}}
{\int_{-1}^{1}d\mu e^{M(\mu)}}
\int_{-1}^{1}d\mu e^{M(\mu)}\int_{-1}^{\mu} d\nu \frac{e^{-M(\nu)}}
{D_{\nu\nu}(\nu)}\int_{-1}^{\nu}d\rho \beta_{(n-1)}(\rho)
-\frac{v}{2}\int_{-1}^{1}d\mu \mu e^{M(\mu)}\int_{-1}^{\mu}d\nu
\frac{e^{-M(\nu)}}{D_{\nu\nu}(\nu)}\int_{-1}^{\nu}d\rho \beta_{(n-1)}(\rho)
\label{kntz}\nonumber\\
\end{eqnarray}
with
\begin{eqnarray}
\beta_{(n-1)}&=&e^{M(\mu)}\int_{-1}^{\mu} d\nu \frac{e^{-M(\nu)}}{D_{\nu\nu}(\nu)}
\int_{-1}^{\nu}d\rho \beta_{(n-2)}(\rho)
-\frac{e^{M(\mu)}}{\int_{-1}^{1}d\mu e^{M(\mu)}}\int_{-1}^{1}d\mu  e^{M(\mu)}\int_{-1}^{\mu}d\nu
\frac{e^{-M(\nu)}}{D_{\nu\nu}(\nu)}\int_{-1}^{\nu}d\rho \beta_{(n-2)}(\rho),
\label{betan}\\
\beta_{(n-2)}&=&e^{M(\mu)}\int_{-1}^{\mu} d\nu \frac{e^{-M(\nu)}}{D_{\nu\nu}(\nu)}
\int_{-1}^{\nu}d\rho \beta_{(n-3)}(\rho)
-\frac{e^{M(\mu)}}{\int_{-1}^{1}d\mu e^{M(\mu)}}\int_{-1}^{1}d\mu  e^{M(\mu)}\int_{-1}^{\mu}d\nu
\frac{e^{-M(\nu)}}{D_{\nu\nu}(\nu)}\int_{-1}^{\nu}d\rho \beta_{(n-3)}(\rho),
\label{betan-1}\\
&&\hspace{5cm}\cdots\hspace{3cm}\cdots\nonumber\\
&&\hspace{5cm}\cdots\hspace{3cm}\cdots\nonumber\\
\beta_{(3)}&=&e^{M(\mu)}\int_{-1}^{\mu} d\nu \frac{e^{-M(\nu)}}{D_{\nu\nu}(\nu)}
\int_{-1}^{\nu}d\rho \beta_{(2)}(\rho)
-\frac{e^{M(\mu)}}{\int_{-1}^{1}d\mu e^{M(\mu)}}\int_{-1}^{1}d\mu  e^{M(\mu)}\int_{-1}^{\mu}d\nu
\frac{e^{-M(\nu)}}{D_{\nu\nu}(\nu)}\int_{-1}^{\nu}d\rho \beta_{(2)}(\rho),\\
\label{beta3}
\beta_{(2)}&=&e^{M(\mu)}\int_{-1}^{\mu} d\nu \frac{e^{-M(\nu)}}{D_{\nu\nu}(\nu)}
\int_{-1}^{\nu}d\rho \beta_{(1)}(\rho)
-\frac{e^{M(\mu)}}{\int_{-1}^{1}d\mu e^{M(\mu)}}\int_{-1}^{1}d\mu  e^{M(\mu)}\int_{-1}^{\mu}d\nu
\frac{e^{-M(\nu)}}{D_{\nu\nu}(\nu)}\int_{-1}^{\nu}d\rho \beta_{(1)}(\rho),\\
\label{beta2}
\beta_{(1)}&=&e^{M(\mu)}\int_{-1}^{\mu} d\nu \frac{e^{-M(\nu)}}{D_{\nu\nu}(\nu)}
\left(2\frac{\int_{-1}^{\nu}d\rho e^{M(\rho)}}{\int_{-1}^{1}d\mu e^{M(\mu)}}
-1\right)
-\frac{e^{M(\mu)}}{\int_{-1}^{1}d\mu e^{M(\mu)}}\int_{-1}^{1}d\mu  e^{M(\mu)}\int_{-1}^{\mu}d\nu
\frac{e^{-M(\nu)}}{D_{\nu\nu}(\nu)}
\left(2\frac{\int_{-1}^{\nu}d\rho e^{M(\rho)}}{\int_{-1}^{1}d\mu e^{M(\mu)}}
-1\right).\nonumber\\
\label{beta1}
\end{eqnarray}

Employing the evaluating method of \citet{wq2018}, we can find
\begin{eqnarray}
\kappa_{ttz}&=&-\frac{13}{108}\frac{v}{D^2}\xi,
\label{kttz-value}\\
\kappa_{3tz}&=&\frac{5}{81}\frac{v}{D^3}\xi,\\
\kappa_{4tz}&=&-\frac{121}{3888}\frac{v}{D^4}\xi,\\
\kappa_{5tz}&=&\frac{91}{5832}\frac{v}{D^5}\xi,\\
\kappa_{6tz}&=&-\frac{1093}{139968}\frac{v}{D^6}\xi,\\
\kappa_{7tz}&=&\frac{205}{52488}\frac{v}{D^6}\xi,\\
&&\cdots\hspace{1cm}\cdots,
\end{eqnarray}
from which we can obtian
\begin{eqnarray}
\frac{\kappa_{3tz}}{\kappa_{ttz}}D&=&-0.5125205128,\\
\frac{\kappa_{4tz}}{\kappa_{3tz}}D&=&-0.5041666667,\\
\frac{\kappa_{5tz}}{\kappa_{4tz}}D&=&-0.5013774105,\\
\frac{\kappa_{6tz}}{\kappa_{5tz}}D&=&-0.5004578755,\\
\frac{\kappa_{7tz}}{\kappa_{6tz}}D&=&-0.5001524855,\\
&&\cdots\hspace{1cm}\cdots.\nonumber
\end{eqnarray}
Thus, we can find the following formula
\begin{eqnarray}
\lim_{n\rightarrow\infty}\frac{\kappa_{ntz}}{\kappa_{(n-1)tz}}D&=&-0.5,
\end{eqnarray}
thereafter, the following formula can be obtained
\begin{eqnarray}
\kappa_{ntz}\approx\left(\frac{1}{2D}\right)^{n-2}\kappa_{2tz}.
\label{kntz-k2tz}
\end{eqnarray}

\end{appendices}

\clearpage
\begin{table}[ht]
\begin{center}
\caption{The table of the acronyms}
\label{The table of the acronyms}
 \begin{tabular}{|l|l|}
 \hline

 \hline
the Spatial Parallel Diffusion Coefficient  & SPDC    \\
 \hline
the Equation of the Isotropic Distribution Function $F(z,t)$ & EIDF \\
 \hline
the Derivative Iterative Operation  & DIO  \\
 \hline
Taylor-Green-Kubo & TGK  \\
\hline
the Derivation Method of the EIDF & DME\\
\hline

\end{tabular}
\end{center}
\end{table}

\end{document}